\documentclass[11pt,a4paper]{article}
\usepackage{amsfonts,amssymb,amsmath,amsthm,cite}
\usepackage{graphicx}
\evensidemargin=\oddsidemargin
\DeclareMathOperator{\Ker}{Ker}           
\DeclareMathOperator{\Res}{Res}         
\DeclareMathOperator{\Tr}{Tr}                 

\newtheorem{assumption}{Assumption}[section]
\newtheorem{theorem}[assumption]{Theorem}
\newtheorem{corollary}[assumption]{Corollary}

\newtheorem{lemma}[assumption]{Lemma}
\newtheorem{definition}[assumption]{Definition}
\newtheorem{prop}[assumption]{Proposition}
\newtheorem{remark}[assumption]{Remark}

\newcommand{\A}{\mathcal{A}}              
\newcommand{\Abb}{\mathbb{A}}          
\renewcommand{\a}{\alpha}                    
\newcommand{\B}{\mathcal{B}}              
\newcommand{\C}{\mathbb{C}}              
\newcommand{\CR}{\mathcal{R}}           
\newcommand{\DD}{\mathcal{D}}           
\newcommand{\eps}{\varepsilon}          
\newcommand{\ga}{\gamma}                 
\renewcommand{\H}{\mathcal{H}}          
\newcommand{\half}{{\mathchoice{\thalf}{\thalf}{\shalf}{\shalf}}}
\newcommand{\hideqed}{\renewcommand{\qed}{}} 
\newcommand{\K}{\mathcal{K}}             
\newcommand{\la}{\lambda}                   
\newcommand{\N}{\mathbb{N}}            
\newcommand{\ox}{\otimes}                  

\newcommand{\R}{\mathbb{R}}             

\newcommand{\set}[1]{\{\,#1\,\}}              
\newcommand{\shalf}{{\scriptstyle\frac{1}{2}}} 
\renewcommand{\SS}{\mathcal{S}}        
\newcommand{\thalf}{\tfrac{1}{2}}            
\newcommand{\U}{\mathcal{U}}              
\newcommand{\wh}{\widehat}                  
\newcommand{\wt}{\widetilde}                 
\newcommand{\Z}{\mathbb{Z}}                 
\newcommand{\Uq}{{\mathcal U}_q(su(2))}  

\def\<#1,#2>{\langle#1\,,\,#2\rangle}      

\newcommand{\be}{\begin{enumerate}}
\newcommand{\ee}{\end{enumerate}}

\newbox\ncintdbox \newbox\ncinttbox
\setbox0=\hbox{$-$} \setbox2=\hbox{$\displaystyle\int$}
\setbox\ncintdbox=\hbox{\rlap{\hbox
    to \wd2{\kern-.1em\box2\relax\hfil}}\box0\kern.1em}
\setbox0=\hbox{$\vcenter{\hrule width 4pt}$}
\setbox2=\hbox{$\textstyle\int$}
\setbox\ncinttbox=\hbox{\rlap{\hbox
    to \wd2{\kern-.14em\box2\relax\hfil}}\box0\kern.1em}
\newcommand{\ncint}{\mathop{\mathchoice{\copy\ncintdbox}
    {\copy\ncinttbox}{\copy\ncinttbox}
    {\copy\ncinttbox}}\nolimits}
\newcommand{\up}{{\mathord{\uparrow}}}  
\newcommand{\dn}{{\mathord{\downarrow}}}
\newcommand{\ooh}{{\tfrac{3}{2}}}       
\newcommand{\oh}{{\tfrac{1}{2}}}          
\newcommand{\ssesq}{{\scriptstyle\frac{3}{2}}} 
\newcommand{\sesq}{{\mathchoice{\ooh}{\ooh}{\ssesq}{\ssesq}}} 
\newcommand{\kett}[1]{|#1\rangle\!\rangle}  
\newcommand{\piappr}{\underline{\pi}}         
\newcommand{\ul}[1]{\underline{#1}}             
\newcommand{\ket}[1]{|#1\rangle}                  
\newcommand{\sg}{\sigma}                              

\begin{document}

\thispagestyle{empty}

\begin{center}

CENTRE DE PHYSIQUE TH\'EORIQUE$\,^1$\\
CNRS--Luminy, Case 907\\
13288 Marseille Cedex 9\\
FRANCE\\

\vspace{3cm}

{\Large\textbf{Spectral action on $SU_{q}(2)$}} \\
\vspace{0.5cm}

{\large  B. Iochum$^{1, 2}$, C. Levy$^{1, 2}$
and A. Sitarz$^{3, 4}$} \\

\vspace{1.5cm}

{\large\textbf{Abstract}}
\end{center}

\begin{quote}
The spectral action on the equivariant real spectral triple
over $\A\big(SU_q(2)\big)$ is computed explicitly. Properties of the
differential calculus arising from the Dirac operator are
studied and the results are compared to the commutative case
of the sphere $\mathbb{S}^3$.
\end{quote}

\vspace{2cm}



CPT-P06-2007

\vspace{5cm}

{\small
\noindent $^1$ UMR 6207

-- Unit\'e Mixte de Recherche du CNRS et des
Universit\'es Aix-Marseille I, Aix-Marseille II et de l'Universit\'e
du Sud Toulon-Var

-- Laboratoire affili\'e \`a la FRUMAM -- FR 2291\\
$^2$ Also at Universit\'e de Provence,
iochum@cpt.univ-mrs.fr, levy@cpt.univ-mrs.fr\\
$^3$ Institute of Physics, Jagiellonian University,
Reymonta 4, 30-059 Krak\'ow, Poland, sitarz@if.uj.edu.pl\\
$^4$ Partially supported by Polish Government grants
189/6.PRUE/2007/7;
115/E-343/SPB/6.PR UE/DIE and N 201 1770 33
}

\newpage
\tableofcontents
\section{Introduction}

The quantum group $SU_{q}(2)$ has already a rather long history
of studies \cite{Klimyk} being one of the finest examples of quantum
deformation. This includes an approach via the noncommutative notion
of spectral triple introduced by Connes \cite{Book,ConnesMarcolli}
and various notions of Dirac operators were introduced in
\cite{Bibikov, Goswani, Pal1, Cindex, Pal3}. Finally, a real
spectral triple, which was exhibited in \cite{DLSSV}, is invariant
by left and right action of $\U_{q}(su(2))$ and satisfies almost
all postulated axioms of triples except the commutant and first-order
properties. These, however, remain valid only up to infinitesimal
of arbitrary high order. The last presentation generalizes in
a straightforward way all geometric construction details of the
spinorial spectral triple for the classical three-sphere. In
particular, both the equivariant representation and the symmetries
have a $q \to 1$ proper classical limit.

The goal of this article is to obtain the spectral action defined in
\cite{CC} by
\begin{equation}
\label{action}
\SS(\DD_{A},\Phi,\Lambda):=\Tr \big( \Phi( \DD_{A} /\Lambda) \big)
\end{equation}
where $\DD$ is the Dirac operator, $A$ is a selfadjoint one-form,
$\DD_A= \DD+A+JAJ^{-1}$ and $J$ is the reality operator.
Here, $\Phi$ is any even positive cut-off function which could
be replaced by a step function up to some mathematical difficulties
investigated in \cite{Odysseus}. This means that $\SS$ counts the
spectral values of $\vert \DD_{A} \vert$ less than the mass scale
$\Lambda$. Actually, as shown in \cite{CC1}
\begin{align}
    \SS(\DD_{A},\Phi,\Lambda)  &= \,\sum_{0<k\in Sd^+} \Phi_{k}\,
    \Lambda^{k} \ncint \vert D_{A}\vert^{-k} + \Phi(0) \,
    \zeta_{D_{A}}(0)
+\mathcal{O}(\Lambda^{-1}),\label{formuleaction1}\\
\zeta_{D_{A}}(0)&=\zeta_{D}(0) +\sum_{k=1}^d
\tfrac{(-1)^k}{k}\ncint  (A \DD^{-1})^k. \label{Zeta}
\end{align}
where $D_A = \DD_A + P_A$, $P_A$ the projection on $\Ker \DD_A$,
$\Phi_{k}= \half\int_{0}^{\infty} \Phi(t) \, t^{k/2-1} \, dt$, $d$ is
the spectral dimension of the triple and
$Sd^+$ is the strictly positive part of the dimension
spectrum $Sd$ of $(\A,\H,\DD)$. Here, $Sd^+=Sd=\set{1,2,3}$, so
\begin{align}
\label{formuleaction}
    \SS(\DD_{A},\Phi,\Lambda) \, = \,\sum_{1\leq k \leq 3}
\Phi_{k}\,
    \Lambda^{k} \ncint \vert D_{A}\vert^{-k} + \Phi(0) \,
    \zeta_{D_{A}}(0).
\end{align}
Recall that the tadpole of order $\Lambda^{k}$ is the linear term in
$A\in \Omega^1_{D}(\A)$ in the $\Lambda^{k}$ part of
\eqref{formuleaction}.

Note that there are no terms in $\Lambda^{-k}$, $k>0$ because the
dimension spectrum is bounded below
by 1. This spectral action has been computed on few examples:
\cite{Carminati,CC1,CCM,ConnesMarcolli,MCC,GI2002,
GIV,GIVas,GW,Knecht}.

Here, we compute \eqref{formuleaction} with the main difficulty which
is to control the differential calculus generated by the Dirac
operator. To proceed, we introduce two presentations of one-forms.
The main ingredient is $F=$ sign ($\DD$) which appears to be a
one-form up to $OP^{-\infty}$.

In section 2, we discuss the spectral action of an arbitrary
3-dimensional spectral triple using cocycles.

In sections 3 and 4 we recall the main results on $SU_q(2)$ of
\cite{DLSSV} and show that the full spectral action with reality
operator given by \eqref{formuleaction} is completely determined by
the terms
$$
\ncint A^{q} |\DD|^{-p}, \quad 1\leq q\leq p\leq 3\,.
$$
This question of computation of spectral action was addressed in the
epilogue of \cite{Walter}.

In section 5, we establish a differential calculus up to some ideal
in pseudodifferential operators and apply these results to the
precise computation of previous noncommutative integrals.

Section 6 is devoted to explicit examples, while in next section are
given different comparisons with the commutative case of the 3-sphere
corresponding to $SU(2)$.

\section{Spectral action in $3$-dimension}

\subsection{Tadpole and cocycles}

Let $(\A,\H, \DD)$ be a spectral triple of dimension 3. For
$n\in\N^*$ and
$a_{i} \in \A$, define
\begin{align*}
\phi_n(a_0, \cdots,a_n) & :=\ncint a_0 [\DD, a_1]
\DD^{-1}\cdots[\DD, a_n] \DD^{-1}.
\end{align*}
We also use notational integrals on the universal $n$-forms
$\Omega_u^n(\A)$
defined by
$$
\int_{\phi_{n}}a_{0}da_{1}\cdots
da_{n}:=\phi_{n}(a_{0},a_{1},\cdots,a_{n}).
$$
and the reordering fact that
$(da_{0})a_{1}=d(a_{0}a_{1})-a_{0}da_{1}$.

We use the $b-B$ bicomplex defined in \cite{Book}: $b$ is the
Hochschild coboundary map (and $b'$ is truncated one) defined on
$n$-cochains $\phi$ by
\begin{align*}
b\phi(a_{0},\ldots,a_{n+1})&:=b'\phi(a_{0},\ldots,a_{n+1})+(-1)^{n+1}
\phi(a_{n+1}a_{0},a_{1},\ldots,a_{n}), \\
 b'\phi(a_{0},\ldots,a_{n+1})&:=\sum_{j=0}^n (-1)^{j}
\phi(a_{0},\ldots,a_{j}a_{j+1},\ldots,a_{n+1}).
\end{align*}
Recall that $B_{0}$ is defined on the normalized cochains $\phi_{n}$
by
$$
B_{0}
\phi_{n}(a_{0},a_{1},\ldots,a_{n-1}):=\phi_{n}(1,a_{0},
\ldots,a_{n-1}), \text{ thus }
\int_{\phi_{n}}d\omega=\int_{B_{0}\phi_{n}}\omega \text{ for }
\omega \in \Omega_u^{n-1}(\A).
$$
Then $B:=NB_{0}$, where $N:=1+\lambda+\ldots \lambda^n$ is the cyclic
skewsymmetrizer on the $n$-cochains and $\lambda$ is the cyclic
permutation $\lambda  \phi(a_{0},\ldots,a_{n}):=(-1)^n
\phi(a_{n},a_{0},\ldots,a_{n-1})$.

We will also encounter the cyclic 1-cochain $N\phi_{1}$:
\begin{align*}
N\phi_{1}(a_0,a_{1}) := \phi_1(a_0,a_{1}) - \phi_1(a_{1},a_0) \text{
and }
\int_{N\phi_{1}}a_{0}da_{1}:=N\phi_{1}(a_{0},a_{1}).
\end{align*}

\begin{remark}
    \label{commut}
Assume the integrand of $\ncint$ is in $OP^{-3}$. Since
$[\DD^{-1},a]=-\DD^{-1}[\DD,a]\DD^{-1}\in OP^{-2}$, this commutator
introduces a integrand in $OP^{-4}$ so has a vanishing integral:
under the integral, we can commute $\DD^{-1}$ with all $a\in \A$
and all one-forms.
\end{remark}

\begin{lemma}
\label{cocycle}
We have
\begin{align*}
&\hspace{-5cm}(i) &&  \hspace{-5cm}b \phi_1 = - \phi_2.\\
&\hspace{-5cm}(ii) && \hspace{-5cm}b \phi_2 =0.\\
&\hspace{-5cm}(iii) && \hspace{-5cm}b \phi_3  = 0.\\
&\hspace{-5cm}(iv) && \hspace{-5cm}B \phi_1 =0.\\
&\hspace{-5cm}(v)  && \hspace{-5cm}B_0 \phi_2 =- (1-\lambda) \phi_1.\\
&\hspace{-5cm}(vi) && \hspace{-5cm}bB_{0}\phi_{2}=2
\phi_{2}+B_{0}\phi_{3}.\\
&\hspace{-5cm}(vii) && \hspace{-5cm}B \phi_{2}=0.\\
&\hspace{-5cm}(viii) && \hspace{-5cm}B_{0} \phi_{3}=Nb' \phi_{1}.\\
&\hspace{-5cm}(ix) &&  \hspace{-5cm}B \phi_{3}=3B_{0} \phi_{3}.
\end{align*}
\end{lemma}

\begin{proof}
$(i)$
\begin{align*}
b \phi_1( a_0,a_1,a_2) &= \ncint a_0 a_1 [\DD,a_2] \DD^{-1}
- \ncint a_0 \left(a_1 [\DD,a_2] + [\DD,a_1] a_2 \right) \DD^{-1}
+ \ncint a_2 a_0 [\DD,a_1] \DD^{-1} \\
&= \ncint a_0 [\DD,a_1] \left( \DD^{-1} a_2 - a_2  \DD^{-1} \right)
= - \ncint a_0 [\DD,a_1] \DD^{-1} [\DD, a_2] \DD^{-1} \\
&= - \phi_2(a_0, a_1, a_2)
\end{align*}
where we have used the trace property of the noncommutative integral.

$(ii)$  $b \phi_2( a_0,a_1,a_2,a_3)$
\begin{align*}
\qquad&= \ncint a_0 a_1 [\DD,a_2] \DD^{-1} [\DD,
a_3] \DD^{-1}
- \ncint a_0 \left(a_1 [\DD,a_2] + [\DD,a_1] a_2 \right) \DD^{-1}
[\DD, a_3] \DD^{-1} \\
&\quad + \ncint a_0 [\DD,a_1] \DD^{-1} (a_2 [\DD,a_3] + [\DD,a_2] a_3
) \DD^{-1}
- \ncint a_3 a_0 [\DD,a_1] \DD^{-1} [\DD, a_2] \DD^{-1} \\
&= \ncint a_0 [\DD,a_1]
\left( \DD^{-1} a_2 - a_2 \DD^{-1} \right) [\DD, a_3]  \DD^{-1}
+ \ncint a_0 [\DD,a_1] \DD^{-1} [\DD,a_2]
\left( a_3  \DD^{-1} - \DD^{-1} a_3 \right) \\
&= - \ncint a_0 [\DD,a_1] \DD^{-1} [\DD, a_2] \DD^{-1} [D, a_3]
\DD^{-1} + \ncint a_0 [\DD,a_1] \DD^{-1} [\DD, a_2] \DD^{-1} [D, a_3]
\DD^{-1} \\
&=  0.
\end{align*}

$(iii)$ Using Remark \ref{commut}, we get
$\phi_{3}(a_{0},a_{1},a_{2},a_{3})=\ncint
a_{0}[\DD,a_{1}][\DD,a_{2}][\DD,a_{3}]\vert \DD \vert^{-3}$,
so similar computations as for $\phi_{2}$ gives $b\phi_{3}=0$.

$(iv)$ \quad $B_0 \phi_1( a_0 ) = \ncint  [\DD,a_0] \DD^{-1} =
\ncint \left( \DD a_0 \DD^{-1} -a_0 \right) = 0.$
\begin{align*}
\hspace{-1.8cm} (v) \quad B_0 \phi_2( a_0,a_1) =& \ncint  [\DD,a_0]
\DD^{-1} [\DD,a_1] \DD^{-1}
= \ncint a_0 \DD^{-1} [\DD,a_1] - \ncint a_0 [\DD,a_1] \DD^{-1} \\
=& \ncint a_0 a_1 - \ncint a_0 \DD^{-1} a_1 \DD - \ncint a_0
[\DD,a_1] \DD^{-1}\\
=& - \ncint a_1 [ \DD, a_0] \DD^{-1} - \ncint a_0 [\DD,a_1] \DD^{-1}
= -  \phi_1(a_1,a_0) - \phi_1(a_0, a_1) .
\end{align*}

$(vi)$ Since $-b \lambda
\phi_{1}(a_{0},a_{1},a_{2})=\phi_{1}(a_{2},a_{0}a_{1})-
\phi_{1}(a_{1}a_{2},a_{0})+\phi_{1}(a_{1},a_{2}a_{0})$, one obtains
that
$$
-b \lambda  \phi_{1}(a_{0},a_{1},a_{2})=\ncint
a_{0}a_{1}\DD^{-1}a_{2}\DD+a_{0}\DD^{-1}a_{1}\DD a_{2}
-a_{0}\DD^{-1}a_{1}a_{2}\DD-a_{0}a_{1}a_{2}.
$$
So by direct expansion,
this is equal to $-\ncint a_{0}\DD^{-1}
[\DD,a_{1}]\DD^{-1}[\DD,a_{2}]$ which means that
\begin{align*}
-b\lambda \phi_{1}(a_{0},a_{1},a_{2})&=\ncint [\DD^{-1},a_{0}]
[\DD,a_{1}]\DD^{-1}[\DD,a_{2}]
-a_{0}[\DD,a_{1}]\DD^{-1}[\DD,a_{2}]\DD^{-1}\\
&=-B_{0}\phi_{3}(a_{0},a_{1},a_{2})
-\phi_{2}(a_{0},a_{1},a_{2}).
\end{align*}
Now the result follows from $(i),(v)$.

$(vii)$ $B \phi_{2}=NB_{0} \phi_{2}=-N(1-\lambda) \phi_{1}=0$ since
$N(1-\lambda)=0$.

\begin{align*}
\hspace{-0.6cm} (viii)\quad B_0 \phi_3( a_0,a_1,a_2) =
& \ncint  [\DD,a_0] \DD^{-1} [\DD,a_1] \DD^{-1} [\DD,a_2] \DD^{-1} \\
=& \ncint a_0 \DD^{-1} [\DD,a_1] \DD^{-1} [\DD,a_2]
-  \ncint a_0 [\DD,a_1] \DD^{-1} [\DD,a_2] \DD^{-1} \\
=& \ncint a_0 a_1 \DD^{-1} [\DD,a_2]
-  \ncint a_0 \DD^{-1} a_1 [\DD,a_2]
-  \ncint a_0 [\DD,a_1] \DD^{-1} [\DD,a_2] \DD^{-1}\\
=& \ncint  a_0 a_1 a_2 - \ncint a_0 a_1 \DD^{-1} a_2 \DD
-  \ncint  a_0 \DD^{-1} a_1 \DD a_2 + \ncint a_0 \DD^{-1} a_1 a_2
\DD\\
&\quad -  \ncint  a_0 [\DD,a_1] \DD^{-1} [\DD,a_2] \DD^{-1}\\
=& \ncint  a_0 a_1 a_2 - a_{2}\DD a_{1}a_{0}\DD^{-1}+a_{1}a_{2}\DD
a_{0}\DD^{-1}+a_2 \DD a_0 a_1 \DD^{-1}\\
&\quad - \left( a_{0}\DD a_{1}a_{2}\DD^{-1}-a_{0}\DD
a_{1}\DD^{-1}-a_{0}a_{1}\DD a_{2}\DD^{-1}+a_{0}a_{1}a_{2}\right).
\end{align*}
Expanding $(id + \lambda+\lambda^2)b'\phi_{1}(a_{0},a_{1},a_{2})$, we
recover previous expression.

$(ix)$ Consequence of $(viii)$.
\end{proof}

\subsection{Scale-invariant term of the spectral action}

We know from \cite{CC1} that the scale-invariant term of the action
can be written as
\begin{align}
    \label{constanttterm}
\zeta_{D_{A}}(0)-\zeta_{D}(0) = - \ncint  A \DD^{-1} + \tfrac{1}{2}
\ncint  A \DD^{-1} A
\DD^{-1}
- \tfrac{1}{3} \ncint  A \DD^{-1}  A \DD^{-1}  A \DD^{-1}.
\end{align}
In fact, this action can be expressed in dimension 3 as
contributions corresponding to tadpole and the Yang--Mills
and Chern--Simons actions in dimension 4:
\begin{prop}
    \label{prop:action}
For any one-form $A$,
\begin{align}
    \label{constanttterm1}
\zeta_{D_{A}}(0)-\zeta_{D}(0) =  - \tfrac{1}{2} \int_{N\phi_{1}}A
+ \tfrac{1}{2} \int_{\phi_2} (dA + A^2)
- \tfrac{1}{2} \int_{\phi_3}( A dA + \tfrac{2}{3} A^3).
\end{align}
\end{prop}

To prove this, we calculate now each terms of the action.
\begin{lemma}
      \label{cocycleandoneform}
For any one-form $A$, we have

(i) $\int_{\phi_{2}}dA=\int_{B_{0}\phi_{2}} A=-\int_{\phi_{1}}A
-\int_{\lambda\phi_{1}}A$.

(ii) $\ncint A\DD^{-1}=\int_{\phi_{1}} A=\tfrac 12
\int_{N\phi_{1}}A-\tfrac 12\int_{\phi_{2}}dA$.

(iii) $\ncint  A \DD^{-1} A \DD^{-1} = - \int_{\phi_3} A dA +
\int_{\phi_2} A^2.$

(iv) $\ncint  A \DD^{-1} A \DD^{-1} A \DD^{-1} = \int_{\phi_3} A^3.$
\end{lemma}

\begin{proof}
$(i)$ and $(ii)$ follow directly from Lemma \ref{cocycle} $(v)$.

$(iii)$ With the shorthand $A=a_{i}db_{i}$ (summation on $i$)
\begin{align*}
\ncint  A \DD^{-1} A \DD^{-1} =& \ncint a_0 [\DD,b_0] \DD^{-1}
a_1 [\DD,b_1] \DD^{-1} \\
=&- \int_{\phi_3} A d A + \ncint a_0 [\DD, b_0] a_1 b_1 \DD^{-1}
- \ncint a_0 [\DD, b_0] a_1 \DD^{-1} b_1.
\end{align*}
We calculate further the remaining terms
\begin{align*}
\ncint a_0 [\DD, b_0] a_1 b_1 \DD^{-1} - \ncint a_0 [\DD, b_0] a_1
\DD^{-1} b_1
&= \ncint a_0 \DD b_0 a_1 b_1 \DD^{-1} - \ncint a_0 b_0 \DD a_1 b_1
\DD^{-1} \\
&\qquad- \ncint a_0 \DD b_0 a_1 \DD^{-1} b_1
+ \ncint a_0 b_0 \DD a_1 \DD^{-1} b_1,
\end{align*}
which are compared with
$\int_{\phi_2}A^2=\int_{\phi_{2}}a_{0}(db_{0})a_{1}db_{1}=
\int_{\phi_{2}}a_{0}d(b_{0}a_{1})db_{1}-a_{0}b_{0}da_{1}db_{1}$:
\begin{align*}
\int_{\phi_2} A^2 &= \ncint a_0 [\DD, b_0 a_1] \DD^{-1} [\DD, b_1]
\DD^{-1}
- \ncint a_0  b_0 [\DD, a_1] \DD^{-1} [\DD, b_1] \DD^{-1} \\
&= \ncint a_0 \DD b_0 a_1 b_1 \DD^{-1}
 - \ncint a_0 \DD b_0 a_1 \DD^{-1} b_1 - \ncint a_0 b_0 a_1 \DD b_1
\DD^{-1}
  + \ncint a_0 b_0 a_1 b_1 \\
&\qquad - \ncint a_0  b_0 \DD a_1 b_1 \DD^{-1}
  + \ncint a_0  b_0 \DD a_1 \DD^{-1} b_1 + \ncint a_0  b_0 a_1 \DD
b_1 \DD^{-1}
 - \ncint a_0  b_0 a_1 b_1 \\
&=  \ncint a_0 \DD b_0 a_1 b_1 \DD^{-1}  - \ncint  b_1 a_0 \DD b_0
a_1 \DD^{-1}
- \ncint a_0  b_0 \DD a_1 b_1 \DD^{-1} + \ncint  b_1 a_0  b_0 \DD a_1
\DD^{-1}.
\end{align*}

$(iv)$ Note that
\begin{align*}
\int_{\phi_{3}}A^3&=\int_{\phi_{3}}a_{0}(db_{0})a_{1}(db_{1})a_{2}db_{2}=
\int_{\phi_{3}}a_{0}d(b_{0}a_{1})d(b_{1}a_{2})db_{2}
-a_{0}b_{0}da_{1}d(b_{1}a_{2})db_{2}\\
&\qquad -a_{0}d(b_{0}a_{1}b_{1})d(a_{2}db_{2} +a_{0}b_{0}
d(a_{1}b_{1})da_{2}db_{2}\\
&=\ncint
a_{0}[\DD,b_{0}a_{1}]\DD^{-1}[\DD,b_{1}a_{2}]\DD^{-1}[\DD,b_{2}]\DD^{-1}
-a_{0}b_{0}[\DD,a_{1}]\DD^{-1}[\DD,b_{1}a_{2}]\DD^{-1}[\DD,b_{2}]\DD^{-1}\\
&\qquad-a_{0}[\DD,b_{0}a_{1}b_{1}]\DD^{-1}[\DD,a_{2}]\DD^{-1}[\DD,b_{2}]\DD^{-1}
+a_{0}b_{0}[\DD,a_{1}b_{1}]\DD^{-1}[\DD,a_{2}]\DD^{-1}
[\DD,b_{2}]\DD^{-1}.
\end{align*}
Summing up the first two terms and the last two ones gives
$$
\int_{\phi_{3}}A^3=\ncint a_{0}[\DD,b_{0}]a_{1}
\DD^{-1}[\DD,b_{1}a_{2}]\DD^{-1}[\DD,b_{2}]\DD^{-1}
-a_{0}[\DD,b_{0}]a_{1}b_{1}\DD^{-1}[\DD,a_{2}]\DD^{-1}[\DD,b_{2}]\DD^{-1}.
$$
Using Remark \ref{commut}, we can commute
under the integral $\DD^{-1}$ with all $a\in \A$ and similarly
$$
\ncint  A \DD^{-1} A \DD^{-1} A \DD^{-1} =\ncint
a_{0}[\DD,b_{0}]a_{1}\DD^{-1}[\DD,b_{1}]a_{2}\DD^{-1}[\DD,b_{2}]\DD^{-1}
$$
which proves $(iv)$.
\end{proof}

We deduce Proposition \ref{prop:action} from (\ref{constanttterm})
using the previous lemma.


\section{The $SU_{q}(2)$ triple}

\subsection{The spectral triple}

We briefly recall the main facts of the real spectral triple
$\big(\A(SU_q(2)),\H,\DD\big)$ introduced in \cite{DLSSV}, see also
\cite{Cindex,Pal1,Pal2}.
\vspace{0.5cm}

{\it The algebra:}

Let $\A := \A(SU_q(2))$ be
the $^*$-algebra generated polynomially by $a$ and~$b$,
subject to the following commutation rules with $0 < q < 1$:
\begin{eqnarray}
    \label{defrule}
&ba = q \,ab,  \qquad  b^*a = q\,ab^*, \qquad bb^* = b^*b,
&a^*a + q^2 \,b^*b = 1,  \qquad  aa^* + bb^* = 1\, .
\end{eqnarray}

\begin{lemma}
For any representation $\pi$ of $\A$, 
\begin{align*}
& Spect\big(\pi(bb^*)\big) = \set{0,q^{2k} \,\vert \,\vert k\in \N}
\text{ or } \pi(b)=0,\\
& Spect \big(\pi(aa^*)\big) =  \set{1,1-q^{2k} \,\vert \, k\in \N}
\text{ or $\pi(b)=0$ and $\pi(a)$ is a unitary}.
\end{align*}
\end{lemma}

\begin{proof} \cite{HL} Since
$\set{0}\cup\sigma\big(\pi(aa^*)\big)=\set{0} \cup
\sigma\big(\pi(a^*a)\big)$,  we get 
\begin{align}
   \label{spectre}
\set{1} \cup B=\set{1} \cup q^2B
\end{align} 
if $B:=\sigma\big(\pi(bb^*)\big)$. Since $0 \leq \pi(bb^*) \leq 1$,
so $B$ is a closed subset of $ [0,1]$. Assume $b \neq 0$. 

Let $s:=\sup (B)$ and suppose $s\neq 1$. Then $s=q^2x$ where $x\in
B$. Thus $s=q^2x<x\leq s$ gives $s=0$ and the contradiction $b=0$,
thus $1\in B$. Similar argument for $\inf (B)$ implies $0 \in B$.

Let $C:=\set{0, q^{2k} \, \vert \, k \in \N} \subset B$ and assume $B
\backslash C \neq \emptyset$. Then $B \backslash C=(q^2B) \backslash
C$ by \eqref{spectre} and this is equal to $q^2(B\backslash C)$ since
$q^{-2}>1$. If $s:= \sup (B \backslash C)$, then $s= \lim_n (q^2
x_n)$ where $x_n \in B \backslash C$ and $s=q^2 \lim_n x_n \leq q^2
s$ implying $s=0$. But $B \backslash C \subset \set{0}$ yields a
contradiction, so $B \backslash C=\emptyset$.
\end{proof}
This lemma is interesting since it shows the appearance of
discreteness for $0 \leq q<1$ while for $q=1$, $SU_q(2)=SU(2) \simeq
\mathbb{S}^3$ and the spectrum of the commuting operator $\pi(aa^*)$
and $\pi(bb^*)$ are equal to $[0,1]$. Moreover, all foregoing results
on noncommutative integrals will involve $q^2$ and not $q$. 

\vspace{0.5cm}
Any element of $\A$ can be
uniquely decomposed as a
linear combination of terms of the form
$a^\a b^\beta {b^{*}}^\gamma$ where $\a\in \Z$,
$\beta, \gamma \in \N$, with the convention
$$
a^{-\vert \a\vert} := {a^*}^{\vert \a \vert}.
$$

{\it The spinorial Hilbert space:}

$\H=\H^{\up} \oplus \H^{\dn}$ has an orthonormal
basis consisting of vectors $\ket{j\mu n\up}$
with $j = 0,\half,1,\dots$,
$\mu = -j,\dots,j$ and $n = -j^+,\dots,j^+$, together with
$\ket{j\mu n\dn}$ for $j=\half,1,\dots$, $\mu = -j,\ldots,j$ and
$n = -j^-,\dots,j^-$ (here $x^{\pm}:=x\pm \half$).

It is convenient to use a vector notation, setting:
\begin{equation}
    \label{eq:kett-defn}
\kett{j\mu n} :=\genfrac{(}{)}{0pt}{1}{\, \ket{j\mu n\up}\,}
{\,\ket{j\mu n\dn}\,}
\end{equation}
and with the convention that the lower component is zero when
$n = \pm(j + \half)$ or $j = 0$.
\vspace{0.5cm}

{\it The representation $\pi$ and its approximate $\ul \pi$:}

It is known that representation theory of $SU_q(2)$ is similar to
that of $SU(2)$ \cite{Woronowicz}. The representation $\pi$ given in
\cite{DLSSV} is:
\begin{align}
  \label{eq:reprexact}
\pi(a)\, \kett{j\mu n } &:= \alpha^+_{j\mu n}\,\kett{j^+ \mu^+ n^+}
+\alpha^-_{j\mu n}\,\kett{j^- \mu^+ n^+},\nonumber \\
\pi(b)\, \kett{j\mu n } &:= \beta^+_{j\mu n}\,\kett{j^+ \mu^+ n^-}
+\beta^-_{j\mu n}\,\kett{j^- \mu^+ n^-},\nonumber\\
\pi(a^*)\, \kett{j\mu n } &:= \tilde{\alpha}^+_{j\mu n}\,\kett{j^+
\mu^- n^-}+\tilde{\alpha}^-_{j\mu n}\,\kett{j^- \mu^- n^-},\nonumber
\\
\pi(b^*)\, \kett{j\mu n } &:= \tilde{\beta}^+_{j\mu n}\,\kett{j^+
\mu^- n^+}+\tilde{\beta}^-_{j\mu n}\,\kett{j^- \mu^- n^+}
\end{align}
where
\begin{align*}
\alpha^+_{j\mu n}&:=\sqrt{ q^{\mu +n-1/2} [j+\mu+1] }
\left( \begin{array}{cc}
q^{-j-1/2}\tfrac{ \sqrt{[j+n+3/2]}}{[2j+2]} &   0\\
q^{1/2}\tfrac{ \sqrt{[j-n+1/2]}}{[2j+1][2j+2]} &
q^{-j}\tfrac{\sqrt{[j+n+1/2]}}{[2j+1]}  
\end{array} \right),\\
\alpha^-_{j\mu n}&:=\sqrt{ q^{\mu +n+1/2} [j-\mu] }
\left( \begin{array}{cc}
q^{j+1}\tfrac{ \sqrt{[j-n+1/2]}}{[2j+1]} &   -q^{1/2}
\tfrac{\sqrt{[j+n+1/2]}}{[2j][2j+1]}\\
0 & q^{j+1/2}\tfrac{\sqrt{[j-n-1/2]}}{[2j]}  \\
\end{array} \right),\\
\beta^+_{j\mu n}&:=\sqrt{ q^{\mu +n-1/2} [j+\mu+1] }
\left( \begin{array}{cc}
\tfrac{ \sqrt{[j-n+3/2]}}{[2j+2]} &   0\\
-q^{-j-1}\tfrac{ \sqrt{[j+n+1/2]}}{[2j+1][2j+2]} &
q^{-1/2}\tfrac{\sqrt{[j-n+1/2]}}{[2j+1]}  
\end{array} \right),\\
\beta^-_{j\mu n}&:=\sqrt{ q^{\mu +n-1/2} [j-\mu] }
\left( \begin{array}{cc}
- q^{-1/2}\tfrac{\sqrt{[j+n+1/2]}}{[2j+1]} &   -q^{j}\tfrac{
\sqrt{[j-n+1/2]}}{[2j][2j+1]} \\
0 & -\tfrac{\sqrt{[j+n-1/2]}}{[2j]}  
\end{array} \right)
\end{align*}
with $\tilde{\alpha}^{\pm}_{j\mu n}:=(\alpha^{\mp}_{j^{\pm}\mu^-
n^-})^*$, $\tilde{\beta}^{\pm}_{j\mu n}:=(\beta^{\mp}_{j^{\pm}\mu^-
n^+})^*$ and with the q-number of $\alpha \in \R$ be defined as
$$
[\a]:=\tfrac{q^\a-q^{-\a}}{q-q^{-1}}.
$$
For the purpose of this paper it is sufficient to use the approximate
spinorial $^*$-representation $\piappr$ of $SU_q(2)$ presented in
\cite{SDLSV,DLSSV} instead of the full spinorial one $\pi$. 

This approximate representation is
$$
\piappr(a) := {a}_+ + {a}_-, \quad \piappr(b) := {b}_+ + {b}_-
$$
with the following definitions:
\begin{align}
    \label{eq:rpnappr}
{a}_+ \,\kett{j\mu n}
&:= q_{j^++\mu^+} \,
\genfrac{(}{)}{0pt}{1}{q_{j^+ +n^+ +1} \qquad
0}{\quad 0 \quad \quad\quad  q_{j^+ +n}}\,\kett{j^+ \mu^+
n^+}
,\nonumber\\
{a}_- \,\kett{j\mu n}
&:= q^{2j+\mu+n+\half} \, \genfrac{(}{)}{0pt}{1}{q \quad  0}{0\quad 1
}\,\kett{j^- \mu^+ n^+},\nonumber\\
{b}_+ \,\kett{j\mu n} &:= q^{j+n-\half} q_{j^+ +\mu^+} \,
\genfrac{(}{)}{0pt}{1}{q
\quad  0}{0\quad 1 }\,\kett{j^+ \mu^+ n^-},\nonumber\\
{b}_- \,\kett{j\mu n}
&:= -q^{j+\mu} \,
\genfrac{(}{)}{0pt}{1}{q_{j^+ + n}  \quad
0\quad}{\quad0  \quad q_{j^- + n}}\,\kett{j^- \mu^+
n^-}.
\end{align}
All disregarded terms are trace-class and do not influence residue
calculations. More precisely, $\pi(x)-\piappr(x) \in \K_{q}$ where
$\K_{q}$ is the principal ideal generated by the operators
\begin{align}
\label{J_q}
J_{q}\, \kett{j\mu n}:= q^j\,\kett{j\mu n}.
\end{align}
Actually, $\K_{q}$ is independent of $q$ and is contained in all
ideals of operators such that $\mu_n=o(n^{-\a})$  (infinitesimal of
order $\a$) for any $\a>0$, and $\K_{q} \subset OP^{-\infty}$.

\vspace{0.5cm}

We define the alternative orthonormal basis $v^{j\up}_{m,l}$ and
$v^{j\dn}_{m,l}$ and the vector notation 
$$
v^j_{m,l}:=\genfrac{(}{)}{0pt}{1}{\,v^{j\up}_{m,l}\,}{\,v^{j\dn}_{m,l}\,}
\text{ where } v^{j\up}_{m,l}:=\ket{j,m-j,l-j^{+},\up},\quad
v^{j\dn}_{m,l}:=\ket{j, m-j, l-j^{-},\dn}.
$$
Here $j\in \half \N$, $0\leq m \leq 2j$, $0\leq l\leq 2j+1$
and $v^{\dn,j}_{m,l}$ is zero whenever $j=0$ or $l = 2j$ or $2j+1$.
The interest is that now, the operators $a_{\pm}$ and $b_{\pm}$
satisfy the simpler relations
\begin{align}
a_{+}\, v^j_{m,l} &= q_{m+1} \,q_{l+1}\, v^{j^+}_{m+1,l+1}\,,\quad
a_-\,v^j_{m,l} = q^{m+l+1}\, v^{j^-}_{m,l} \,,\nonumber\\
b_+\, v^j_{m,l} &= q^{l}\,q_{m+1}\, v^{j^+}_{m+1,l} \,,\qquad \quad
b_-\, v^j_{m,l} = -q^{m}\,q_{l}\,  v^{j^-}_{m,l-1}\,.\label{defonv}
\end{align}
Thus
\begin{align}
a_{+}^*\, v^j_{m,l} &= q_{m} \,q_{l}\, v^{j^-}_{m-1,l-1}\,,\qquad
a_{-}^*\,v^j_{m,l} = q^{m+l+1}\, v^{j^+}_{m,l} \,,\nonumber\\
b_{+}^*\, v^j_{m,l} &= q^{l}\,q_{m}\, v^{j^-}_{m-1,l} \,,\qquad \quad
b_{-}^*\, v^j_{m,l} = -q^{m}\,q_{l+1}\,  v^{j^+}_{m,l+1}\,
\label{defadjointonv}.
\end{align}
Moreover, we have
\begin{align}
    \label{eq:commutrulefora,b}
&a_- a_+ = q^{2}\, a_+ \, a_-\,, \quad b_-b_{+}=q^2\,b_+b_-\,, \qquad
b_+a_+ = q\,a_+b_+\,,  \quad b_-a_{-} = q \, a_-b_-\,,\nonumber\\
&a_{-}^*a_{+}=q^2\, a_{+}a_{-}^*\,, \quad a_{-}^*a_{-}=
a_{-}a_{-}^*\,, \ \ \qquad a_{-}^*b_{+}=q\, b_{+}a_{-}^*\,,
\quad a_{-}^*b_{-}=q\,b_{-}a_{-}^*\,,\nonumber\\
&a_{+}^*a_{-}=q^2\,a_{-}a_{+}^*\,, \quad b_{-}^{*}b_{+}=
b_{+}b_{-}^*\,,\ \ \   \qquad b_{-}^*a_{+}=q\,a_{+}b_{-}^*\,,
\quad a_{-}b_{+}=q\, b_{+}a_{-}\,.
\end{align}
Note for instance that
\begin{align*}
&a_{+}a_{+}^* \, v^j_{m,l}=q_{m}^2q_{l}^2\, v^j_{m,l}\,, \quad
a_{+}^*a_{+}\, v^j_{m,l}=q_{m+1}^2q_{l+1}^2\, v^j_{m,l}\,,\\
& b_{+}b_{+}^* \, v^j_{m,l}=q^{2l}q_{m}^2\, v^j_{m,l}\,,
\quad b_{+}^*b_{+} \, v^j_{m,l}=q^{2l}q_{m+1}^2\, v^j_{m,l}\,,
\end{align*}
so applied to $v^j_{m,l}$, we get the first relation (and similarly
for the others)
\begin{align}
& a_{+}^{*}a_{+}-q^{2}\,a_{+}a_{+}^{*}+q^2\,(b_{+}^{*}b_{+}
-b_{+}b_{+}^{*})=1-q^2,\label{astuce1}\\
& a_{+}a_{+}^*+a_{-}a_{-}^*+b_{+}b_{+}^*+b_{-}b_{-}^*=1,
\label{astuce1'}\\
& a_{+}^*a_{+}+a_{-}^*a_{-}+q^2\,(b_{+}^*b_{+}+b_{-}^*b_{-})=1,
\label{astuce1''}\\
& a_{-}^{*}a_{-}-q^{2}\,a_{-}a_{-}^{*}+q^2\,b_{-}^{*}b_{-}
-q^2\, b_{-}b_{-}^{*}=0,\label{astuce2}\\
& a_{+}a_{-}^*+b_{-}^*b_{+}=0,
\hspace{3cm} a_{-}^*a_{+}+q^2\,b_{-}^*b_{+}=0, \label{astuce3}\\
& a_{-}a_{+}^*+b_{+}^*b_{-}=0,
\hspace{3cm} a_{+}^*a_{-}+q^2\,b_{+}^*b_{-}=0,\label{astuce4}\\
& b_{+}b_{+}^{*}-b_{+}^{*}b_{+}+b_{-}b_{-}^{*}
-b_{-}^{*}b_{-}=0, \label{astuce5}\\
& q\,a_{+}b_{-}-b_{-}a_{+}+q\, a_{-}b_{+}-b_{+}a_{-}=0.
\label{astuce6}
\end{align}

\vspace{0.5cm}

{\it And two others:}

Note that we also use two other infinite
dimensional
$^*$-representations $\pi_{\pm}$ of $\A$ on $\ell^2(\N)$ defined as
follows
on the orthonormal basis $\set{\eps_{n}:n\in\N}$ of $\ell^2(\N)$ by
\begin{align}
 \label{eq:pi-pm}
\pi_\pm(a) \,\eps_n &:= q_{n+1}\,\eps_{n+1},  \qquad
\pi_\pm(b) \,\eps_n := \pm q^n \,\eps_n,\\
q_{n}&:=\sqrt{1-q^{2n}}.\nonumber
\end{align}
These representations are irreducible but not faithful since for
instance $\pi_{\pm}(b-b^*)=0$.

\vspace{0.5cm}

{\it The Dirac operator:}

It is chosen the same as in the
classical case of a 3-sphere with the round metric:
\begin{align}
\DD \,\kett{j\mu n} :=
\genfrac{(}{)}{0pt}{1}{2j + \sesq \quad\,\,\,0}{\,0 \quad -2j
- \half} \, \kett{j\mu n},  \label{eq:dirac}
\end{align}
which means, with our convention, that
$\DD\,v^j_{ml}=\genfrac{(}{)}{0pt}{1}{2j + \sesq \quad\,\,\,0}{\,0
\quad -2j- \half} \, v^j_{ml}$. 
Note that this operator is asymptotically diagonal with linear
spectrum and

\centerline{ the eigenvalues $2j+\tfrac12$ for $j\in \tfrac12 \N$,
have
multiplicities $(2j+1)(2j+2)$,}
\centerline{the eigenvalues $-(2j+\tfrac12)$ for $j\in \tfrac12
\N^*$, have
multiplicities $2j(2j+1)$.}

So this Dirac operator coincide exactly with the classical one on the
3-sphere (see \cite{Bar, Homma}) with a gap around 0.

Let $\DD=F\vert \DD \vert$ be the polar decomposition of $\DD$, thus
\begin{align}
\vert \DD \vert \,\kett{j\mu n} &=
\genfrac{(}{)}{0pt}{1}{d_{j^+}\,\,\,0}{\,0 \quad d_j} \,
\kett{j\mu n}, \quad d_j:=2j+\half \,, \label{eq:absdirac}
\\
F \,\kett{j\mu n} &=\genfrac{(}{)}{0pt}{1}{1 \quad\,\,\,0}
{\,0 \quad -1} \, \kett{j\mu n}, \label{eq:defF}
\end{align}
and it follows from \eqref{eq:rpnappr} and \eqref{eq:defF} that
\begin{align}
    \label{Fcommutes}
F \text{ commutes with } a_{\pm},\, b_{\pm}.
\end{align}

\quad

{\it The reality operator:}

This antilinear operator $J$ is defined on the basis of $\H$ by
\begin{align}
  \label{DefJ}
J \,\vert j,\mu, n,\up \rangle:=i^{2(2j+\mu+n)}\,  \vert j,-\mu,- n,
\up\rangle, \qquad
J \,\vert j,\mu, n,\dn \rangle:=i^{2(2j-\mu-n)}\,  \vert j,-\mu,- n,
\dn\rangle
\end{align}
thus it satisfies
\begin{align*}
&J^{-1} = -J=J^* \text{ and } \DD J=J\DD, \\
&J  \, v^{j\up}_{m,l}= i^{2(m+l)-1}v^{j\up}_{2j-m,2j+1-l} \,,\qquad
J  \, v^{j\dn}_{m,l}= i^{-2(m+l)+1}v^{j\dn}_{2j-m,2j-1-l}\,.
\end{align*}
\vspace{0.5cm}

{\it The Hopf map $r$}

For the explicit calculations of residues, we need
a $^*$-homomorphism $r:X \rightarrow \pi_{+}(\A) \ox \pi_{-}(\A)$
defined by the tensor product in the sense of Hopf algebras of
representations $\pi_{+}$ and $\pi_{-}$:
\begin{align}
    \label{eq:r}
r(a_+) &:= \pi_+(a) \ox\pi_-(a), &
r(a_-) &:= -q\,\pi_+(b) \ox \pi_-(b^*),\nonumber\\
r(b_+) &:= - \pi_+(a) \ox \pi_-(b), &
r(b_-) &:= - \pi_+(b) \ox \pi_-(a^*).
\end{align}
In fact, $\A$ is a Hopf $^*$-algebra under the coproduct
$\Delta(a):=a\otimes a -q\, b\otimes b^{*},\, \Delta(b):= a\otimes
b+b\otimes a^*$. These homomorphisms appeared in \cite{Woronowicz}
with the translation $\alpha \leftrightarrow a^{*},\, \gamma
\leftrightarrow -b$.
In particular, if
$U:=\genfrac{(}{)}{0pt}{1}{a \quad \,\,\,\,\,\,b}{-qb^* \quad a^*}$
is the canonical generator of the $K_{1}(\A)$-group
$(\Delta a,\, \Delta b)=
(a,\,b) \dot{\otimes}\, U$ where the last $\dot{\otimes}$
means the matrix product of tensors of components.

\vspace{0.5cm}

{\it The grading:}

According to the shift $j\rightarrow j^{\pm}$
appearing in formulae
\eqref{defonv}, \eqref{defadjointonv}, we get a $\Z$-grading on $X$
defined by the
degree $+1$ on
$a_{+},b_{+},{a_{-}}^*,{b_{-}}^*$ and $-1$ on
$a_{-},b_{-},{a_{+}}^*,{b_{+}}^*$.

Any operator $T\in X$ can be (uniquely) decomposed as
$T=\sum_{j\in J\subset \Z} T_j$ where $T_j$ is homogeneous of degree
$j$.

For $T \in X$, $T^\circ$ will denote the $0$-degree part
of $T$ for this grading and by a slight abuse of notations, we write
$r(T)^\circ$ instead of $r(T^\circ)$.
\vspace{0.5cm}

{\it The symbol map:}

We also use the $^*$-homomorphism
$\sg$: $\pi_{\pm}(\A) \to C^{\infty}(S^1)$ defined for $z \in S^1$ on
the generators
by
$$
\sg\big(\pi_{\pm}(a)\big)(z) := z, \;\;\;
\sg\big(\pi_{\pm}(a^*)\big)(z) := \bar z, \;\;\;
\sg\big(\pi_{\pm}(b)\big) (z)=
\sg\big(\pi_{\pm}(b^*)\big)(z):=0.
$$
The application $(\sg \otimes \sg) \circ r $ is defined on $X$ (and
so on $\B$) with values in
$C^{\infty}(S^1) \otimes C^{\infty}(S^1)$.
\vspace{0.5cm}

We define
$$
dT:=[\DD,T] \text{ and } \delta(T):=[\vert \DD \vert, T].
$$
\begin{lemma}
    \label{commutateur}
$a_{\pm}$, $b_{\pm}$ are bounded operators on $\H$ such that for all
$p\in \N$,

(i) $\delta(a_\pm) = \pm a_\pm \, , \quad \delta(b_\pm)=\pm b_\pm\, ,$

(ii) $\delta^p(\ul\pi(a))= a_+ + (-1)^p a_- \, , \quad
\delta^p(\ul\pi(b))= b_+ + (-1)^p b_- \, , $

(iii) $\delta(a_\pm^p) = \pm p\, a_\pm^p \, , \quad \delta(b_\pm^p) =
\pm p\, b_\pm ^p\, .$
\end{lemma}

\begin{proof}
$(i)$ By definition,
$a_\pm \, \kett{j\mu n} =\genfrac{(}{)}{0pt}{1}{\alpha_{\pm}\quad
0\,}{\,0 \quad\beta_{\pm}} \,\kett{j^\pm \mu^+ n^+}$ where the numbers
$\a_\pm$ and $\beta_\pm$ depend on $j$, $\mu$, $n$ and $q$, so we get
by \eqref{eq:absdirac}
\begin{align*}
\delta(a_\pm)  \kett{j\mu n} &=\genfrac{(}{)}{0pt}{1}{(d_{j^{+\pm}})
\a_\pm\quad 0\,}{\quad0 \qquad d_{j^\pm}\,\beta_{\pm}} \,
\kett{j^\pm \mu^+ n^+} -
\genfrac{(}{)}{0pt}{1}{(d_{j^{+}})
\a_\pm\quad 0\,}{\quad0 \qquad d_{j}\,\beta_{\pm}} \,\kett{j^\pm \mu^+
n^+} \\
&=  \genfrac{(}{)}{0pt}{1}{\pm\alpha_{\pm}\quad
0\,}{\,0 \quad \pm \beta_{\pm}} \, \kett{j^\pm \mu^+
n^+}=\pm a_\pm  \,\kett{j\mu n}
\end{align*}
and similar proofs for $b_\pm$.

$(ii)$ and $(iii)$ are straightforward consequences of $(i)$ and
definition of $\ul \pi$.
\end{proof}

We note

$\quad \B$ the $^*$-subalgebra of $\B(\H)$ generated
by the operators in $\delta^k(\pi(\A))$ for all $k\in \N$,

$\quad \Psi_{0}^{0}(\A)$ the algebra generated by $\delta^{k}\big(
\pi(\A)\big)$ and $\delta^{k}([\DD,\pi(\A)])$ for all $k\in \N$,

$\quad X$ the $^*$-subalgebra of $\B(\H)$ algebraically generated
by the set $\set{a_{\pm},b_{\pm}}$.

\begin{remark}
\label{pseudodiff}
By Lemma \ref{commutateur}, we see that, modulo $OP^{-\infty}$, $X$
is equal to $\B$ and in particular contains $\pi(\A)$.

Using \eqref{Fcommutes}, we get that
$\B\subset \Psi_{0}^0(\A) \subset \text{algebra generated by
$\B$ and $\B F$}$.
\end{remark}
Note that, despite the last inclusion, $F$ is not a priori
in $\Psi_{0}^0(\A)$.

\quad

\subsection{The noncommutative integrals}

Recall that for any pseudodifferential operator $T$,
$\ncint T:=\underset{s=0}{\Res} \,\zeta_\DD^T(s)$ where
$\zeta_\DD^T(s):=\Tr(T \vert \DD \vert^{-s})$.

\begin{theorem}
    \label{Theo}
The dimension spectrum (without reality structure given by $J$) of
the spectral triple
$\big(\A(SU_q(2)),\H,D\big)$ is simple and equal to $\{1,2,3\}$. 

Moreover,  the corresponding residues for
$T \in \B$ are
\begin{align*}
&\ncint T |\DD|^{-3} = 2(\tau_1 \ox \tau_1) \bigl(r(T)^\circ \bigr),\\
&\ncint T |\DD|^{-2}= 2 \bigl(\tau_1 \ox \tau_0 +
\tau_0 \ox \tau_1 \bigr) \bigl(r(T)^\circ\bigr),\\
&\ncint T |\DD|^{-1}= \big(2 \, \tau_0 \ox \tau_0 - \tfrac{1}{2}
\, \tau_1 \ox \tau_1\big)\bigl(r(T)^\circ\bigr),\\
&\ncint F\,T |\DD|^{-3} =0,\\
&\ncint F\,T |\DD|^{-2}= 0,\\
&\ncint F\,T |\DD|^{-1}=\big(\tau_{0} \ox \tau_{1}-\tau_{1} \ox
\tau_{0}\big)\big(r(T)^\circ\bigr),
\end{align*}
where the functionals $\tau_0$, $\tau_1$ are defined for $x \in
\pi_{\pm}(\A)$ by
\begin{align*}
\tau_0(x) := \lim_{N\to\infty} \big(\Tr_N x -(N+1) \,
\tau_1(x)\big), \qquad
\tau_1(x) := \tfrac{1}{2\pi} \int_{0}^{2\pi} \sg(x)(e^{i\theta})
\,d\theta,
\end{align*}
with $\Tr_N x=\sum_{n=0}^N \langle \eps_{n},x\,\eps_{n}\rangle$.
\end{theorem}

\begin{proof}
Consequence of \cite[Theorem 4.1 and (4.3)]{SDLSV}.
\end{proof}

\begin{remark}
Since $F$ is not in $\B$, the equation of Theorem \ref{Theo} are not
valid for all $T\in\Psi_{0}^0(\A)$.\\
But when $T \in \Psi_{0}^0(\A)$, $\ncint T|\DD|^{-k}=0  \text{ for k}
\notin \set{1,2,3}$ since the dimension spectrum is $\set{1,2,3}$
\cite{SDLSV}.
\end{remark}
Compared to \cite{SDLSV} where we had
\begin{align*}
\tau_0^\up(x) := \lim_{N\to\infty} \Tr_N x - (N+\sesq) \,
\tau_1(x),\;\;\;\;\;
\tau_0^\dn(x) := \lim_{N\to\infty} \Tr_N x - (N+\half) \, \tau_1(x),
\end{align*}
we replaced them with $\tau_0$:
$$
\tau_0^\up = \tau_0 - \tfrac{1}{2} \tau_1, \;\;\;\;\;
\tau_0^\dn = \tau_0 + \tfrac{1}{2} \tau_1.
$$
Note that $\tau_{1}$ is a trace on $\pi_{\pm}(\A)$ such that
$\tau_{1}(1)=1$, while $\tau_{0}$ is not since $\tau_{0}(1)=0$ and
\begin{align}
    \label{eq:tau0}
\tau_{0}\big(\pi_{\pm}(aa^*)\big)=\lim_{N\rightarrow
\infty}\sum_{n=0}^\infty (1-q^{2n}) -(N+1)=-\tfrac{1}{1-q^2},
\end{align}
so, because of the shift, the replacement $a\leftrightarrow a^*$ gives
\begin{align}
    \label{nontrace}
\tau_{0}\big(\pi_{\pm}(a^*a)\big)=q^2 \,
\tau_{0}\big(\pi_{\pm}(aa^*)\big).
\end{align}

\subsection{The tadpole}

\begin{lemma}
For $SU_q(2)$, the condition of the vanishing tadpole (see
\cite{ConnesMarcolli}) is not satisfied.
\end{lemma}

\begin{proof}
For example, an explicit calculation gives
$ \ncint \pi(b) [\DD, \pi(b^*)] \DD^{-1} = \tfrac{2}{1 - q^2} $:

Let $x,y \in \ul \pi(\A)$. Since $[F,x]=0$, we have
$$ \ncint x [\DD,y] D^{-1} = \ncint x \delta(y) |\DD|^{-1}
= \tau'\big(r(x \delta(y)\big)^0) $$
where $\tau' :=2 \, \tau_0 \ox \tau_0 - \tfrac{1}{2} \, \tau_1 \ox
\tau_1$.

By Lemma \ref{commutateur}, $\ul \pi(b) \delta \big(\ul \pi(b^*)\big)
= (b_+ + b_-)\big((b_-)^* -
(b_+)^*\big)=-b_{+}{b_{+}}^*
+b_{-}{b_{-}}^*+b_{+}{b_{-}}^*-b_{-}{b_{+}}^*$.
Since only the first two terms have degree $0$, we get, using the
formulae from
Theorem \ref{Theo}
\begin{align*}
\tau'\big(r(-b_{+}{b_{+}}^*)\big)&=-\tau'\big(\pi_{+}(aa^*)\otimes
\pi_{-}(bb^*)\big)\\
&=-2\tau_{0}\big(\pi_{+}(aa^*)\big)
\tau_{0}\big(\pi_{-}(bb^*)\big)+\tfrac12\tau_{1}\big(\pi_{+}(aa^*)\big)
\tau_{1}\big(\pi_{-}(bb^*)\big)
\end{align*}
and $\tau_{1}\big(\pi_{+}(aa^*)\big)=\tfrac{1}{2\pi}
\int_{0}^{2\pi}1\,d\theta=1$,
$\tau_{1}\big(\pi_{-}(bb^*)\big)=0$. Similarly, using \eqref{nontrace}
$$
\tau'\big(r(-b_{-}{b_{-}}^*)\big)=2\tau_{0}\big(\pi_{+}(bb^*)\big)
\tau_{0}\big(\pi_{-}(a^*a)\big)=
2q^2\tau_{0}\big(\pi_{-}(aa^*)\big)\tau_{0}\big(\pi_{+}(bb^*\big).
$$
Since $\tau_{0}\big(\pi_{\pm}(bb^*)\big)=\Tr\big(\pi_{\pm}(bb^*)\big)=
\sum_{n=0}^\infty q^{2n}=\tfrac{1}{1-q^2}$ and \eqref{eq:tau0},
\begin{align*}
\ncint \pi(b) [\DD,\pi(b^*)] \DD^{-1} =
2\tfrac{1}{1-q^2}\tfrac{1}{1-q^2}
+2q^2 \tfrac{-1}{1-q^2}  \tfrac{1}{1-q^2}
= \tfrac{2}{1-q^2}.
\tag*{\qed}
\end{align*}
\hideqed
\end{proof}
In particular the pairing of the tadpole cyclic cocycle
$\phi_{1}$ with the generator of $K_{1}$-group is nontrivial:

\begin{remark} Other examples: with the shortcut $x$ instead of
$\piappr(x)$,
\begin{align*}
&(\tau_1\otimes \tau_1)\, r\big(a\delta(a^*)^\circ \big) =-1 ,&&
(\tau_1\otimes \tau_1) r\big(a^*\delta(a)^\circ\big) =1,\\
&(\tau_0\otimes\tau_0)\, r\big(a\delta(a^*)^\circ \big) =
\tfrac{1}{q^2-1}\,, &&
(\tau_0\otimes\tau_0)\, r\big(a^*\delta(a)^\circ\big) =
\tfrac{q^2}{q^2-1}\,, \\
&\ncint a\delta(a^*)|\DD|^{-1} = \tfrac{q^2+3}{2(q^2-1)}\,,&&
\ncint a^*\delta(a)|\DD|^{-1} =  \tfrac{3q^2+1}{2(q^2-1)}\,,\\
&\ncint b\delta(b) |\DD|^{-1} =0,&&  \ncint b^*\delta(b^*)
|\DD|^{-1}=0,\\
&\ncint b\delta(b^*) |\DD|^{-1} =\tfrac{-2}{q^2-1}\,,&&  \ncint
b^*\delta(b) |\DD|^{-1}=\tfrac{-2}{q^2-1}\,.
\end{align*}

In particular, $N\Phi_1$ does not vanish on 1-forms since
$\int_{N\Phi_{1}}ada^*=N\Phi_1(a,a^*)=-1$.
\end{remark}

Let U be the canonical generator of the $K_{1}(\A)$-group,
$U=\genfrac{(}{)}{0pt}{1}{a \quad \,\,\,\,\,\,b}{-qb^* \quad a^*}$
acting on $\H \ox \C^2$. Then for $A_{U}:=\sum_{k,l=1}^2
\pi(U_{kl})\,d\pi({U^*}_{kl})$, using above remark,
$\int_{\phi_{1}}A_{U}=-2$ as obtained in \cite[page 391]{SDLSV}: in
fact, with $P:=\tfrac 12 (1+F)$,
$$\psi_{1}(U,U^*):=2\sum_{k,l}\ncint U_{kl}\delta(U^*_{kl}) P\vert
\DD\vert^{-1}
-\ncint U_{kl}\delta^2(U^*_{kl}) P\vert \DD\vert^{-2}
+\tfrac 23 \ncint U_{kl}\delta^3(U^*_{kl}) P\vert \DD\vert^{-3}$$
satisfies $\psi_{1}(U,U^*)=2\sum_{k,l}\ncint U_{kl}\delta(U^*_{kl})
P\vert
\DD\vert^{-1}=\int_{\phi_{1}}A_{U}$.

\section{Reality operator and spectral action on $SU_q(2)$}

\subsection{Spectral action in dimension 3 with $[F,\A]\in
OP^{-\infty}$} 
Let $(\A,\H,\DD)$ a be real spectral triple of dimension 3. Assume 
 that $[F,\A]\in OP^{-\infty}$, where $F:=\DD
|\DD|^{-1}$ (we suppose $\DD$ invertible).
Let $\Abb$ be a selfadjoint one form, so $\Abb$ is  of the form
$\Abb=\sum_{i}a_i d b_i$ where
$a_{i},b_{i} \in \A$.

Thus, $\Abb \simeq AF \mod OP^{-\infty}$ where
$A:=\sum_{i}a_i\delta(b_i)$ is the $\delta$-one-form associated to
$\Abb$. Note that $A$ and
$F$ commute modulo $OP^{-\infty}$.

We define
\begin{align*}
D_\Abb &:= \DD_\Abb + P_\Abb,
\quad P_\Abb \text{ the projection on } \Ker \DD_\Abb,\\
\DD_\Abb&:= \DD+\wt \Abb, \quad \wt \Abb := \Abb + J\Abb J^{-1}.
\end{align*}

\begin{theorem}
\label{coeffsASJ}
The coefficients of the full  spectral action (with reality operator)
on any real spectral triple $(\A,\H,\DD)$ of dimension 3 such that
$[F,\A]\in OP^{-\infty}$ are
\begin{align*}
&(i) \hspace{1cm} \ncint \vert D_{\Abb}\vert^{-3}  = \ncint
|\DD|^{-3}.\\
&(ii) \hspace{.9cm} \ncint \vert D_{\Abb}\vert^{-2}  = \ncint
|\DD|^{-2} - 4\ncint A|\DD|^{-3}.\\
&(iii) \hspace{.8cm}\ncint \vert D_{\Abb}\vert^{-1}  = \ncint
|\DD|^{-1}-2\ncint A|\DD|^{-2} +2\ncint A^2 |\DD|^{-3}
+2\ncint AJAJ^{-1}|\DD|^{-3}  .\\
&(iv)  \hspace{.9cm} \zeta_{D_{\Abb}}(0)  = \zeta_{D}(0)-2\ncint
A|\DD|^{-1} +
\ncint A(A+JAJ^{-1})|\DD|^{-2} +  \ncint
\delta(A)(A+JAJ^{-1})|\DD|^{-3}\\
&\hspace{4cm}-\tfrac{2}{3}\ncint
A^3|\DD|^{-3} - 2 \ncint A^2
JAJ^{-1}|\DD|^{-3}.
\end{align*}
\end{theorem}

\begin{proof}
$(i)$ We apply \cite[Proposition 4.9]{MCC}.

$(ii)$ By \cite[Lemma 4.10 $(i)$]{MCC}, we have  $\ncint \vert
D_{\Abb}\vert^{-2}=\ncint |\DD|^{-2}-\ncint (\wt \Abb \DD+\DD \wt
\Abb +\wt \Abb^2)|\DD|^{-4}$. By the trace property of the
noncommutative integral and the fact that
$\wt \Abb^2 |\DD|^{-4}$ is trace-class, we get $\ncint \vert
D_{\Abb}\vert^{-2}= \ncint |\DD|^{-2}-2\ncint \wt \Abb \DD
|\DD|^{-4}=\ncint |\DD|^{-2}-4\ncint \Abb \DD |\DD|^{-4}$. Since
$\Abb \DD \sim A|\DD| \mod OP^{-\infty}$, we get the result.

$(iii)$ By \cite[Lemma 4.10 $(ii)$]{MCC}, we have
$$\ncint \vert
D_{\Abb}\vert^{-1}=\ncint |\DD|^{-1}-\half \ncint (\wt \Abb \DD+\DD
\wt \Abb +\wt \Abb^2)|\DD|^{-3} + \tfrac{3}{8}\ncint (\wt \Abb
\DD+\DD \wt \Abb +\wt \Abb^2)^2 |\DD|^{-5}.$$
Following arguments of $(ii)$, we get
\begin{align*}
\ncint (\wt \Abb \DD+\DD \wt \Abb +\wt \Abb^2)|\DD|^{-3}&= 4\ncint
A|\DD|^{-2} +2 \ncint A^2 |\DD|^{-3} + 2 \ncint AJAJ^{-1}
|\DD|^{-3},\\
\ncint (\wt \Abb \DD+\DD \wt \Abb +\wt \Abb^2)^2|\DD|^{-5}&=8\ncint
A^2 |\DD|^{-3} + 8\ncint AJAJ^{-1}|\DD|^{-3},
\end{align*}
and the result follows.

$(iv)$ By \cite[Lemma 4.5]{MCC} gives
$\zeta_{D_{\Abb}}(0)  =  \sum_{j=1}^{3} \tfrac{(-1)^j}{j}\ncint (\wt
\Abb\DD^{-1})^j\, .$

Moreover, we have
$\ncint \wt \Abb \DD^{-1} = 2\ncint A|\DD|^{-1}$ and
$\ncint  (\wt\Abb \DD^{-1})^2= 2\ncint (A|\DD|^{-1})^2 + 2\ncint
A|\DD|^{-1}JAJ^{-1}|\DD|^{-1}$.
Since $\delta(A)\in OP^{0}$, we can check that $\ncint
(A|\DD|^{-1})^2 = \ncint A^2 |\DD|^{-2}+\ncint \delta(A)A|\DD|^{-3}$
and, with the same argument, that $\ncint
A|\DD|^{-1}JAJ^{-1}|\DD|^{-1}=\ncint A JAJ^{-1}|\DD|^{-2}+\ncint
\delta(A)
JAJ^{-1} |\DD|^{-3}$. Thus, we get
\begin{equation}\label{ncintAD2}
\ncint  (\wt\Abb \DD^{-1})^2 = 2\ncint A(A+JAJ^{-1})|\DD|^{-2} +2
\ncint \delta(A)(A+JAJ^{-1})|\DD|^{-3}.
\end{equation}
The third term to be computed is
$$
\ncint  (\wt\Abb \DD^{-1})^3= 2\ncint (A|\DD|^{-1})^3 + 4\ncint
(A|\DD|^{-1})^2 JAJ^{-1}|\DD|^{-1}+2\ncint
A|\DD|^{-1}JAJ^{-1}|\DD|^{-1} A|\DD|^{-1}.
$$
Any operator in $OP^{-4}$ being trace-class here, we get
\begin{equation}
    \label{ncintAD3}
\ncint  (\wt\Abb \DD^{-1})^3 = 2\ncint A^3|\DD|^{-3} + 4\ncint A^2
JAJ^{-1}|\DD|^{-3}+2\ncint AJAJ^{-1} A|\DD|^{-3}.
\end{equation}
Since $\ncint AJAJ^{-1} A|\DD|^{-3}=\ncint A^2
JAJ^{-1}|\DD|^{-3}$
by trace property and the fact that $\delta(A)\in OP^{0}$,
the result follows then from (\ref{ncintAD2}) and (\ref{ncintAD3}).
\end{proof}

\begin{corollary} For the spectral action of $\Abb$ without
the reality operator (i.e. $\DD_{\Abb}=\DD+\Abb$), we get
\begin{align*}
&\ncint \vert D_{\Abb}\vert^{-2}  = \ncint |\DD|^{-2} - 2\ncint
A|\DD|^{-3},\\
&\ncint \vert D_{\Abb}\vert^{-1}  = \ncint
|\DD|^{-1}-\ncint A|\DD|^{-2} +\ncint A^2 |\DD|^{-3},\\
&\zeta_{D_{\Abb}}(0)  = \zeta_{\DD}(0) -\ncint A|\DD|^{-1} +\half
\ncint A^2|\DD|^{-2} +  \half \ncint
\delta(A)A|\DD|^{-3}-\tfrac{1}{3}\ncint
A^3|\DD|^{-3}.
\end{align*}
\end{corollary}

\subsection{Spectral action on $SU_q(2)$: main result}

On $SU_q(2)$, since $F$ commutes with $a_\pm$ and $b_\pm$, the
previous lemma can be used for the spectral action computation.

Here is the main result of this section

\begin{theorem}
    \label{mainThmJ}
In the full spectral action \eqref{formuleaction}
(with the reality operator) of $SU_q(2)$ for a one-form $\Abb$ and
$A$ its associated $\delta$-one-form, the coefficients are:
\begin{align*}
&\ncint \vert D_{\Abb}\vert^{-3}  = 2\,, \\
& \ncint\vert D_{\Abb}\vert^{-2}  =  - 4\, \ncint A |\DD|^{-3},\\
&\ncint \vert D_{\Abb}\vert^{-1}  = -\half +2\big( \ncint
A^2|\DD|^{-3}- \ncint A|\DD|^{-2} \big)+ \big\vert\ncint
A|\DD|^{-3}\big\vert^2 ,\\
&\zeta_{D_{\Abb}}(0) =-2 \ncint A|\DD|^{-1} +\ncint A^2 |\DD|^{-2}
-\tfrac{2}{3} \ncint A^3 |\DD|^{-3}\\
&\hspace{3cm}+\overline{\ncint A |\DD|^{-3}}\big(\half\ncint
A|\DD|^{-2} -\ncint A^2 |\DD|^{-3}\big) +\half \ncint
A|\DD|^{-3}\overline{\ncint A|\DD|^{-2}}\, .
\end{align*}
\end{theorem}

In order to prove this theorem, we will use a decomposition of
one-forms in the Poincar\'e-Birkhoff--Witt basis of $\A$ with an
extension of previous representations to operators like $T JT'
J^{-1}$ where $T$ and $T'$ are in $X$. 

\subsection{Balanced components and Poincar\'e--Birkhoff--Witt basis
of $\A$}
  \label{balancedcomponent}

Our objective is to compute all integrals in term of $A$ and the
computation will lead to functions of $A$ which capture certain
symmetries on $A$.

For convenience, let us introduce
now these functions:

Let $\Abb=\sum_i \pi(x^i) d \pi(y^i) $ on $SU_q(2)$ be one-form and
$A$ the associated $\delta$-one-form. The $x^i$ and $y^i$ are in $\A$
and as such
they can be uniquely written as finite sums $x^i=\sum_\a x^i_\a m^\a$
and $y^i = \sum_\beta y^i_\beta m^{\beta}$ where
$m^\a:=a^{\a_1}b^{\a_2}{b^*}^{\a_3}$ is the canonical monomial of
$\A$ with $\a,\beta \in \Z \times  \N \times  \N$ based on a fixed
Poincar\'e--Birkhoff--Witt type basis of $\A$.

\begin{remark}
Any one-form $\Abb=\sum_i \pi(x^i) d \pi(y^i) $ on $SU_q(2)$ is
characterized by a complex valued
matrix $A_{\a}^\beta=\sum_{i}x_\a^i\,y_\beta^i$ where $\a,\beta \in
\Z\times\N\times\N$. This matrix is such that
$$
A= A_{\a}^\beta\, M^{\a}_\beta
$$
where $M^{\a}_\beta:= \pi(m^\a)\delta \big(\pi(m^\beta) \big)$.

In the following, we note 
$$
\bar A:= \bar A^\beta_\a\, M^\a_\beta
$$
so for any $p\in \N$, $\ncint \bar A |\DD|^{-p} = \overline{ \ncint A
|\DD|^{-p}}$. 
\end{remark}
This presentation of one-forms is not unique modulo $OP^{-\infty}$
since, as we will see in section 5, 
$F = \sum_i x_i dy_i$ where $x_i,y_i \in\A$, thus for any generator
$z$,
$[F, z] = \sum_i x_i d( y_i z) - x_i y_i dz - z x_i dy_i = 0 \mod
OP^{-\infty}$.
We do not know however if this presentation is unique when the
$OP^{-\infty}$ part is taken into account.

\quad

The $\delta$-one-forms $M^\a_\beta$ are said to be {\itshape
canonical}. Any product of $n$ canonical $\delta$-one forms,
where $n\in \N^*$, is called a {\itshape canonical
$\delta^n$-one-form}. Thus, if $A$ is a $\delta$-one-form,
$A^n = (A^n)_{\bar \a}^{\bar \beta}\, M^{\bar\a}_{\bar \beta}$ where
$\bar \a=(\a,\a',\cdots,\a^{(n-1)})$,  $\bar
\beta=(\beta,\beta',\cdots,\beta^{(n-1)})$ are in $\Z^n\times \N^n
\times \N^n$,
$(A^n)_{\bar \a}^{\bar\beta} := A_{\a}^{\beta} \cdots
A_{\a^{(n-1)}}^{\beta^{(n-1)}}$ and $M^{\bar\a}_{\bar\beta}$ is the
canonical $\delta^n$-one form equals to $
M^{\a}_{\beta} \cdots M^{\a^{(n-1)}}_{\beta^{(n-1)}}$.

\begin{definition}
A canonical $\delta^n$-one-form is $a$-balanced if it is of the form
$$
a^{\a_1}\delta(a^{\beta_1})\cdots
a^{\a^{(n-1)}_1}\delta(a^{\beta^{(n-1)}_1})
$$
where $\sum_{i=0}^{n-1} \a^{(i)}_1+\beta^{(i)}_1 = 0$.

For any $\delta$-one-form $A$, the $a$-balanced components of $A^n$
are noted $B_a(A^n)_{\bar \a}^{\bar\beta}$.
\end{definition}

Note that
$$
B_a(A)_{\bar \a}^{\bar\beta} = A_{-\beta_1 0 0}^{\beta_1 0 0}\,
\delta_{\a_1+\beta_1,0} \,\delta_{\a_2+\a_3+\beta_2+\beta_3,0}.
$$
\begin{definition}
A canonical $\delta^n$-one-form is balanced if it is of the form
$$
m^{\a}\delta(m^{\beta})\cdots m^{\a^{(n-1)}}\delta(m^{\beta^{(n-1)}})
$$
where $\sum_{i=0}^{n-1} \a^{(i)}_1+\beta^{(i)}_1 = 0$ and
$\sum_{i=0}^{n-1} \a^{(i)}_2+\beta^{(i)}_2  = \sum_{i=0}^{n-1}
\a^{(i)}_3+\beta^{(i)}_3$.

For any $\delta$-one-form $A$, the balanced components of $A^n$ are
noted $B(A^n)_{\bar \a}^{\bar\beta}$.
\end{definition}
Note that
$$
B(A)_{\bar \a}^{\bar\beta} = A_{-\beta_1 \a_2 \a_3}^{\beta_1 \beta_2
\beta_3}\,\delta_{\a_1+\beta_1,0}\,
\delta_{\a_2+\beta_2,\a_3+\beta_3}.
$$
As we will show, a contribution to the $k^{th}$-coefficient in the
spectral action, is only brought by one-forms $\Abb$ such that $A^k$
is balanced (and even $a$-balanced
in the case $k=1$).

Note also that if $A$ is balanced, then $A^{k}$ for $k\geq 1$ is also
balanced, whereas the converse is false.

\subsection{The reality operator $J$ on $SU_q(2)$}
For any $n,p\in \N$,
\begin{align*}
 &q_n:=\sqrt{1-q^{2n}},
&q_{-n}&:= 0 \text{ if } n>0 ,\\
&q_{n,p}^\up := q_{n+1}\cdots q_{n+p}\,,
&q_{n,p}^\dn &:= q_n \cdots q_{n-(p-1)}\, ,
\end{align*}
with the convention $q_{n,0}^\up= q_{n,0}^\dn := 1$.  Thus, we have
the relations
\begin{align*}
&\pi_{\pm}(a^p)\ \eps_n = q_{n,p}^\up\  \eps_{n+p}\,, &
\pi_{\pm}({a^*}^p)&\ \eps_n  = q_{n,p}^\dn\ \eps_{n-p}\,,\\
&\pi_{\pm}(b^p)\  \eps_n  =( \pm
q^{n})^p\ \eps_n\,, &\pi_{\pm}({b^*}^p)&\  \eps_n=( \pm
q^{n})^p\ \eps_n \,,
\end{align*}
where $\eps_{k} := 0$ if $k<0$.

The sign of $x\in \R$ is noted $\eta_x$. By convention, $a_j := a$,
$a_{\pm,j}:=a_{\pm}$ if $j\geq0$ and $a_j:=a^*$,
$a_{\pm,j}:=a_{\pm}^*$
if $j<0$.
Note that, with convention
$$
q_{n,p}^{\up_{\a_1}}:= q_{n,p}^\up
\text{ if } {\a_1}>0, \quad
q_{n,p}^{\up_{\a_1}}:= q_{n,p}^\dn \text{ if } {\a_1}<0
, \text{ and } q_{n,p}^{\up_0}:=1,
$$
we have for any ${\a_1} \in \Z$ and $p\leq
{\a_1}$,
$\pi_{\pm}(a_{\a_1}^p)\, \eps_n = q_{n,p}^{\up_{\a_1}}\
\eps_{n+\eta_{\a_1} p}$.

Recall that the reality operator $J$ is defined by
\begin{align*}
J  \, v^{j\up}_{m,l}= i^{2(m+l)-1}v^{j\up}_{2j-m,2j+1-l} \,,\qquad
J  \, v^{j\dn}_{m,l}= i^{-2(m+l)+1}v^{j\dn}_{2j-m,2j-1-l}\,,
\end{align*}
thus the real conjugate operators
$$
\wh a_\pm:=Ja_{\pm}J^{-1}, \quad \wh b_\pm:=J b_\pm J^{-1}
$$
satisfy
\begin{align*}
\wh a_+ \, v^j_{m,l} &:=
-q_{2j+1-m} \, \genfrac{(}{)}{0pt}{1}{q_{2j+2-l}\quad 0 \quad}
{\quad0 \qquad q_{2j-l}}  \, v^{j^+}_{m,l}\,, \quad
&\wh a_- \, v^j_{m,l} &:=
-q^{2j-m} \, \genfrac{(}{)}{0pt}{1}{q^{2j+2-l}\quad 0\quad}{\quad 0
\quad
q^{2j-l}} \, v^{j^-}_{m-1,l-1},\\
\wh b_+ \, v^j_{m,l} &:=
q_{2j+1-m} \, \genfrac{(}{)}{0pt}{1}{q^{2j+1-l} \quad 0 \quad}{\quad
0\quad q^{2j-1-l}} \, v^{j^+}_{m,l+1}\,, \quad
&\wh b_-  \,v^j_{m,l} &:=
-q^{2j-m} \,\genfrac{(}{)}{0pt}{1}{q_{2j+1-l}\quad 0 \quad}{\quad 0
\quad q_{2j-1-l}} \, v^{j^-}_{m-1,l}\,.
\end{align*}
So the real conjugate operator behave differently on the
up and down part of the Hilbert space. The difference comes from the
fact that the index $l$ is not treated uniformly by $J$ on up and
down parts.

We note $\wh X$ the algebra generated by $\set{\wh a_\pm, \wh
b_\pm}$, $\wt X$ the algebra generated by $\set{a_{\pm},b_{\pm}, \wh
a_\pm, \wh b_\pm}$ and
$\H':=\ell^2(\N) \otimes \ell^2(\Z)$ and we construct
two $*$-representations $\wh \pi_{\pm}$ of $\A$:

The representation $\wh \pi_+$ gives bounded operators
on $\H'$ while $\wh \pi_-$ represents $\A$ into $\B(\H'\otimes \C^2)$.

The representation $\wh \pi_+$ is defined on the generators by:
$$
\wh \pi_{+}(a)\, \eps_m \otimes \eps_{2j}:=q_{2j+1-m}\, \eps_m\otimes
\eps_{2j+1},
\qquad \wh \pi_{+} (b)\, \eps_m \otimes \eps_{2j}:=-q^{2j-m}\,
\eps_{m+1}\otimes \eps_{2j+1}
$$
while $\wh \pi_-$ is defined by:
\begin{align*}
 \wh \pi_{-} (a) \, \eps_l \otimes \eps_{2j}\otimes \eps_{\up\dn}&:=
-q_{2j+1\pm1-l}\, \eps_l \otimes \eps_{2j+1}\otimes \eps_{\up\dn}\,
,\\
  \wh \pi_{-} (b)\, \eps_l \otimes \eps_{2j}\otimes \eps_{\up\dn} &:=
-q^{2j\pm1-l}\, \eps_{l+1}\otimes \eps_{2j+1}\otimes \eps_{\up\dn}\, ,
 \end{align*}
where $\eps_{\up\dn}$ is the canonical basis of $\C^2$ and the $+$ in
$\pm$ corresponds to $\up$ in $\up\dn$.

The link between $\wh \pi_{\pm}$ and $\pi_{\pm}$ which explains the
notations about these intermediate objects
and the fact that $\wh \pi_{\pm}$ are representations on
different Hilbert spaces, is in the parallel between equations
\eqref{eq:r}, \eqref{eq:hat} and \eqref{eq:prime}.

Let us give immediately a few properties ($x_\beta$ equals $x$ if
the sign $\beta$ is positive and equals $x^*$ otherwise)
\begin{align*}
\wh \pi_+(a_\beta)^p \,\eps_m\otimes \eps_{2j} &=
q^{\up_\beta}_{2j-m,p} \, \eps_m\otimes\eps_{2j+\eta_\beta p}\,,\\
\wh \pi_-(a_\beta)^p\, \eps_l\otimes \eps_{2j}\otimes\eps_{\up\dn} &=
(-1)^p \, q^{\up_\beta}_{2j\pm1-l,p} \,
\eps_l\otimes\eps_{2j+\eta_\beta p}\otimes\eps_{\up\dn}\,,\\
\wh \pi_+(b_\beta)^p \,\eps_m\otimes \eps_{2j}&= (-1)^p\,
q^{(2j-m)p}\,
\eps_{m+\eta_\beta p}\otimes \eps_{2j+\eta_\beta p}\,, \\
\wh \pi_-(b_\beta)^p \,\eps_l\otimes \eps_{2j}\otimes\eps_{\up\dn}&=
(-1)^p\, q^{(2j\pm1-l)p}\, \eps_{l+\eta_\beta p}\otimes
\eps_{2j+\eta_\beta p}\otimes\eps_{\up\dn}\,.
\end{align*}

Note that the $\wh \pi_\pm$ representations still contain the shift
information, contrary to representations  $\pi_\pm$.
Moreover, $\wh \pi_\pm(b)\neq \wh \pi_\pm (b^*)$ while
$\pi_\pm(b) = \pi_\pm (b^*)$.

The operators $\wh a_\pm, \wh b_\pm$ are coded on $\H'\otimes
\H'\otimes \C^2$ as the correspondence
\begin{align}
    \label{eq:hat}
&\wh a_+ \longleftrightarrow  \wh \pi_+ (a) \otimes \wh \pi_-(a)
,\quad
&\wh a_- &\longleftrightarrow  -q \,\wh \pi_+(b^*) \otimes
\wh\pi_-(b^*) ,\nonumber\\
&\wh b_+\longleftrightarrow -\wh \pi_+(a)\otimes \wh \pi_-(b)
,\quad
&\wh b_-& \longleftrightarrow -\wh \pi_+(b^*) \otimes \wh
\pi_-(a^*).
\end{align}

We now set the following extension to $\B(\H')$ of  $\pi_+$ and to
$\B(\H'\otimes \C^2)$ of $\pi_-$ by
\begin{align*}
\pi'_{+}(a)&:=\pi_{+}(a)\otimes V, \quad \pi'_{+}(b) := \pi_+
(b)\otimes V \quad\text{ (V is the shift of }\ell^2(\Z)),\\
\pi'_{-}(a)&:=\pi_{-}(a)\otimes V \otimes 1_2, \quad \pi'_{-}(b) :=
\pi_- (b)\otimes V \otimes 1_2.
\end{align*}
So, we can define a canonical algebra morphism $\wt \rho\,$ from $\wt
X$
into the bounded operators on
$\H' \otimes\H'\otimes \C^2$.  This morphism is defined on the
generators part $\set{\wh a_{\pm},\wh b_{\pm}}$ of $\wt X$
by preceding correspondence and on the
generators part $\set{ a_{\pm}, b_{\pm}}$ by $-$see \eqref{eq:r}:
\begin{align}
    \label{eq:prime}
 & a_+ \longleftrightarrow   \pi'_+ (a) \otimes \pi'_-(a) ,\quad
 & a_- & \longleftrightarrow     -q \,\pi'_+(b^*) \otimes \pi'_-(b^*)
 ,\nonumber\\
 & b_+\longleftrightarrow       - \pi'_+(a)\otimes  \pi'_-(b) ,\quad
 & b_- & \longleftrightarrow     - \pi'_+(b^*) \otimes  \pi'_-(a^*).
\end{align}

We note $S$ the canonical surjection from $\H'\otimes \H'\otimes\C^2$
onto $\H$. This surjection is associated to the parameters
restrictions on $m,j,l,j'$. In  particular, the index $j'$ associated
to the second $\ell^2(\N)$ in $\H'\otimes \H'\otimes \C^2$ is set to
be equal to $j$.
Any vector in $\H'\otimes \H'\otimes\C^2$ not
satisfying these restrictions is sent to 0 in $\H$.

Denote by $I$ the canonical injection
of $\H$ into $\H'\otimes \H'\otimes \C^2$ (the index $j$ is doubled).
Thus, $S\wt \rho(\cdot) I$ is the identity on
$\wt X$.

In the computation of residues of $\zeta_{\DD}^T$ functions,
we can therefore replace the operator $T$ by  $S\wt \rho(T) I$.

We now extend $\tau_0$ on $\pi'_\pm (\A) \wh \pi_\pm(\A)$: For $x,y
\in \A$, we set
\begin{align*}
\Tr_{N} \big(\pi'_+(x)\wh \pi_+(y) \big) & := \sum_{m=0}^{N}
\,\langle \eps_m\otimes \eps_{N} \,,\, \pi'_+(x)\wh
\pi_+(y)\,\eps_m\otimes \eps_{N} \rangle\, , \\
\Tr_{N}^{\up}  \big(\pi'_-(x)\wh \pi_-(y) \big) & := \sum_{l=0}^{N}
\,\langle \eps_l\otimes \eps_{N-1}\otimes \eps_{\up}\,,\,
\pi'_-(x)\wh \pi_-(y)\,\eps_l\otimes \eps_{N-1}\otimes
\eps_{\up}\rangle\, ,  \\
\Tr^\dn_{N}  \big(\pi'_-(x)\wh \pi_-(y) \big) & := \sum_{l=0}^{N}
\,\langle \eps_l\otimes \eps_{N+1}\otimes \eps_{\dn} \,,\,
\pi'_-(x)\wh \pi_-(y)\,\eps_l\otimes \eps_{N+1}\otimes\eps_{\dn}
\rangle \, .
\end{align*}
Actually, a computation on monomials of $\A$ shows that $\Tr^\dn_{N}
\big(\pi'_-(x)\wh \pi_-(y) \big)=
\Tr^\up_{N}  \big(\pi'_-(x)\wh \pi_-(y) \big)$.
For convenience, we shall note $\Tr_N\big(\pi'_-(x)\wh \pi_-(y) \big)$
this functional.

\begin{lemma}
    \label{tau0ext}
Let $x,y \in \A$. Then,

(i) $\tau_{0} \big(\pi'_\pm(x)\wh \pi_\pm(y) \big):=\lim_{N
\to \infty} U_N$ exists where $$U_{N}:=\Tr_{N}
\big( \pi'_\pm(x)\wh \pi_\pm(y)\big)
-(N+1)\tau_1\big(\pi_\pm(x)\big)\,  \tau_1\big(\pi_\pm(y)\big).$$

(ii) $U_N=\tau_{0} \big(\pi'_\pm(x)\wh \pi_\pm(y) \big) +
\mathcal{O}(N^{-k})$ for all $k>0$.
\end{lemma}

\begin{proof}
$(i)$
We can suppose that $x$ and $y$ are monomials, since the result will
follow by linearity. We will give a proof for the case of the $\pi_+$
representations, the case $\pi_-$ being similar, with minor changes.

We have $\wh \pi_+ (y) = (\wh\pi_+ a_{\beta_1})^{|\beta_1|} \,
(\wh\pi_+ b)^{\beta_2} \,(\wh\pi_+ b^*)^{\beta_3} $.
A computation gives
$$
\wh \pi_+(y) \,\eps_m\otimes \eps_{2j} = (-1)^{\beta_2+\beta_3}\,
q^{(2j-m)(\beta_2+\beta_3)}\, q^{\up_{\beta_1}}_{2j-m,|\beta_1|}\,
\eps_{m-\beta_3+\beta_2}
\otimes \eps_{2j-\beta_3+\beta_2+\beta_1}
$$
and with the notation $t_{2j,m}:=\langle \eps_m\otimes \eps_{2j}
\,,\, \pi'_\pm(x)\wh \pi_\pm(y)\,\eps_m\otimes \eps_{2j} \rangle $
and $T_{2j}:=\sum_{m=0}^{2j} t_{2j,m}$, we get
\begin{align*}
t_{2j,m}&=(-1)^{\beta_2+\beta_3}\, q^{(2j-m)(\beta_2+\beta_3)}\,
q^{\up_{\beta_1}}_{2j-m,|\beta_1|}\,
q^{\up_{\a_1}}_{m-\beta_3+\beta_2,|\a_1|}\,
q^{(m+\beta_2-\beta_3)(\a_2+\a_3)}\, \delta_{\a_1+\beta_2-\beta_3,0}\\
& \hspace{2cm} \times  \delta_{-\a_3+\a_2+\beta_1,0}\\
 &=(-1)^{\a_1}\, q^{(2j-m)(\beta_2+\beta_3)+(m-\a_1)(\a_2+\a_3)}
 \, q^{\up_{\beta_1}}_{2j-m,|\beta_1|}\,
q^{\up_{\a_1}}_{m-\a_1,|\a_1|}\, \delta_{\a_1+\beta_2-\beta_3,0} \,
\delta_{\a_2-\a_3+\beta_1,0}\\
&=: f_{\a,\beta} \ q^{2j\la }\,  t'_{2j,m} =: f_{\a,\beta}
\ q^{2j\kappa }\,  t''_{2j,2j-m}
\end{align*}
where
\begin{align}
t'_{2j,m}&:=q^{m(\kappa-\la)}\, q^{\up_{\beta_1}}_{2j-m,|\beta_1|}\,
q^{\up_{\a_1}}_{m-\a_1,|\a_1|}\label{t'2jm}\,,\\
t''_{2j,m}&:=q^{m(\la-\kappa)}\, q^{\up_{\beta_1}}_{m,|\beta_1|}\,
q^{\up_{\a_1}}_{2j-m-\a_1,|\a_1|}\label{t''2jm}\,,
\end{align}
with $\la:=\beta_2+\beta_3 \geq0$ and $\kappa:=\a_2+\a_3\geq0$. We
will now prove that if $\la\neq\kappa$, then $(T_{2j})$
is a convergent sequence.
Suppose $\kappa>\la$. Let us note $U'_{2j}:= \sum_{m=0}^{2j}
t'_{2j,m}$.
Since the $t'_{2j,m}$ are positive and $t'_{2j+1,m}\geq t'_{2j,m}$
for all $j,m$,
$U'_{2j}$ is an increasing real sequence.
The estimate
$$
U'_{2j}\leq \sum_{m=0}^{2j} q^{m(\kappa-\la)} \leq
\tfrac{1}{1-q^{\kappa-\la}}<\infty
$$
proves then that $U'_{2j}$ is a convergent sequence.
With $T_{2j}=f_{\a,\beta}\, q^{2j\la}\, U'_{2j}$, we obtain our
result.

Suppose now that $\la>\kappa$.
Let us note $U''_{2j}:= \sum_{m=0}^{2j} t''_{2j,m}$.
Since the $t''_{2j,m}$ are positive and $t''_{2j+1,m}\geq t''_{2j,m}$
for all $j,m$,
$U''_{2j}$ is an increasing real sequence.
The estimate
$$
U''_{2j}\leq \sum_{m=0}^{2j} q^{m(\la-\kappa)} \leq
\tfrac{1}{1-q^{\la-\kappa}}<\infty
$$
proves then that $U''_{2j}$ is a convergent sequence.
With $T_{2j}=f_{\a,\beta}\, q^{2j\kappa}\, U''_{2j}$, we have
again our result. Moreover, note that if $\la$
and $\kappa$ are both different from zero, the limit of $(T_{2j})$
is zero and more precisely,
\begin{align}
T_{2j}&=\mathcal{O}(q^{2j\lambda}) \text{ if } \kappa>\lambda>0
\label{ka>la},\\
T_{2j}&=\mathcal{O}(q^{2j\kappa}) \text{ if }
\lambda>\kappa>0\label{la>ka}.
\end{align}
Suppose now that $\la = \kappa \neq 0$.
In that case, $(T_{2j})$ also converges rapidly to zero.
Indeed, let us fix $q< \eps <1$. we have $\eps^{-2j\la}\,T_{2j}
= \sum_{m=0}^{2j} c_{m} d_{2j-m} =c*d(2j)$ where $c_m:=f_{\a,\beta}\,
(q/\eps)^{\la m} \, q^{\up_{\a_1}}_{m-\a_1,|\a_1|}$ and
$d_m:=(q/\eps)^{\la m}
q^{\up_{\beta_1}}_{m,|\beta_1|}$. Since both $\sum_m c_m$
and $\sum_m d_m$ are absolutely convergent series, their
Cauchy product $\sum_{2j} \eps^{-2j\la}\,T_{2j}$ is
convergent. In particular, $\lim_{j\to\infty} \eps^{-2j \la}\,
T_{2j}=0$, and
\begin{equation}
T_{2j}= \mathcal{O}(\eps^{2j\lambda})\label{la=ka}.
\end{equation}
Finally, $T_{2j}$ has a finite limit in all cases
except possibly when $\la=\kappa=0$, which is the case
when $\a_1=\a_2=\a_3=\beta_1=\beta_2=\beta_3=0$.
In that case, $t_{2j,m}=1$.

A straightforward computation gives $\tau_1 \big(\pi_\pm(x)\big)
\, \tau_1\big(\pi_{\pm}(y)\big) =
\delta_{\a_1,0}\,\delta_{\beta_1,0}\,\delta_{\a_2,0}\,
\delta_{\beta_2,0}\,\delta_{\a_3,0}\,\delta_{\beta_3,0}$.

Thus,
$$
U_{2j}= T_{2j}-(2j+1) \delta_{\a_1,0}\,\delta_{\beta_1,0}
\,\delta_{\a_2,0}\,\delta_{\beta_2,0} \, \delta_{\a_3,0}
\,\delta_{\beta_3,0}
$$
has always a finite limit when $j\to \infty$.

$(ii)$ The result is clear if $\la=\kappa=0$ (in that case
$U_N=\tau_0=0$).
Suppose $\la$ or $\kappa$ is not zero.
In that case $U_{2j} = T_{2j}$. By (\ref{la>ka}), (\ref{ka>la})
and (\ref{la=ka}), we see that if $\la>\kappa >0$ or $\kappa>\la >0$
or $\kappa=\lambda$, $(T_{2j})$ converges to 0 with a rate
in $\mathcal{O}(\eps^{2j\a})$ where $\a>0$ and $q\leq \eps<1$.
Thus, it only remains to check the cases $(\kappa>0,\lambda=0)$
and $(\kappa=0, \lambda>0)$. In the first one, we get
from (\ref{t'2jm}), $U_{2j}=f_{\a,\beta} \sum_{m=0}^{2j}
q^{m\kappa} q^{\up_{\beta_1}}_{2j-m,|\beta_1|}$. If $\beta_1 = 0$, we
are done.

Suppose $\beta_1> 0$. We have
$q^{\up_{\beta_1}}_{2j-m,|\beta_1|}=\sum_{p=0}^{\infty} l_p \,
q^{r_p}\, q^{2|p|_1(2j-m)}$ where $p=(p_1,\cdots,p_{\beta_1})$ and
$l_p=(-1)^{|p|_1}\genfrac{(}{)}{0pt}{1}{{\half}}{p}$,
$r_p:=2p_1+\cdots +2\beta_1 p_{\beta_1}$.
Thus, cutting the sum in two, we get, noting $L_{2j}:=
f_{\a,\beta}\sum_{m=0}^{2j}q^{m\kappa}$,
$$
U_{2j}-L_{2j}=f_{\a,\beta}\sum_{|p|_1>\kappa/2}l_p \,
q^{r_p}\, \tfrac{q^{4|p|_1j}-q^{(2j+1)\kappa-2|p|_1}}
{1-q^{\kappa-2|p|_1}} + f_{\a,\beta}\sum_{0\neq |p|_1
\leq \kappa/2}l_p \, q^{r_p}\,q^{4|p|_1 j}\sum_{m=0}^{2j}
q^{m(\kappa-2|p|_1)}.
$$
Since $\sum_{0\neq |p|_1\leq \kappa/2}l_p \, q^{r_p}\,q^{4|p|_1 j}
\sum_{m=0}^{2j} q^{m(\kappa-2|p|_1)}$
is in $\mathcal{O}_{j\to \infty}(jq^{4j})$,
we have, modulo a rapidly decreasing sequence,
$$
U_{2j}-L_{2j}\sim f_{\a,\beta}\sum_{|p|_1>\kappa/2}l_p \, q^{r_p}\,
\tfrac{q^{4|p|_1j}-q^{(2j+1)\kappa-2|p|_1}}
{1-q^{\kappa-2|p|_1}}=:f_{\a,\beta}q^{2\kappa j} V_{2j}
$$
with
$$
V_{2j}=\sum_{|p|_1>\kappa/2}l_p \, q^{r_p}\,
\tfrac{1-q^{(2|p|_1-\kappa)(2j+1)}}
{1-q^{2|p|_1-\kappa}} = \sum_{|p|_1>\kappa/2}\
\sum_{m=0}^{2j}\,  l_p \, q^{r_p}\,  q^{(2|p|_1-\kappa)m}.
$$
The family $v_{m,p}:= (l_p \, q^{r_p}\,
q^{(2|p|_1-\kappa)m})_{(p,m)\in I}$, where $I=\set{(p,m)
\in \N^{\beta_1}\times \N  \  :
\ |p|_1>\kappa/2 }$ is (absolutely) summable.
Indeed $|v_{m,p}|\leq |l_p| q^{r_p}\, q^{m}$
so $|v_{m,p}|$ is summable as the product of two summable families.
As a consequence, $\lim_{j\to \infty} V_{2j}$
exists and is finite, which proves that $(q^{2\kappa j} V_{2j})$,
and thus $(U_{2j}-L_{2j})$ converge rapidly to 0.

Suppose now that $\beta_1<0$. In that case,
$q^{\up_{\beta_1}}_{2j-m,|\beta_1|} =
q^{\dn}_{2j-m,|\beta_1|} = q^{\up}_{2j-(m+|\beta_1|),|\beta_1|}$
and by (\ref{t'2jm}), we get $U_{2j}=f_{\a,\beta}
\sum_{m=0}^{2j}q^{m\kappa} q^{\up}_{2j-(m+|\beta_1|),|\beta_1|} =
f_{\a,\beta}\, q^{-|\beta_1|\kappa}\,
\sum_{m=|\beta_1|}^{2j+|\beta_1|}  q^{m\kappa}
q^\up_{2j-m,|\beta_1|}$,
so the same arguments as in case $\beta_1>0$ apply here,
the summation on $m$ simply shifted
of $|\beta_1|$.

The same proof can be applied for the other case
$(\kappa=0, \lambda>0)$. This time, we only
need to use (\ref{t''2jm}) instead of (\ref{t'2jm})
and the preceding arguments follow by replacing $\kappa$
by $\la$ and $\beta_1$ by $\a_1$.
\end{proof}

\begin{remark}
Contrary to the preceding $\tau_0$, the new
functional contains the shift information.
In particular, it filters the parts of nonzero degree.
\end{remark}

If $T\in X \wh X$, $\wt \rho(T) \in \pi_+(\A)\wh
\pi_+(\A)\otimes\pi_-(\A)\wh \pi_-(\A) $.

For notational convenience, we define $\tau_1$ on $\pi'_\pm(\A)\wh
\pi_\pm(\A)$ as
$$
\tau_1\big(\pi'_\pm(x)\wh \pi_\pm(y)\big):=
\tau_1\big(\pi_\pm(x)\big)\, \tau_1\big(\pi_\pm(y)\big).
$$

In the following, the symbol $\sim_e$ means equals modulo a entire
function.

\begin{theorem}
    \label{ncinttau}
Let $T\in  X \wh X$. Then
\begin{align*}
&\hspace{-2cm}(i) \hspace{.7cm} \zeta_\DD^{T}(s)\, \sim_e
2(\tau_1\otimes \tau_1) \, \big(\wt \rho(T)\big)\,
\zeta(s-2) + 2(\tau_0\otimes \tau_1 + \tau_1\otimes  \tau_0)\,
\big(\wt\rho(T)\big)\,\zeta(s-1) \\
& \hspace{2cm}+2(\tau_0\otimes  \tau_0 -
\tfrac{1}{2}\tau_1\otimes \tau_1)\, \big(\wt \rho(T)\big)\,\zeta(s)
,\\
&\hspace{-2cm}(ii) \hspace{.5cm} \ncint T |\DD|^{-3} = 2(\tau_1 \ox
\tau_1) \bigl(\wt \rho(T)\bigl),\\
&\hspace{-2cm}(iii) \hspace{.4cm} \ncint T |\DD|^{-2}=2(\tau_0\otimes
\tau_1 + \tau_1\otimes
\tau_0)\, \big(\wt \rho(T)\big) ,\\
&\hspace{-2cm}(iv) \hspace{.5cm} \ncint T|\DD|^{-1}= 2(\tau_0\otimes
 \tau_0- \tfrac{1}{2}\tau_1\otimes \tau_1)\, \big(\wt \rho(T)\big).
\end{align*}
\end{theorem}

\begin{proof}
$(i)$ Since $T\in X \wh X$, $\wt \rho(T)$ is a linear combination of
terms
like
$\pi'_+(x)\wh \pi_+(y)\otimes \pi'_-(z)\wh\pi_-(t)$,
where $x,y,z,t \in \A$. Such a term is noted in the
following $T_+\otimes T_-$. Linear combination of
these term is implicit.
With the shortcut $T_{c_1,\cdots,c_p}:=\langle\eps_{c_1}\otimes
\cdots \otimes \eps_{c_p}, T \eps_{c_1}\otimes
\cdots \otimes \eps_{c_p}\rangle$, recalling that $v^{j,\dn}_{m,l}$
is 0
when $j=0$, or $l\geq 2j$, we get
\begin{align*}
\zeta_{\DD}^T(s)=&
\sum_{2j=0}^{\infty}\sum_{m=0}^{2j}\sum_{l=0}^{2j+1} \langle \,
\genfrac{(}{)}{0pt}{1}{{v^{j,\up}_{m,l}}}{0}\,,\, S\wt \rho(T)I\,
\genfrac{(}{)}{0pt}{1}{{v^{j,\up}_{m,l}}}{0}\rangle \,d_{j^+}^{-s} +
\langle \genfrac{(}{)}{0pt}{1}{{0}}{v^{j,\dn}_{m,l}}\,,\, S\wt
\rho(T)I\,
 \genfrac{(}{)}{0pt}{1}{{0}}{v^{j,\dn}_{m,l}}\rangle \,d_{j}^{-s} \\
 =&\sum_{2j=0}^{\infty}\sum_{m=0}^{2j}\sum_{l=0}^{2j+1}
\wt\rho(T)_{m,2j,l,2j,\up}
\,d_{j^+}^{-s}+\sum_{2j=1}^{\infty}\sum_{m=0}^{2j}\sum_{l=0}^{2j-1}
 \wt\rho(T)_{m,2j,l,2j,\dn} \, d_{j}^{-s}  \\
 =&\sum_{2j=0}^{\infty}\big( \Tr_{2j}(T_+)\, \Tr_{2j+1}^{\up}(T_-)
 + \Tr_{2j+1}(T_+)\, \Tr_{2j}^\dn(T_-)\big) \, d_{j^+}^{-s} \,.
 \end{align*}
By Lemma \ref{tau0ext} $(ii)$, for all $k>0$,
\begin{align*}
\Tr_{2j}(T_\pm) &= (2j+\tfrac{3}{2})\tau_1(T_\pm) +
\tau_{0}(T_\pm)-\half \tau_1(T_\pm) +
\mathcal{O}\big((2j)^{-k}\big),\\
\Tr_{2j+1}(T_\pm) &= (2j+\tfrac{3}{2})\tau_1(T_\pm) +
\tau_{0}(T_\pm) +\half \tau_1(T_\pm) +
\mathcal{O}\big((2j)^{-k}\big)\, .
\end{align*}
The result follows by noting that the difference of the
Hurwitz zeta function $\zeta(s,\tfrac{3}{2})$ and Riemann zeta
function $\zeta(s)$ is
an entire function.

$(ii,iii,iv)$ are direct consequences of $(i)$.
\end{proof}

\subsection{The smooth algebra $C^{\infty}(SU_q(2))$}

In \cite{Cindex,SDLSV}, the smooth algebra $C^\infty(SU_q(2)$ is
defined by pulling back the smooth structure $C^\infty(D_{q^\pm}^2)$
into the $C^*$-algebra generated by $\A$, through the morphism $\rho$
and the application $\la$ (the compression which gives an operator on
$\H$ from an operator on $l^2(\N)\ox l^{2}(\N)\ox l^2(\Z)\ox \C^2$).
The important point is that with \cite[Lemma 2, p. 69]{Cindex}, this
algebra is stable by holomorphic calculus. By defining $\rho:=\wt
\rho\circ c$ and $\la(\cdot):= S (\cdot) I$, the same lemma (with
same notation) can be applied to our setting, with $c:= \pi(x)
\mapsto \ul\pi(x)$ and
\begin{align*}
&\mathcal{C}:=C^\infty(D_{q^+}^2)\ox C^\infty(S^1) \ox
C^\infty(D_{q^+}^2)\ox C^\infty(S^1) \ox \mathcal{M}_2(\C)
\end{align*}
as algebra stable by holomorphic calculus containing the image of
$\wt\rho$. Here, we use Schwartz sequences to define the smooth
structures. We finally obtain $C^\infty(SU_q(2))$ with real structure
as a subalgebra stable by holomorphic calculus of the $C^*$-algebra
generated by $\pi(\A) \cup J\pi(\A) J^{-1}$ and containing $\pi(\A)
\cup J \pi(\A) J^{-1}$ .

\begin{corollary}
The dimension spectrum of the real spectral triple
$\big(C^{\infty}(SU_q(2)),\H,D\big)$ is simple and given by
$\{1,2,3\}$. Its KO-dimension is $3$.
\end{corollary}

\begin{proof}
Since $F$ commutes with $\ul\pi(\A)$, the pseudodifferential
operators of order 0 (without the real structure and in the sense of
\cite{MCC}) are exactly (modulo $OP^{-\infty}$) the operators in $\B
+ \B F$. From Theorem \ref{Theo} we see that the dimension spectrum
of $SU_q(2)$ without taking into account the reality operator $J$ is
$\set{1,2,3}$, in other words, the possible poles of $\zeta^b_{\DD}:
s\mapsto \Tr(b F^{\eps}|\DD|^{-s})$ (with $\eps\in \set{0,1}$, $b\in
\B$) are in $\set{1,2,3}$. 
Theorem \ref{ncinttau} $(i)$ shows that the possible poles are still
$\set{1,2,3}$ when we take into account the real structure of
$SU_q(2)$, that is to say, when $\B$ is enlarged to $\B J\B J^{-1}$.
Indeed, any element of  $\B J\B J^{-1}$ is in $X\wh X$ and it is
clear from the preceding proof that adding $F$ in the previous zeta
function do not add any pole to $\set{1,2,3}$.

All arguments goes true from the polynomial algebra $\A(SU_q(2))$ to
the smooth pre-C$^*$-algebra $C^{\infty}(SU_q(2))$.

KO-dimension refers just to $J^2=-1$ and $\DD\,J=J\,\DD$ since there
is no chirality because spectral dimension is 3.
\end{proof}

\subsection{Noncommutative integrals with reality operator and
one-forms on $SU_q(2)$}

The goal of this section is to obtain the following suppression of
$J$:

\begin{theorem}
\label{ncintJ}
Let $A$ and $B$ be $\delta$-one-forms. Then
\begin{align*}
&  \hspace{-3cm} (i)\hspace{0.5cm} \ncint AJBJ^{-1} |\DD|^{-3}=
\tfrac 12  \ncint  A  \vert \DD \vert ^{-3} \overline{ \ncint B
\vert \DD \vert ^{-3}},\\
&   \hspace{-3cm}  (ii)\hspace{0.4cm}\ncint AJBJ^{-1} |\DD|^{-2}=
\half  \ncint  A  \vert \DD \vert
^{-2}\,\overline{\ncint  B  \vert \DD \vert ^{-3}} + \half\ncint   A
\vert \DD \vert ^{-3}\, \overline{\ncint B \vert \DD \vert
^{-2}},\\
&   \hspace{-3cm}  (iii)\hspace{0.3cm} \ncint A^2 JBJ^{-1}
|\DD|^{-3}=\tfrac 12 \ncint  A^2  \vert \DD
\vert ^{-3} \,\overline{\ncint B  \vert \DD \vert ^{-3}},\\
&    \hspace{-3cm} (iv)\hspace{0.4cm} \ncint \delta(A)A |\DD|^{-3}=
\ncint \delta(A)JAJ^{-1}|\DD|^{-3} = 0.
\end{align*}
\end{theorem}

We gather at the beginning of this section the main notations for
technical lemmas which will follow.

For any pair $(k,p)\in \N^3\times \N^3$ such that $k_i\leq |\a_i|$,
$p_i\leq |\beta_i|$, where $\a,\beta \in \Z\times\N \times \N$, we
define
\begin{align*}
v_{k,p} &:=  g(p)\
\genfrac(){0pt}{1}{|{\a_1}|}{k_1}_{q^{2\eta_{\a_1}}}
\genfrac(){0pt}{1}{{\a_2}}{k_2}
\genfrac(){0pt}{1}{{\a_3}}{k_3}\genfrac(){0pt}{1}{|{\beta_1}|}{p_1}_{q^{2\eta_{\beta_1}}}
\genfrac(){0pt}{1}{{\beta_2}}{p_2}
\genfrac(){0pt}{1}{{\beta_3}}{p_3} \,
(-1)^{k_1+p_1+{\a_2}+{\a_3}+{\beta_2}+{\beta_3}}\, q^{\sigma_{k,p}},\\
h_{k,p} &:= {\a_1}+{\a_2}-{\a_3} -2 (\eta_{\a_1} k_1 +k_2 -k_3)
+g(p)\, , \\
g(p) &:=  {\beta_1}+{\beta_2} -{\beta_3}-2(\eta_{\beta_1} p_1 +p_2
-p_3)\, ,\\
\sigma_{k,p}&:= k_1+p_1 +\sigma^t_{k,p} +\sigma^u_{k,p}\, , \\
\sigma^t_{k,p}&:= k_1\wh k_2-\wh k_3(k_1+k_2)+\eta_{\beta_1} \wh
p_1|k|_1 +\wh p_2(|k|_1+p_1) -\wh p_3(|k|_1+p_1+p_2)\, , \\
\sigma^u_{k,p}&:=(k_3+\eta_{\beta_1} \wh p_1 -p_2 +p_3)(k_1+\wh k_2
+\wh k_3) -k_2(k_1+\wh k_2) + (p_1+\wh p_2)(-p_2+p_3) +\wh p_3p_3 \,
,\\
t_{k,p}&= a_{\a_1}^{\wh k_1} \,a^{\wh k_2}\, {a^*}^{\wh k_3}\,
a_{\beta_1}^{\wh p_1}\, a^{\wh p_2} \,{a^*}^{\wh p_3}
\,b^{|k|_1+|p|_1}    \, ,\\
u_{k,p}&= a_{\a_1}^{\wh k_1} \,{a^*}^{k_2} \,a^{k_3}
\,a_{\beta_1}^{\wh p_1} \,{a^*}^{p_2} \,a^{p_3}\,b^{|\wt k|_1+|\wt
p|_1}   \, .
\end{align*}
where we used the notation
$$
\wh k_i:= |{\a_i}|-k_i, \,  \wh
p_i:=|{\beta_i}|-p_i,
$$
so $0\leq \wh k_i \leq |\a_i|,\, 0\leq  \wh
p_i \leq |\beta_i|$. We will also use the shortcut $\wt k:=(k_1,\wh
k_2,\wh k_3)$.

For ${\beta_1}\in \Z$ and $j\in \N$, we define
\begin{align*}
w_1({\beta_1},j) &:= \sum_{n=0}^\infty \big( q^{2jn}
(q_{n,|{\beta_1}|}^{\up_{\beta_1}})^{2}-\delta_{j,0}\big),\\
w_{\beta}^\a&:= 2{\beta_1}\,
q^{{\beta_1}(2{\a_3}+{\beta_3}-{\beta_2})}
\,w_1({\beta_1},{\a_3}+{\beta_3}).
\end{align*}

We introduce the following notations:
\begin{align*}
q^+_{k,p,n}&:=q^{n(|k|_1+|p|_1)} q_{n+r^+_{k,p} -\eta_{\a_1} \wh
k_1,\wh k_1}^{\up_{\a_1}} q_{n-\wh k_3+
\eta_{\beta_1} \wh p_1+\wh p_2-\wh p_3,\wh k_2}^\up
q_{n+\eta_{\beta_1} \wh p_1+\wh p_2-\wh p_3
,\wh k_3}^\dn q_{n+\wh p_2-\wh p_3,\wh p_1}^{\up_{\beta_1}} q_{n-\wh
p_3,\wh p_2}^{\up} q_{n,\wh p_3}^\dn\, ,\\
q^-_{k,p,n}&:=q^{n(|\wt k|_1+|\wt p|_1)} q_{n+r^-_{k,p}-\eta_{\a_1}
\wh k_1,\wh k_1}^{\up_{\a_1}}
q_{n+k_3+\eta_{\beta_1} \wh p_1-p_2 +p_3 ,k_2}^\dn
q_{n+\eta_{\beta_1} \wh p_1-p_2+p_3,k_3}^\up q_{n-p_2+p_3,\wh
p_1}^{\up_{\beta_1}} q_{n+p_3,p_2}^\dn q_{n,p_3}^\up\\
&\hspace{1cm}\times (-1)^{|\wt k|_1+|\wt p|_1},\\
r^+_{k,p}&:=\eta_{\a_1} \wh k_1+\wh k_2-\wh k_3+\eta_{\beta_1} \wh
p_1  +\wh p_2  -\wh p_3 \, , \\
r^-_{k,p}&:=\eta_{\a_1} \wh k_1 -k_2+k_3+\eta_{\beta_1} \wh
p_1-p_2+p_3\, .
\end{align*}
Thus, $\pi_+(t_{k,p})\eps_n = q^+_{k,p,n} \eps_{n+ r^+_{k,p}}$ and
$\pi_-(u_{k,p})\eps_n = q^-_{k,p,n} \eps_{n+ r^-_{k,p}}$.

\begin{lemma}
    \label{technicalxdy}
We have
$$
r\left( (M_{\beta}^\a)^\circ \right) =
\sum_{k,p}\delta_{h_{k,p},0}\, v_{k,p}\, \pi_+(t_{k,p}) \otimes
\pi_-(u_{k,p})
$$
where the summation is done on $k_i, p_i$ in $\N$ such that $k_i\leq
{|\a_i|}, p_i \leq {|\beta_i|}$ for $i\in \set{1,2,3}$.
\end{lemma}

\begin{proof}
Since $\ul\pi(m^\a)=
(a_++a_-)^{\a_1}(b_++b_-)^{\a_2}(b_+^*+b_-^*)^{\a_3}$, with
$v_k:=\genfrac(){0pt}{1}{|{\a_1}|}{k_1}_{q^{2\eta_{\a_1}}}
\genfrac(){0pt}{1}{{\a_2}}{k_2}
\genfrac(){0pt}{1}{{\a_3}}{k_3}$,
$$
\ul \pi(m^\a) = {\sum}_k v_k\, c_k
\text{ where }
c_k:=a_{+,{\a_1}}^{|{\a_1}|-k_1}\, a_{-,{\a_1}}^{k_1}
\,b_+^{{\a_2}-k_2} \,b_-^{k_2}\, {b_+^*}^{{\a_3}-k_3}
\,{b_-^*}^{k_3}.
$$
By Lemma \ref{commutateur} $(iii)$ we see that
$\delta(\ul \pi(m^\beta)) = \sum_p w_p\, d_p$ where we introduce

$w_p:=\genfrac(){0pt}{1}{|{\beta_1}|}{p_1}_{q^{2\eta_{\beta_1}}}\genfrac(){0pt}{1}{{\beta_2}}{p_2}
\genfrac(){0pt}{1}{{\beta_3}}{p_3}$ and
$d_p:=g(p)\,a_{+,{\beta_1}}^{|{\beta_1}|-p_1}\, a_{-,{\beta_1}}^{p_1}
\,b_+^{{\beta_2}-p_2} \,b_-^{p_2}\, {b_+^*}^{{\beta_3}-p_3}
\,{b_-^*}^{p_3} $.

As a consequence, $(M^{\a}_\beta)^\circ = \sum_{k,p}
\delta_{h(k,p),0}\  g(p)\  v_k \,w_p \,c_{k,p}$ where
\begin{equation}
    \label{ckp}
c_{k,p}=a_{+,{\a_1}}^{\wh k_1} \,a_{-,{\a_1}}^{k_1} \,b_+^{\wh k_2}\,
b_-^{k_2} \,{b_+^*}^{\wh k_3}\, {b_-^*}^{k_3}\,
a_{+,{\beta_1}}^{\wh p_1} \,a_{-,{\beta_1}}^{p_1}\,
b_+^{\wh p_2} \,b_-^{p_2}\, {b_+^*}^{\wh p_3} \,{b_-^*}^{p_3}
\end{equation}
With (\ref{ckp}), we get
$r(c_{k,p})= (-1)^{k_1+p_1+{\a_2}+{\a_3}+{\beta_2}+{\beta_3}}\,
q^{k_1+p_1}\, \pi_+(t'_{k,p}) \otimes \pi_-(u'_{k,p})$ where
\begin{align*}
t'_{k,p}&=  a_{\a_1}^{\wh k_1}\,b^{k_1}\,a^{\wh k_2}\, b^{k_2}\,
{a^*}^{\wh k_3}\,b^{k_3}
a_{\beta_1}^{\wh p_1}\,b^{p_1}\,a^{\wh p_2} \,b^{p_2} \,{a^*}^{\wh
p_3}\,b^{p_3}\, ,\\
u'_{k,p}&=  a_{\a_1}^{\wh k_1} \,b^{k_1}\,b^{\wh k_2}\,{a^*}^{k_2}
\,{b}^{\wh k_3}a^{k_3} \,a_{\beta_1}^{\wh p_1}\,b^{p_1}\,b^{\wh
p_2}\, {a^*}^{p_2} \,{b}^{\wh p_3}\,a^{p_3}\,  .
\end{align*}
A recursive use of relation $b a_j =q^{\eta_j} a_j b$ yields the
result.
\end{proof}

\begin{lemma}
    \label{ncintLin}
We have

(i) $(\tau_1 \otimes \tau_1) \big(r(M^{\a}_\beta)^\circ\big) =
{\beta_1}\
\delta_{{\a_1},-{\beta_1}}\,\delta_{{\a_2},0}\,\delta_{{\a_3},0}\,
\delta_{{\beta_2},0}\, \delta_{{\beta_3},0}\, .$

(ii) $(\tau_1 \otimes \tau_0 + \tau_0 \otimes \tau_1)
\big(r(M^{\a}_\beta)^\circ\big)
=\delta_{{\a_1},-{\beta_1}} \,
\delta_{{\a_2}+{\beta_2},{\a_3}+{\beta_3}}
\, w_\beta^\a.$

In particular, if $A$ is a $\delta$-one-form, we have
\begin{align*}
\ncint A |\DD|^{-3} &= 2 {\beta_1 } \, A_{-\beta_1 0 0 }^{\beta_1 0 0
} \,, \\ 
\ncint A |\DD|^{-2} &= 2 w^{\a}_\beta\, B(A)^{\beta}_\a  \,.
\end{align*}
where we implicitly summed on all $\a,\beta$ indices.
\end{lemma}

\begin{proof}
$(i)$ Using same notations of Lemma \ref{technicalxdy}, we obtain by
definition of $\tau_1$,
\begin{align}
\tau_1\big(\pi_+(t_{k,p})\big) &= \delta_{k,0} \,\delta_{p,0}\
\delta_{{\a_1}+{\a_2}-{\a_3} +{\beta_1} +{\beta_2}-{\beta_3},0} \, ,
\label{tau1t} \\
\tau_1\big(\pi_-(u_{k,p})\big) &= \delta_{\wt k,0} \,\delta_{\wt
p,0}\  \delta_{{\a_1}-{\a_2}+{\a_3} +{\beta_1}
-{\beta_2}+{\beta_3},0}\, .\label{tau1u}
\end{align}
We get  $\tau_1\big(\pi_+(t_{k,p})\big)
\,\tau_1\big(\pi_-(u_{k,p})\big) = \delta_{k,0}\,\delta_{p,0}\,
\delta_{{\a_2},0} \,\delta_{{\a_3},0} \,
\delta_{{\beta_2},0} \, \delta_{{\beta_3},0} \,
\delta_{{\a_1},-{\beta_1}}$,
so Lemma  \ref{technicalxdy} gives the result.

$(ii)$ Since $\pi_+(t_{k,p})\eps_n = q^+_{k,p,n} \eps_{n+ r^+_{k,p}}$
and $\pi_-(u_{k,p})\eps_n = q^-_{k,p,n} \eps_{n+ r^-_{k,p}}$, we get,
\begin{align}
\tau_0\big(\pi_+(t_{k,p})\big)&=\delta_{r^+_{k,p},0}\,
\sum_{n=0}^\infty \big(q^+_{k,p,n}-\, \delta_{k,0} \, \delta_{p,0}\,
\delta_{{\a_1}+{\a_2}-{\a_3} +{\beta_1} +{\beta_2}-{\beta_3},0}\big)
\, ,\label{tau0tkp}\\
\tau_0\big(\pi_-(u_{k,p})\big)&=\delta_{r^-_{k,p},0}\,
\sum_{n=0}^\infty \big(q^-_{k,p,n}-\, \delta_{\wt k,0}\, \delta_{\wt
p,0} \,
\delta_{{\a_1}-{\a_2}+{\a_3} +{\beta_1} -{\beta_2}+{\beta_3},0}\big)
\, .\label{tau0ukp}
\end{align}
With (\ref{tau1t}) and (\ref{tau0ukp}) we get
\begin{align*}
\tau_1\big(\pi_+(t_{k,p})\big) \, \tau_0\big(\pi_-(u_{k,p})\big) &=
\delta_{k,0}\,\delta_{p,0}\,
 \delta_{{\a_2}+{\beta_2},{\a_3}+{\beta_3}}
\,\delta_{{\a_1},-{\beta_1}} \sum_{n=0}^\infty \big( \delta_{k,0}
\,\delta_{p,0}\, q^-_{k,p,n}-\delta_{\a_3+\beta_3,0}\big)\\ &=
\delta_{k,0}\,\delta_{p,0}\,
 \delta_{{\a_2}+{\beta_2},{\a_3}+{\beta_3}}
\,\delta_{{\a_1},-{\beta_1}} w_1(\beta_1,\a_3+\beta_3).
 \end{align*}

Using (\ref{tau1u}) and (\ref{tau0tkp}),
\begin{align*}
\tau_0\big(\pi_+(t_{k,p})\big) \, \tau_1\big(\pi_-(u_{k,p})\big) &=
\delta_{\wt k,0}\,\delta_{\wt p,0}\,
 \delta_{{\a_2}+{\beta_2},{\a_3}+{\beta_3}}
\,\delta_{{\a_1},-{\beta_1}} \sum_{n=0}^\infty \big( \delta_{\wt k,0}
\,\delta_{\wt p,0}\, q^+_{k,p,n}-\delta_{\a_3+\beta_3,0}\big)\\ &=
\delta_{\wt k,0}\,\delta_{\wt p,0}\,
 \delta_{{\a_2}+{\beta_2},{\a_3}+{\beta_3}}
\,\delta_{{\a_1},-{\beta_1}} w_1(\beta_1,\a_3+\beta_3).
 \end{align*}
Lemma \ref{technicalxdy} yields the result.
\end{proof}

With notations of Lemma \ref{technicalxdy}, it is direct to check
that for given $\bar\a=(\a,\a',\cdots ,\a^{(n-1)})$
and $\bar\beta=(\beta,\beta',\cdots ,\beta^{(n-1)})$,
\begin{equation}
    \label{rproduct}
r\big((M^{\bar\a}_{\bar\beta} )^\circ \big)= \sum_{K,P}
\delta_{h_{K,P},0}\,v_{K,P}\, \pi_+(t_{K,P}) \otimes
\pi_-(u_{K,P})
\end{equation}
where $K=(k,k',\cdots k^{(n-1)})$, $P=(p,p',\cdots,p^{(n-1)})$ with
$0\leq
k^{(j)}_i \leq |\a^{(j)}_i|$, $0\leq p^{(j)}_i\leq |\beta^{(j)}_i|$,
\begin{align*}
t_{K,P}&:=t_{k,p} \,t_{k',p'}\cdots t_{k^{(n-1)},p^{(n-1)}} \,\,,&&
u_{K,P}:=u_{k,p}u_{k',p'}\cdots u_{k^{(n-1)},p^{(n-1)}} \,\,,\\
v_{K,P}&:=v_{k,p} \,v_{k',p'}\cdots v_{k^{(n-1)},p^{(n-1)}} \,\,,&&
h_{K,P}:= h_{k,p} + h_{k',p'}+\cdots h_{k^{(n-1)},p^{(n-1)}}\,\,.
\end{align*}

In the following, we will use the shortcuts
$A_i:=\a_i+\a'_i+\cdots +\a^{(n-1)}_i$,
$B_i:=\beta_i+\beta'_i+\cdots+\beta^{(n-1)}_i$.

In the case $n=2$, we also note $r^{\pm}_{K,P}:=  r^{\pm}_{k,p}+
r^{\pm}_{k',p'}$ and
$q^{\pm}_{K,P,n}:=q^{\pm}_{k',p',n}
\,q^{\pm}_{k,p,n+r^{\pm}_{k',p'}}$.

Thus,  we have $\pi_+(t_{K,P})\, \eps_m = q^+_{K,P,m} \, \eps_{m+
r^+_{K,P}}$ and $\pi_-(u_{K,P}) \, \eps_m = q^-_{K,P,n} \, \eps_{m+
r^-_{K,P}}$.

We also introduce, still for $n=2$,
\begin{align*}
v_{\beta_1,\a'_1,\beta'_1}(l,j)&:=\sum_{n=0}^\infty\big(q^{l+2nj}\,
q^{\up_{-\beta'_1-\a'_1-\beta_1}}_{n+\beta'_1+\a'_1+\beta_1,|\beta'_1+\a'_1+\beta_1|}
q^{\up_{\beta_1}}_{n+\beta'_1+\a'_1,|\beta_1|}
q^{\up_{\a'_1}}_{n+\beta'_1,|\a'_1|}
q^{\up_{\beta'_1}}_{n,|\beta'_1|}-\delta_{j,0}\big),
\\
V^{\bar \a}_{\bar \beta}
&:=2[\beta_1\beta'_1+(\beta_2-\beta_3)(\beta'_2-\beta'_3)]\,
q^{2{\beta_1}({\a_2}+{\a_3})+2{\beta'_1}({\a'_2}+{\a'_3})}\,\\
&\hspace{2cm}\times
v_{\beta_1,\a'_1,\beta'_1}((\a_2+\beta_2+\a_3+\beta_3)(\a'_1+\beta'_1),A_3+B_3).
\end{align*}

\begin{lemma} 
\label{ncintA2} 
We have

$(i) \quad (\tau_1 \otimes \tau_1)\, \big(r(M^{\a}_\beta
M^{\a'}_{\beta'})^\circ\big) =
{\beta_1}\beta'_1\
\delta_{A_1,-B_1}\,\delta_{A_2,0}\,\delta_{A_3,0}\,
\delta_{B_2,0} \delta_{B_3,0\, .}$

$(ii) \quad (\tau_1 \otimes \tau_0+\tau_0 \otimes \tau_1)\,  \big(
r(M^{\a}_\beta M^{\a'}_{\beta'})^\circ\big)
=\delta_{A_2+B_2,A_3+B_3} \delta_{A_1,-B_1} V^{\bar \a}_{\bar
\beta}$\,.

$(iii) \quad (\tau_1 \otimes \tau_1)\, \big(r(M^{\a}_\beta
M^{\a'}_{\beta'}M^{\a''}_{\beta''})^0 \big) =
{\beta_1}\beta'_1\beta''_1\
\delta_{A_1,-B_1}\,\delta_{A_2,0}\,\delta_{A_3,0}\,
\delta_{B_2,0} \delta_{B_3,0}.$

$(iv) \quad (\tau_{1}\otimes \tau_1)\  \big(r(\delta(M^{\a}_\beta)
M^{\a'}_{\beta'})^0 \big)
=-(\a'_1+\beta'_1){\beta_1}\beta'_1\
\delta_{A_1,-B_1}\,\delta_{A_2,0}\,\delta_{A_3,0}\,
\delta_{B_2,0} \delta_{B_3,0} .$

$(v)$ In particular, if $A$ is a $\delta$-one-form,
\begin{align*}
&\ncint A^2|\DD|^{-3}= 2\beta_1\beta'_1\ B_a(A^2)^{\bar \beta}_{\bar
\a} \,,\\
&\ncint A^2|\DD|^{-2} = 2 V_{\bar \beta}^{\bar\a}\  B(A^2)^{\bar
\beta}_{\bar \a}\, ,\\
&\ncint A^3|\DD|^{-3}= 2\beta_1\beta'_1\beta''_1\ B_a(A^3)^{\bar
\beta}_{\bar \a}\,,\\
&\ncint \delta(A) A |\DD|^{-3} =\ncint A\delta(A) |\DD|^{-3}=0.
\end{align*}
\end{lemma}

\begin{proof}
We have
\begin{align}
\tau_1(\pi_+(t_{K,P})) &= \delta_{K,0}\,\delta_{P,0}\,
\delta_{A_1+A_2-A_3+B_1+B_2-B_3,0}\label{tau1tKP}\,,\\
\tau_1(\pi_-(u_{ K, P})) &= \delta_{\wt K,0}\,\delta_{\wt P,0}\,
 \delta_{A_1-A_2+A_3+B_1-B_2+B_3,0}\label{tau1uKP} \,.
\end{align}
and
 \begin{align}
\tau_0(\pi_+(t_{K,P})) &= \delta_{r^+_{K,P},0} \sum_{n=0}^\infty
\big(q^+_{K,P,n}  -\delta_{K,0}\,\delta_{P,0} \,
\delta_{A_1+A_2-A_3+B_1+B_2-B_3,0}\big)\label{tau0tKP}\,,\\
\tau_0(\pi_-(u_{ K, P})) &=  \delta_{r^-_{K,P},0} \sum_{n=0}^\infty
\big(q^-_{K,P,n}  -\delta_{\wt K,0} \, \delta_{\wt P,0} \,
 \delta_{A_1-A_2+A_3+B_1-B_2+B_3,0}\big)\label{tau0uKP}\,.
\end{align}
$(i)$ Equations (\ref{tau1tKP}) and (\ref{tau1uKP}) give $(\tau_1
\otimes \tau_1)\, r(\ul A\, \ul A')^0
=\delta_{A_1,-B_1}\delta_{A_2,0}\delta_{A_3,0}\delta_{B_2,0}\delta_{B_3,0}\,
\la_{0,0}$. A computation of $v_{0,0}$ with
$\delta_{A_1,-B_1}\,\delta_{A_2,0}\,\delta_{A_3,0}\,\delta_{B_2,0}\,\delta_{B_3,0}=1$
gives the result.

$(ii)$ Equations (\ref{tau1tKP}) and (\ref{tau0uKP}) yield
\begin{align*}
\tau_1(\pi_+(t_{K,P}))\,\tau_0(\pi_-(u_{K,P}))&=\,
\delta_{K,0}\,\delta_{P,0} \,\delta_{A_2+B_2,A_3+B_3}
\,\delta_{A_1,-B_1} \\ &\hspace{2cm}\times
v_{\beta_1,\a'_1,\beta'_1}((\a_2+\beta_2+\a_3+\beta_3)(\a'_1+\beta'_1),A_3+B_3).
\end{align*}

Equations (\ref{tau0tKP}) and (\ref{tau1uKP}) yield
\begin{align*}
\tau_0(\pi_+(t_{K,P}))\,\tau_1(\pi_-(u_{K,P}))&=\, \delta_{\wt
K,0}\,\delta_{\wt P,0} \,\delta_{A_2+B_2,A_3+B_3} \,\delta_{A_1,-B_1}
\\ &\hspace{2cm}\times
v_{\beta_1,\a'_1,\beta'_1}((\a_2+\beta_2+\a_3+\beta_3)(\a'_1+\beta'_1),A_3+B_3)
\end{align*}
and the result follows.

$(iii)$ With (\ref{rproduct}) a direct computation gives
\begin{align}
\tau_1(\pi_+(t_{K,P})) &= \delta_{K,0}\,\delta_{P,0}
\,\delta_{A_1+A_2-A_3+B_1+B_2-B_3,0}\label{tau1tKP3} \,,\\
\tau_1(\pi_-(u_{ K, P})) &= \delta_{\wt K,0}\,\delta_{\wt P,0}\,
 \delta_{A_1-A_2+A_3+B_1-B_2+B_3,0}\label{tau1uKP3}\,.
\end{align}
Using (\ref{tau1tKP3}) and (\ref{tau1uKP3}), $(\tau_1 \otimes
\tau_1)\, \big(r(\ul A\, \ul A'\,  \ul A'')^\circ \big)
=\delta_{A_1,-B_1}\,\delta_{A_2,0}\,\delta_{A_3,0}\,\delta_{B_2,0}\,\delta_{B_3,0}\,
v_{0,0}$. A computation of $v_{0,0}$ with
$\delta_{A_1,-B_1}\,\delta_{A_2,0}\,\delta_{A_3,0}\,\delta_{B_2,0}\,\delta_{B_3,0}=1$
gives the result.

$(iv)$ We have $\delta(M^{\a}_\beta) M^{\a'}_{\beta'} =
\delta(x)\delta(y)
x'\delta(y') + x\delta^2(y) x'\delta(y')$ where $x,x',y,y'$ are
monomials ($\ul \pi$ omitted). Since
\begin{align*}
\ul \pi(x) = \sum_{k}
\genfrac(){0pt}{1}{\a}{k} a_{+,\a_1}^{\wh k_1} a_{-,\a_1}^{k_1}
b_+^{\wh k_2} b_-^{k_2} {b_+^*}^{\wh k_3} {b_-^*}^{k_3}=:\sum_k
\genfrac(){0pt}{1}{\a}{k} c_k,
\end{align*}
we get $\delta(\ul \pi(x)) =\sum_{k}g(k) \genfrac(){0pt}{1}{\a}{k}
\,c_k$.

Similarly, $\delta(\ul \pi(y)) =\sum_{p}g(p)
\genfrac(){0pt}{1}{\beta}{p}\,c_p$ and $\delta^2(\ul \pi(y))
=\sum_{p}g(p)^2 \genfrac(){0pt}{1}{\beta}{p}c_p$.

Thus, with $c_{K,P} :=c_k \,c_p \,c_{k'} \,c_{p'} $,
\begin{align*}
&\delta(x)\delta(y) x'\delta(y') = \sum_{K,P}
g(k)\,g(p)\,g(p')\genfrac(){0pt}{1}{\a}{K}\genfrac(){0pt}{1}{\beta}{P}\,
c_{K,P}\,,\\
&x\delta^2(y) x'\delta(y') = \sum_{K,P}
g(p)^2g(p')\genfrac(){0pt}{1}{\a}{K}\genfrac(){0pt}{1}{\beta}{P}\,
c_{K,P}\,,\\
& r(\delta(M^{\a}_\beta) M^{\a'}_{\beta'})^0 =
\sum_{K,P}\delta_{h_{K,P},0}
\big(g(k)+g(p)\big)\, g(p)\,
g(p')\,\genfrac(){0pt}{1}{\a}{K}\genfrac(){0pt}{1}{\beta}{P}
\,r(c_{K,P})=: \sum_{K,P} \la_{K,P}\  r(c_{K,P})\, .
\end{align*}
Since
$r(c_k) = (-q)^{k_1} (-1)^{{\a_2}+{\a_3}}\pi_+(t_k) \otimes
\pi_-(u_k)$ with $t_k,u_k$ defined by
$$
t_k:=a_{\a_1}^{\wh k_1}\, b^{k_1}\,a^{\wh k_2}\,b^{k_2}\,{a^*}^{\wh
k_3}\,b^{k_3} \text{  and } u_k:=a_{\a_1}^{\wh k_1}\,b^{k_1}\,b^{\wh
k_2}\,{a^*}^{k_2}\,{b}^{\wh k_3}\,a^{k_3},
$$
we get
$$
r(\delta(M^{\a}_\beta) M^{\a'}_{\beta'})^0 = \sum_{K,P}\la_{K,P}\,
(-q)^{k_1+k'_1+p_1+p'_1} (-1)^{A_2+A_3+B_2+B_3}\pi_+(t_{K,P}) \otimes
\pi_-(u_{K,P})
$$
where $t_{K,P}=t_k t_p t_{k'} t_{p'}$ and  $u_{K,P}=u_k u_p u_{k'}
u_{p'}$. Direct computations yield
\begin{align*}
\tau_1 \big(\pi_+(t_{K,P})\big) &= \delta_{K,0}\,\delta_{P,0}\,
\delta_{A_1+A_2-A_3+B_1+B_2-B_3,0}\,,\\
\tau_1 \big(\pi_-(u_{ K, P})\big) &= \delta_{\wt K,0}\,\delta_{\wt
P,0}\,
 \delta_{A_1-A_2+A_3+B_1-B_2+B_3,0}\,.
\end{align*}
The result follows.

$(v)$ For the last equality, note that by $(iv)$
$$
\ncint \delta(A) A |\DD|^{-3} = -2 \sum_{\a_1,\a'_1,\beta_1,\beta'_1}
(\a'_1+\beta'_1)
{\beta_1}\beta'_1\, A^{\beta_1 0 0}_{\a_1 0 0 0}\,A^{\beta'_1 0
0}_{\a'_1 0 0}\, \delta_{\a_1+\a'_1+\beta_1+\beta'_1,0}.
$$
The following change of variables $\a_1 \leftrightarrow \a'_1$,
$\beta_1\leftrightarrow \beta'_1$, implies by symmetry that this is
equal to zero.
\end{proof}

For a given $\delta$-1-form $A$, we say that $A$ is homogeneous of
degree in $a$ equal to $n\in \Z$ 
if it is a linear combination of $M_\beta^\alpha$ such that
$\a_1+\beta_1=n$. From Lemma \ref{ncintA2} $(iv)$ we get,

\begin{corollary}
    \label{commutation1}
Let $A$, $A'$ be two $\delta$-1-forms, then
\begin{align*}
 &\ncint (A |\DD|^{-1})^2 = \ncint A^2 |\DD|^{-2}\, ,\\
& \ncint  A |\DD|^{-1} A' |\DD|^{-1} = \ncint A A' |\DD|^{-2}
- n \ncint A A' |\DD|^{-3} \, , \text{ when $A'$
homogenous of degree $n$.}\\
\end{align*}
\end{corollary}

\begin{lemma}
\label{ncintJlem}
We have

(i)  $(\tau_1\otimes \tau_1)\,  \wt \rho \big(  M^{\a}_\beta J
M^{\a'}_{\beta'}
J^{-1}\big)={\beta_1}\beta'_1\
\delta_{\a_1,-\beta_1}\,\delta_{\a'_1,-\beta'_1}\,\delta_{A_2,0}\,\delta_{A_3,0}\,
\delta_{B_2,0} \delta_{B_3,0}\,.$

(ii)  $(\tau_0 \otimes \tau_1+\tau_1 \otimes \tau_0)\,  \wt \rho
\big(  M^{\a}_{\beta} J  M^{\a'}_{\beta'}  J^{-1}\big)
=\delta_{\a_1,-\beta_1}\,\delta_{\a'_1,-\beta'_1}
\big(\beta'_1 w^{\a}_{\beta}
\delta_{\a'_2+\beta'_2+\a'_3+\beta'_3,0}\,\delta_{\a_2+\beta_2,\a_3+\beta_3}$

\hspace{8cm} $+\beta_1
w^{\a'}_{\beta'}\delta_{\a_2+\beta_2+\a_3+\beta_3,0}
\,\delta_{\a'_2+\beta'_2,\a'_3+\beta'_3}\big)$.

(iii) $(\tau_1\otimes \tau_1)\,  \wt \rho \big(  M^{\a}_{\beta}
M^{\a'}_{\beta'} J M^{\a''}_{\beta''}
J^{-1}\big)={\beta_1}\beta'_1\beta''_1\
\delta_{\a_1+\a'_1,-\beta_1-\beta'_1}\,\delta_{\a''_1,-\beta''_1}\,
\delta_{A_2,0}\,\delta_{A_3,0}\,
\delta_{B_2,0} \delta_{B_3,0}\,.$

(iv) $(\tau_{1}\otimes \tau_1)\   \wt \rho \big(
\delta(M^{\a}_{\beta})
 J M^{\a'}_{\beta'} J^{-1}\big)
=-(\a'_1+\beta'_1){\beta_1}\beta'_1\
\delta_{\a_1,-\beta_1}\,\delta_{\a'_1,-\beta'_1}\,\delta_{A_2,0}\,
\delta_{A_3,0}\,
\delta_{B_2,0} \delta_{B_3,0}$ .

$(v)$ In particular, if $A$ and $A'$ are $\delta$-one forms, 
\begin{align*}
&\ncint AJA'J^{-1} |\DD|^{-3}= 2  (\beta_1
A_{-\beta_100}^{\beta_100})(\beta_1 \bar
{A'}_{-\beta_100}^{\beta_100}) ,\\
&\ncint AJA'J^{-1} |\DD|^{-2}= 2(\beta_1
\bar {A'}^{\beta_100}_{-\beta_100})(w^\a_\beta B(A)^{\beta}_\a)
+2(\beta_1 A^{\beta_100}_{-\beta_100})(w^\a_\beta B(\bar
{A'})^{\beta}_\a),\\
&\ncint A^2 JA'J^{-1} |\DD|^{-3}=2(\beta_1
\bar {A'}_{-\beta_100}^{\beta_100}) (\beta_1\beta'_1
B_a(A^2)^{\bar\beta}_{\bar\a}), \\
&\ncint \delta(A)JAJ^{-1} = 0.
\end{align*}

\end{lemma}

\begin{proof}
$(i)$ Following notations of Lemma \ref{technicalxdy}, we have
$$
 M^{\a}_\beta J M^{\a'}_{\beta'} J^{-1}= \sum_{K,P} v_{K,P}
\,c_{k,p}Jc_{k',p'}J^{-1}
$$
where $K=(k,k')$, $P=(p,p')$, $\la_{K,P}=
g(p)g(p')v_{k}v_{k'}w_pw_{p'}$. Thus,
$$
 \wt\rho(M^{\a}_\beta J M^{\a'}_{\beta'} J^{-1})=
(-1)^{A_2+A_3+B_2+B_3}\sum_{K,P}(-q)^{k_1+k'_1+p_1+p'_1} \la_{K,P} \,
T^+_{K,P}\otimes T^-_{K,P}
$$
where $T^+_{K,P}:=\pi'_+(t_k t_p)\wh \pi_{+}(t_{k'} t_{p'})$
 and $T^-_{K,P}:=\pi'_-(u_k u_p)\wh \pi_{-}(u_{k'} u_{p'})$ with
\begin{align*}
t_k&:=a_{\a_1}^{\wh k_1}\,{b^{*}}_{\a_1}^{k_1}\,
 a^{\wh k_2}\, {b^*}^{k_2}\, {a^*}^{\wh k_3} b^{k_3}, \\
u_k&:=a_{\a_1}^{\wh k_1}\,{b^{*}}_{\a_1}^{k_1}\,
 b^{\wh k_2}\, {a^*}^{k_2}\, {b^*}^{\wh k_3} a^{k_3} .
\end{align*}
A direct computation leads to
\begin{align*}
\tau_1(T^+_{K,P})&=\delta_{K,0} \,\delta_{P,0}\
\delta_{{\a_1}+{\a_2}-{\a_3} +{\beta_1} +{\beta_2}-{\beta_3},0} \,
\delta_{{\a'_1}+{\a'_2}-{\a'_3} +{\beta'_1}
+{\beta'_2}-{\beta'_3},0}\, ,\\
\tau_1(T^-_{K,P})&=\delta_{\wt K,0} \,\delta_{\wt
P,0}\  \delta_{{\a_1}-{\a_2}+{\a_3} +{\beta_1}
-{\beta_2}+{\beta_3},0} \, \delta_{{\a'_1}-{\a'_2}+{\a'_3} +{\beta'_1}
-{\beta'_2}+{\beta'_3},0}
\end{align*}
which gives the result.

$(ii)$
Using the commutation relations on $\A$, we see that there are real
functions of $(K,P)$, noted $\sigma^t_{K,P}$ and $\sigma^u_{K,P}$
such that
\begin{align*}
T^+_{K,P} &= q^{\sigma_{K,P}^t}\, \pi'_+(t_{k,p})\wh \pi_+(t_{k',p'}),
\\
T^-_{K,P} &= q^{\sigma_{K,P}^u}\, \pi'_-(u_{k,p}) \wh
\pi_-(u_{k',p'}),\\
t_{k,p}&:= a_{\a_1}^{\wh k_1}\,
 a^{\wh k_2}\,  {a^*}^{\wh k_3}\,a_{\beta_1}^{\wh p_1}\,
 a^{\wh p_2}\,  {a^*}^{\wh p_3}\,
{b^{*}}_{\a_1}^{k_1}\,{b^{*}}_{\beta_1}^{p_1}\,
{b^*}^{k_2+p_2}b^{k_3+p_3},\\
u_{k,p}&:=a_{\a_1}^{\wh k_1}\,
 {a^*}^{ k_2}\,  {a}^{k_3}\,a_{\beta_1}^{\wh p_1}\,
 {a^*}^{ p_2}\,  {a}^{p_3}\,
{b^{*}}_{\a_1}^{k_1}\,{b^{*}}_{\beta_1}^{p_1}\, {b}^{\wh k_2+\wh
p_2}{b^*}^{\wh k_3+\wh p_3}.
\end{align*}

We have, under the hypothesis $\tau_1(T^-_{K,P})=1$,
\begin{align*}
&\wh \pi_+(t_{k',p'})\eps_{m,2j} =
(-1)^{\la'}q^{(2j-m)\la'}\,q^{\up_{\a'_1}}_{2j-m-s+\beta'_1,|\a'_1|}\,
q^{\up_{\beta'_1}}_{2j-m-s,|\beta'_1|}\,\eps_{m+s,2j}\, ,\\
&s:=- \a'_2 + \a'_3 -\beta'_2+\beta'_3=\a'_1+\beta'_1 \,,\\
&\la':=\a'_2+\a'_3+\beta'_2+\beta'_3\,,\\
&\la:=\a_2+\a_3+\beta_2+\beta_3 \,\\
&\tau_1(T^+_{K,P})=  \delta_{\la,0}\, \delta_{\la',0}\, .
\end{align*}
and then,
\begin{align*}
&(T^+_{K,P})_{m,2j}=q^{\sigma^t_{K,P}+s\la}(-1)^{\la'}
q^{(2j-m)\la'+m\la}\,F_m \,F'_{2j-m}\,
\delta_{A_1+B_1,0}\,,\\
&F'_{2j-m}:=q^{\up_{\a'_1}}_{2j-m-\a'_1,|\a'_1|}\,
q^{\up_{\beta'_1}}_{2j-m-\a'_1-\beta'_1,|\beta'_1|}\,,\\
&F_{m}:=
q^{\up_{\a_1}}_{m-\a_1,|\a_1|}q^{\up_{\beta_1}}_{m-\beta_1-\a_1,|\beta_1|}\,.
\end{align*}
Following the proof of Lemma \ref{tau0ext}, we see that
$\tau_0(T^+_{K,P})$ is possibly nonzero only in the two cases
$\la'=0$ or $\la=0$.

Suppose first $\la=\la'=0$. In that case, we have
$$
\tau_0(T^+_{K,P})= \lim_{2j\to\infty} \sum_{m=0}^{2j}
\big((q^{\up_{\beta_1}}_{m,|\beta_1|}
q^{\up_{\beta'_1}}_{2j-m,|\beta'_1|})^2 -1\big) =
\sum_{m=0}^{\infty}\big((q^{\up_{\beta_1}}_{m,|\beta_1|})^2 -1\big) +
\sum_{m=0}^{\infty}\big((q^{\up_{\beta'_1}}_{m,|\beta'_1|})^2 -1\big)
$$
where the second equality comes from Lemma \ref{tech-separation}.

In the case $(\la=0,\la'>0)$, we get $\a'_1=-\beta'_1$ and thus,
$$
(T^+_{K,P})_{m,2j}=q^{\sigma^t_{K,P}} q^{m\la}\,
(q^{\up_{\beta_1}}_{m,|\beta_1|}
q^{\up_{\beta'_1}}_{2j-m,|\beta'_1|})^2
\delta_{\a_1+\beta_1,0}.
$$
Let us note $U_{2j} = \sum_{m=0}^{2j} q^{m\la}\,
(q^{\up_{\beta_1}}_{m,|\beta_1|}
q^{\up_{\beta'_1}}_{2j-m,|\beta'_1|})^2 $
and $L_{2j} = \sum_{m=0}^{2j} q^{m\la}\,
(q^{\up_{\beta_1}}_{m,|\beta_1|})^2$.

Suppose $\beta'_1>0$.
Since $(q^{\up_{\beta'_1}}_{2j-m,|\beta'_1|})^2-1 = \sum_{|p|_1 \neq
0,
p_i\in \set{0,1}} (-1)^{|p|_1}q^{r_p}\, q^{2(2j-m)|p|_1}$ where we
have
$r_p=2+\cdots +2\beta'_1$.
As in the proof of Lemma \ref{tau0ext} $(ii)$, we can conclude that
$U_{2j}-L_{2j}$ converges to 0.
The case $\beta'_1\leq 0$ is similar.

In the other case $(\la>0,\la'=0)$, the
arguments are the same, replacing $\la$ by $\la'$ and $\a_1$,
$\beta_1$ by $\a'_1$,
$\beta'_1$. Finally,
\begin{align*}
&\tau_0(T^+_{K,P})\tau_1(T^-_{K,P}) = \delta_{\wt K,0}\,\delta_{\wt
P,0}\,\delta_{\a_1,-\beta_1}\,\delta_{\a'_1,-\beta'_1}
(\delta_{\la',0}\,\delta_{\a_2+\beta_2,\a_3+\beta_3}\,
s_{\a,\beta}+\delta_{\la,0}\,\delta_{\a'_2+\beta'_2,\a'_3+\beta'_3}
\,s_{\a',\beta'}),\\
&s_{\a\beta}:= q^{\beta_1(\a_3-\a_2)}\,\sum_{m=0}^{\infty}\big(
q^{m\la}\, (q^{\up_{\beta_1}}_{m,|\beta_1|})^2
-\delta_{\la,0}\big)\, .
\end{align*}
A similar computation of $\tau_0(T^-_{K,P})$ can be done following
the same arguments. We find
eventually
$$
\tau_1(T^+_{K,P})\tau_0(T^-_{K,P}) = \delta_{ K,0}\,\delta_{
P,0}\,\delta_{\a_1,-\beta_1}\,\delta_{\a'_1,-\beta'_1}
(\delta_{\la',0}\,\delta_{\a_2+\beta_2,\a_3+\beta_3}\,
s_{\a,\beta}+\delta_{\la,0}\,\delta_{\a'_2+\beta'_2,\a'_3+\beta'_3}
\,s_{\a',\beta'})
$$
and the result follows.

$(iii)$ The same arguments of $(i)$ apply here with minor changes.

$(iv)$ follows from a slight modification of the proof of Lemma
\ref{ncintA2} $(iv)$.

$(v)$ is a straightforward consequence of $(i,ii,iii,iv)$. 
\end{proof}

\begin{lemma}
\label{tech-separation}
Let $\beta,\beta' \in \Z$. Then,
$$\lim_{2j\to\infty} \sum_{m=0}^{2j} \big((q^{\up_{\beta}}_{m,|\beta|}
q^{\up_{\beta'}}_{2j-m,|\beta'|})^2 -1\big) =
\sum_{m=0}^{\infty}\big((q^{\up_{\beta}}_{m,|\beta|})^2 -1\big) +
\sum_{m=0}^{\infty}\big((q^{\up_{\beta'}}_{m,|\beta'|})^2 -1\big).$$
\end{lemma}

\begin{proof} We give a proof for $\beta$ and $\beta'>0$, the other
cases being similar. 

Since
$(q^{\up_{\beta}}_{m,|\beta|})^2= \sum_{p_i\in \set{0,1}}
(-1)^{|p|_1}q^{r_p} q^{2|p|_1 m}$ where
$p=(p_1,\cdots,p_{\beta})$ and $r_p:=2(p_1+\cdots+\beta p_{\beta})$,
we get, with the notations
$\la_{p,p'}:=(-1)^{|p+p'|_1}q^{r_p+r_{p'}}$ and $U_{2j}:=
\sum_{m=0}^{2j}\,(q^{\up_{\beta}}_{m,|\beta|}q^{\up_{\beta'}}_{2j-m,|\beta'|})^2
-1$,
\begin{align*}
 U_{2j}&=
\sum_{m=0}^{2j}\sum_{|p+p'|_1>0} \la_{p,p'}
q^{2|p|_1 m +2|p'|_1(2j-m)}\\
&= \sum_{|p|_1\geq |p'|_1, |p|_1>0} \la_{p,p'} V_{2j,p,p'} +
\sum_{|p|_1< |p'|_1, |p'|_1>0}\la_{p,p'}
V'_{2j,p,p'}
\end{align*}
where
$$
V_{2j,p,p'}=q^{4j|p'|_1}\sum_{m=0}^{2j}
q^{2(|p|_1-|p'|_1) m}, \qquad V'_{2j,p,p'}=q^{4j|p|_1}\sum_{m=0}^{2j}
q^{2(|p'|_1-|p|_1) m}.
$$
It is clear that $V_{2j,p,p'}$ has $0$ for limit when $j\to \infty$
when $|p'|_1>0$, and
$V'_{2j,p,p'}$ has $0$ for limit when $j\to \infty$ when $|p|_1>0$.
As a consequence,
\begin{align*}
 U_{2j}&= \sum_{|p|_1>0} \la_{p,0} V_{2j,p,0} + \sum_{
|p'|_1>0}\la_{0,p'}
V'_{2j,0,p'} + o(1).
\end{align*}
The result follows as

$\sum_{m=0}^{2j}\big((q^{\up_{\beta}}_{m,|\beta|})^2 -1\big) =
\sum_{|p|_1>0} \la_{p,0} V_{2j,p,0}$ and
$\sum_{m=0}^{2j}\big((q^{\up_{\beta'}}_{m,|\beta'|})^2 -1\big)
=\sum_{ |p'|_1>0}\la_{0,p'}
V'_{2j,0,p'}$.
\end{proof}

\begin{proof}[Proof of Theorem \ref{ncintJ}]
The result follows from Lemmas \ref{ncintLin}, \ref{ncintA2} $(v)$
and \ref{ncintJlem} $(v)$.
\end{proof}

\subsection{Proof of Theorem \ref{mainThmJ} and corollaries}

\begin{lemma}
\label{ncintD}
We have on $SU_q(2)$,

(i) \hspace{0.3cm} $\ncint  |\DD|^{-3} = 2 $.

(ii) \hspace{0.2cm} $\ncint  |\DD|^{-2} = 0 $.

(iii) \hspace{0.1cm}  $\ncint  |\DD|^{-1} = -\half $.

(iv) \hspace{0.1cm} $\zeta_\DD(0)=0$.
\end{lemma}

\begin{proof} 
$(iv)$ We have by definition
$$
\zeta_\DD(s) := \Tr(\vert \DD
\vert^{-s})=\sum_{2j=0}^{\infty}\sum_{m=0}^{2j}\sum_{l=0}^{2j+1}\,
\langle v^{j}_{m,l}\,,\, |\DD|^{-s}\, v^{j}_{m,l}\rangle.
$$
Since $|\DD|^{-s} v^j_{m,l} =\genfrac{(}{)}{0pt}{1}{d_{j^+}^{-s}
\quad
0\,\,}{\, 0  \quad d_j^{-s}} \, v^j_{m,l}$ where
$d_j:=2j+\half$, we get
$$
\zeta_\DD(s) = \sum_{2j=0}^{\infty} (2j+1)(2j+2)\,d_{j^+}^{-s}+
\sum_{2j=1}^{\infty} (2j+1)(2j) \,d_j^{-s}=2\sum_{2j=0}^{\infty}
(2j+1)(2j) \,d_j^{-s}.
$$
With the equalities $(2j+1)(2j) = d_{j}^2 - \tfrac{1}{4}$ and
$\zeta(s,\half)=(2^s-1)\zeta(s)$
(here $\zeta(s,x):=\sum_{n\in\N}\tfrac{1}{(n+x)^s}$ is the Hurwitz
zeta function and $\zeta(s):=\zeta(s,1)$ is the Riemann zeta
function) we get
\begin{equation}\label{zetaDs}
\zeta_\DD(s) = 2(2^{s-2}-1)\zeta(s-2)-\half(2^s-1)\zeta(s)
\end{equation}
which entails that $\zeta_\DD(0)=0$.

$(i,ii,iii)$ are direct consequences of equation (\ref{zetaDs}). 
\end{proof}
\begin{proof}[Proof of Theorem \ref{mainThmJ}] It is a consequence of
Lemma \ref{ncintD} and Theorems \ref{coeffsASJ}, \ref{ncintJ}.
\end{proof}
As we have seen, the computation of noncommutative integral on
$SU_q(2)$ leads to certain function of $A$ which filter some symmetry
on the degree in $a$, $a^*$, $b$, $b^*$ of the canonical
decomposition.
Precisely, it is the balanced features that appear and the following
functions of $A^n$, $n\in \set{1,2,3}$:
\begin{align}
\label{I_pA^n}
\ncint A^n |\DD|^{-p}
\end{align}
where $1\leq n\leq p \leq 3$.
We will see in the next section a method for the computation of these
integrals.

\begin{corollary}
Let $u$ be a unitary in $C^\infty\big(SU_q(2)\big)$ and
$\ga_{u}(\Abb):=\pi(u)\Abb \pi(u^*)+\pi(u)d\pi(u^{*})$ be a
gauge-variant of $\Abb$. Then the
following term of Theorem \ref{mainThmJ} are gauge invariant
\begin{align*}
&\ncint A |\DD|^{-3}\, ,\qquad
\ncint A^2 |\DD|^{-3} -\ncint A |\DD|^{-2}\,,\qquad 
&-2\ncint A |\DD|^{-1}  + \ncint A^2 |\DD|^{-2} -\tfrac{2}{3} \ncint
A^3 |\DD|^{-3}.
\end{align*}
\end{corollary}

\begin{proof}
It is sufficient to remark that all terms
$\ncint \vert D_{\Abb}\vert ^{-k}$ and $\zeta_{\DD_{\Abb}}(0)$
in spectral action
\eqref{formuleaction} are gauge invariant. This can also be seen
via the computation $D_{\ga_{u}(\Abb)}=V_{u}\DD V_{u}^*
+V_{u}P_{0}V_{u}^*$ where $P_{0}$ is the projection on Ker$\,\DD$ and
$V_{u}=\pi(u)J\pi(u)J^{-1}$ and $\ncint \vert D_{\Abb}\vert ^{n-k}
=\Res_{s=n-k} \, \Tr \, \big(\vert D_{\Abb}\vert ^{n-k}\big)$
(see \cite[Prop. 5.1 (iii) and Prop. 4.8]{MCC}.)
\end{proof}

\begin{corollary}
    \label{sansJ}
In the case of the spectral action without the reality operator (i.e.
$D_{\Abb}=\DD+\Abb$), we get
\begin{align*}
&  \ncint \vert D_{\Abb}\vert^{-3} =2,\qquad \ncint
\vert D_{\Abb}\vert^{-2}  =  - 2\, \ncint A |\DD|^{-3}\, ,\qquad
\ncint \vert D_{\Abb}\vert^{-1}  = -\half - \ncint A |\DD|^{-2} +
\ncint A^2 |\DD|^{-3},\\
&\zeta_{D_{\Abb}}(0) = - \ncint A |\DD|^{-1} + \half \ncint A^2
|\DD|^{-2}   -\tfrac{1}{3} \ncint A^3 |\DD|^{-3} \, .
\end{align*}
As a consequence, if $\Abb$ is a one-form such that $\ncint A
|\DD|^{-3} =0$,
then the scale invariant term of the
spectral action with or without $J$ is exactly the same
modulo a global factor of 2.
\end{corollary}

\section{Differential calculus on $SU_q(2)$ and applications}

\subsection{The sign of $\DD$ }

There are multiple differential calculi on $SU_{q}(2)$, see
\cite{Woronowicz, Klimyk}. Thanks to \cite[Theorem 3]{Schmuedgen},
the $3D$ and $4D_{\pm}$ differential calculi do not coincide with the
one considers here: the right multiplication of one-forms by an
element in
the algebra $A$ is a consequence of the chosen Dirac operator which
was introduced according to some equivariance properties with respect
to the duality between the two Hopf algebras $SU_{q}(2)$ and
$\U_{q}(su(2))$.

It is known that the Fredholm module associated to $(\A,\, \H,\,
\DD)$ is one-summable since $[F,\pi(x)]$ is trace-class for all $x \in
\A$. In fact, more can be said about $F$
\footnote{Note that a similar result for a different spectral triple
over $SU_q(2)$ when $q=0$ was obtained in \cite[eq. (48)]{Cindex}}:

\begin{prop}
Since
\begin{align}
\label{eq:F2}
F=\tfrac{1}{1-q^{2}}\, \left(\ul\pi(a^*) \,d\ul\pi(a)+ q^2
\,\ul\pi(b) \,d\ul\pi(b^*)+q^2 \, \ul\pi(a) \,d\ul\pi(a^*)+ q^2
\,\ul\pi(b^*) \,d\ul\pi(b) \right),
\end{align}
$F$ is a central one-form modulo $OP^{-\infty}$.
\end{prop}

\begin{proof}
Forgetting $\piappr$, this follows from
\begin{align}
a^* \,\delta a+ q^2 \,b \,\delta b^* +q^2 \, &a \,\delta a^* + q^2
\,b^* \,\delta b \nonumber\\
&=(a_{+}^*+a_{-}^*)(a_{+}-a_{-})+q^2\,(b_{+}+b_{-})(b_{-}^*-b_{+}^*)+q^2\,
(a_{+}+a_{-})(a_{-}^*-a_{+}^*)\nonumber\\
&\hspace{3.9cm} +q^2\,(b_{+}^*+b_{-}^*)(b_{+}-b_{-})\nonumber\\
&=[a_{+}^{*}a_{+}-q^{2}\,a_{+}a_{+}^{*}+q^2\,b_{+}^{*}b_{+}
-q^2\,b_{+}b_{+}^{*}]+R=(1-q^2) +R
\label{CalculF}
\end{align}
by \eqref{astuce1} where we check that the remainder $R$ is zero:
\begin{align*}
R=&-[a_{+}^*a_{-}+q^2\,b_{+}^*b_{-}]+[a_{-}^*a_{+}+q^2\,b_{-}^*b_{+}]
-[a_{-}^{*}a_{-}-q^{2}\,a_{-}a_{-}^{*}+q^2\,b_{-}^{*}b_{-}
-q^2\, b_{-}b_{-}^{*}] \\
&\hspace{1.5cm}+(q^2\,a_{+}a_{-}^*+q^2\,q_{-}^*b_{+})
-(a_{+}^*a_{-}+q^2\, b_{+}^*b_{-}),
\end{align*}
thus, applying \eqref{astuce2}, \eqref{astuce3},
\eqref{astuce4}, $R=+(q^2\,a_{+}a_{-}^*+q^2\,q_{-}^*b_{+})
-(a_{+}^*a_{-}+q^2\, b_{+}^*b_{-})=0$ using commutation
relations \eqref{eq:commutrulefora,b}.

Now, replacing $\delta$ by $d$ in \eqref{CalculF} gives \eqref{eq:F2}
since $F$ commute with $a_{\pm},\, b_{\pm}$ and $F$ is central
by \eqref{Fcommutes}.
\end{proof}

\begin{prop}
The one-form in \eqref{eq:F2} is in fact exactly a function of the
Dirac operator $D$:
\begin{align}
    \label{eq:F3}
\pi(a^*) \,d\pi(a)+ q^2 \,\pi(b) \,d\pi(b^*)+q^2
\, \pi(a) \,d\pi(a^*) + q^2 \,\pi(b^*) \,d\pi(b)
= \xi_q(\DD)=F\, \xi_q (\vert \DD \vert),
\end{align}
where $ \xi_q(s) := q \tfrac{[2s] - 2s}{[s+1/2] [s-1/2]}$.

Moreover,  $F=\lim_{q \to 0} \xi_q(D)$.
\end{prop}

\begin{proof}
First, let us observe that the one-form $\omega$ in (\ref{eq:F3}) is
invariant under the action of the $\Uq \times \Uq$: $h \triangleright
\omega=\epsilon(h)\, \omega$ for any $h\in \Uq \times \Uq$. For
instance, using notations of \cite{DLSSV}
\begin{align*}
e \triangleright \omega= q^{\oh} a^* db + q^2 \left( - q^{\oh-1} b
da^*
+ q^{-\oh} b da^* - q^{-1-\oh} a^* db \right) = 0=\epsilon(e) \,
\omega.
\end{align*}
Therefore, since both the representation $\pi$ as well as the
operator $D$ are equivariant, the image of $\omega$ must
be diagonal in the spinorial base.  
A tedious computation with the full spinorial representation $\pi$
given in \eqref{eq:reprexact} yields
\begin{align*}
\langle v^{j \up}_{ml} ,\,\omega\, v^{j \up}_{ml} \rangle &=
\tfrac{q^{8j+8}-q^{8j+6}-(4j+3)\,q^{4j+6}+(8j+6)\,q^{4j+4}-(4j+3)\,q^{4j+2}-q^2+1}{(q^{4j+4}-1)(q^{4j+2}-1)}=\xi_q(2j+\tfrac
32),\\ 
\,\langle v^{j \dn}_{ml} ,\,\omega\, v^{j \dn}_{ml} \rangle
&=\tfrac{-q^{8j+4}+q^{8j+2}+(4j+1)\,q^{4j+4}-(8j+2)\,q^{4j+2}+(4j+1)\,q^{4j}+q^2-1}{(q^{4j+2}-1)(q^{4j}-1)}=-\xi_q(2j+\tfrac
12).
\end{align*}
These expressions have a clear $q=0$ limit equal respectively to 1
and -1, so $\omega \to F$ as $q \to 0$.
\end{proof}

In the $q=1$ limit, these expressions yields identically
$0$, which is confirmed by the fact that all one-forms are central,
it could be expressed as $d( a a^* + b b^*) = d\,1$.

Note that since the invariant one-form we constructed differs
by $OP^{-\infty}$ from $F$, hence any commutator with it will
be itself in $OP^{-\infty}$.

We do not know if a central form $\omega$ is automatically invariant
by the action of both $U_q(su(2))$, that is: 
$h \triangleright \omega = \epsilon(h) \omega$.

\begin{prop}
The order one calculus up to $OP^{-\infty}$ is not universal.
\end{prop}

\begin{proof}
Let us take the one-form $\omega_F$ from (\ref{eq:F2}),
which gives $F$. Then, for any $x \in \A(SU_q(2))$ we
have
$ \ul\pi(x \omega_F - \omega_F x) = 0$.
\end{proof}

\begin{corollary}
Still modulo $OP^{-\infty}$, $1 \in \pi\big(\Omega^2_{u}(\A)\big)$.
\end{corollary}

\begin{proof}
$1=F^2$ is by definition in $\pi\big(\Omega^2_{u}(\A)\big)$.
\end{proof}
In fact, one checks, using \eqref{astuce1}, \eqref{astuce2},
\eqref{astuce5} that
\begin{align}
    \label{eq:1is2form}
q^2\, da\, da^*-da^*\, da=1-q^2
\end{align}
showing again that $1 \in \pi\big(\Omega^2_{u}(\A)\big)$.

Similarly, using
\eqref{eq:commutrulefora,b} and \eqref{astuce1'},
\eqref{astuce5}, \eqref{astuce6}, we get still up to $OP^{-\infty}$
\begin{align}
    \label{2forms}
&q \,da \, db = db \, da, && q \,da \, db^* = db^* \, da, \nonumber\\
&da^* \, db = q \, db \,da^*, && da^* \, db^* = q \, db^* \, da^*
\nonumber\\
&db \, db^* = db^* \,db, &&  da \, da^* +db \, db^* = -1.
\end{align}
The use of the last equality of \eqref{2forms} and
\eqref{eq:1is2form} gives

\begin{prop}
Up to $OP^{-\infty}$, $F$ is not a (universal) closed one-form, as
\begin{align}
\label{dF}
da^* \,da+q^2 \, da \,da^*+ q^2 \,db^* \,db+ q^2 \,db \,db^*  =
-1-q^2.
\end{align}
\end{prop}

\subsection{The ideal $\CR$}

In order to perform explicit calculations of all terms of
the spectral action, we observe that each $\delta$-one-form
could be expressed in terms of $x \delta(z) y$, where $z$ is one
of the generators $a,a^*,b,b^*$ and $x,y$ are some elements of
the algebra $\A(SU_q(2))$.

Then, for the computation of $\ncint x dz y |D|^{-1}$
we can use the trace property of the noncommutative integral to get:
$$\ncint x \delta(z) y \,|D|^{-1} = \ncint yx \delta(z) \,|D|^{-1} +
\ncint x \delta(z) \,|D|^{-1} \delta(y) \, |D|^{-1}.$$

Therefore, the problem of calculating the tadpole-like integral could
be in effect reduced to the calculation of much simpler integrals:
$ \ncint x \delta(z)$  for all generators $z$ and the integrals
of higher order in $|D|^{-1}$.

However, it appears that the calculations of higher-order terms
simplify
a lot, when we further restrict the algebra by introducing an ideal,
which is {\em invisible} to the parts of integral at dimension $2$ and
$3$. For instance, consider the space of
pseudodifferential operators
$T \in \Psi^{0}(\A)$ of order less or equal to zero (see \cite{CM}),
which satisfy
\begin{align}
    \label{ideal}
\ncint T\, t \,|D|^{-2} =\ncint t\,T \,|D|^{-2}
= \ncint T\, t \,|D|^{-3} = \ncint t\, T \,|D|^{-3} = 0,
 \;\forall t \in \Psi^0_{0}(\A).
\end{align}
The elements $a_{-}, \,\, b_- b_+,\,
\, b_{-}b_{+}^*$ and their adjoints
are in this space up to $OP^{-\infty}$: this is due to the fact that
in Theorem \ref{Theo},
$\tau_{1}\otimes \tau_1\big(r(x)\big)=0$ when $r(x)\in
\pi_{\pm}(\A)\otimes \pi_{\pm}(\A)$ mod $OP^{-\infty}$ 
contains tensor products of $\pi_\pm (b)$ or $\pi_\pm(b^*)$ since
these elements are in the kernel of the grading $\sigma$.
\begin{definition}
    \label{Rem:onR}
Let $R$ be the kernel in $X$ of $(\sigma \otimes \sigma)\circ r$
where $r$ is the Hopf-map defined in \eqref{eq:r} and $\sigma$ is the
symbol map and let $\CR$  be the vector space generated by $R$ and
$R\,F$.
\end{definition}

Note that $R$ is a $^*$-ideal in $X$ and $$a_{-}, \,\,b_-
b_+(=q^2\,b_+ b_-),\,
\,b_{-}b_{+}^* \text{ are in }\CR.$$

By construction and Theorem \ref{Theo}, any $T \in \CR$ satisfies
\eqref{ideal} and $\CR$ is invariant by $F$.

Moreover, by \eqref{astuce2}, $[b_{-},b_{-}^*] \in R$, so by
\eqref{astuce1} and \eqref{astuce5},
$a_{+}^*a_{+}-q^2\,a_{+}a_{+}^*-(1-q^2) \in R$ and by
\eqref{astuce6}, $q\,a_{+}b_{-}-b_{-}a_{+} \in R$.

It is interesting to quote, thanks to Theorem \ref{Theo}
that if $x\in R$, then $\ncint F\,x\,\vert\DD\vert^{-1}=0$ while a
priori, $\ncint x\, \vert \DD \vert^{-1}\neq 0$.

Note that $F \in \Psi^0(\A)$ also satisfies \eqref{ideal} by
Theorem \ref{Theo} while $F \notin \CR$ since $F^2=1$.

Moreover other elements are in $\CR$ like for instance
$d(b^*b)= d(bb^*)$:
\begin{align*}
\delta(bb^*)&=-\delta (aa^*)=-\delta a\,a^*-a\,\delta
a^*=-(a_{+}-a_{-})(a_{+}^*+a_{-}^*)
-(a_{+}+a_{-})(a_{-}^*-a_{+}^*)\\
&=2(a_{+}a_{-}^*-a_{-}a_{+}^*)
\end{align*}
is in $R$ since $a_{-}\in R$  yielding $d(bb^*)\in R\,F$.

We do not know if $\CR$ is equal to the subset of the algebra
generated by $\B$ and $\B\, F$ satisfying \eqref{ideal}.

\begin{lemma}
$\CR$ is a $*$-ideal in $\Psi^0(\A)$ which is invariant by
$F$, $d$, $\delta$.
\end{lemma}

\begin{proof}
Since $R$ is an ideal in $X=\B$ mod $OP^{-\infty}$ (see Remark
\ref{pseudodiff}), $\CR$ appears to be an ideal in $\Psi^0(\A)
\subset \text{algebra generated by } \B \text{ and } \B\,F$.
Since $\CR$ is invariant by $F$, its invariance by $d$ follows from
its invariance by $\delta$ which is true on the generators of $R$.
\end{proof}

Note that, according to Theorem \ref{ncintLin},
$\ncint da \,|D|^{-2}= \ncint da \,|D|^{-3}=0$ while
$\ncint a^*da \,|D|^{-3}=2$ which emphasize the role of
"for all $t$" in \eqref{ideal}.

\begin{lemma}
For any $t \in \Psi_{0}^0(\A)$ and $T \in \CR$, we have 
$\ncint t \,T \,|D|^{-1} = \ncint T \,t \,|D|^{-1}$.
\end{lemma}

\begin{proof}
For any $t\in \B$, we have
 $\ncint T \,t \,|D|^{-1} = \ncint t \,T \,|D|^{-1} +
\ncint T \,|D|^{-1} \,\delta(t) \,|D|^{-1}$ and moreover
$\ncint T \,|D|^{-1}\, \delta(t) \, |D|^{-1} = \ncint T \, \delta(t)
\,|D|^{-2}
- \ncint T \, \delta^2(t) \,|D|^{-3}$ which comes from
\begin{align*}
|D|^{-1}\delta(t)|D|^{-1}&=\delta(t)|D|^{-2}
+[|D|^{-1},\delta(t)]|D|^{-1}
=\delta(t)|D|^{-2}-|D|^{-1}\delta^2(t)|D|^{-2}\\
&=\delta(t)|D|^{-2}-\delta^2(t)|D|^{-3}+|D|^{-1}\delta^3(t)|D|^{-3}.
\end{align*}
So we get the result because $T$ satisfies \eqref{ideal}.
\end{proof}

\begin{lemma}
    \label{table:commrule}
If $\simeq$ means equality up to the ideal $\CR$, the following rules
with $d(.)=[\DD,.]$ of the first-order differential
calculus hold (forgetting $\piappr$)
\begin{center}
\begin{tabular}{l l l l}
 $a \,da \simeq da \,a$,
& $a^* \,da \simeq - da^* \,a$,
& $b \,da   \simeq q \, da \,b$,
& $b^* \,da \simeq q \,da \,b^*$, \\
 $a\, da^*\simeq - da\,a^*$,
& $a^*\,da^* \simeq da^*\,a^*$,
& $b\,da^* \simeq q^{-1}\,da^*\,b$,
& $b^*\, da^* \simeq q^{-1}\,da^*\,b^*$,\\
 $a \,db \simeq q^{-1} \,db \,a$,
& $a^* \,db \simeq q \,db \,a^*$,
& $b \,db \simeq db \,b$,
& $b^* \,db \simeq db \, b^*\simeq -b\,db^*$, \\
 $a\,db^* \simeq q^{-1}\,db^* \, a$,
& $a^*\, db^* \simeq q\,db^*\, a^*$,
& $b\,db^* \simeq db^*\,b \simeq -b^*\,db$,
& $b^* \, db^* \simeq db^* \, b^*$.
\end{tabular}
\end{center}
Moreover
\begin{align}
    \label{eq:F}
& a^* \,da - q^2 da \, a^* \simeq (1 - q^2) \, F  ,
& q^2 \,a \,da^* - da^* \,a \simeq (1 - q^2) \, F.
\end{align}
\end{lemma}

\begin{proof}
The table follows from relations
\eqref{defrule} and Lemma \ref{commutateur} with \eqref{Fcommutes}
(one can also use \eqref{eq:commutrulefora,b}.)
For instance, since $a_{-}\in \CR$, using the fact that $\CR$ is
invariant by $F$,
\begin{align*}
b\, da &=(b_{+}+b_{-})(a_{+}-a_{-})\,F\simeq
(b_{+}+b_{-})(a_{+}+a_{-})\,F=ba\,F=q\,ab \,F
\simeq q  \,(a_{+}-a_{-})\,F\,b\\
&=q \,da \,b
\end{align*}
or similarly, $a^*\,da=(a_{+}^*+a_{-}^*)(a_{+}-a_{-})F\simeq
(a_{+}^*-a_{-}^*)(a_{+}+a_{-})F=-da^*\,a$.

The second of equivalence of \eqref{eq:F}
is just the adjoint of the first one that we prove now:
\begin{align*}
a^* \,da - q^2 da \, a^* &=(a_{+}^*+a_{-}^*)(a_{+}-a_{-})F-q^{2}
\,(a_{+}-a_{-})F(a_{+}^*+a_{-}^*) \\
&\simeq
(a_{+}^*+a_{-}^*)(a_{+}+a_{-})F-q^{2}\,(a_{+}+a_{-})
(a_{+}^*+a_{-}^*)F= (a^{*}a-q^2\,aa^{*})\,F\\
&=(1-q^{2})\,F.
\tag*{\qed}
\end{align*}
\hideqed
\end{proof}

\begin{remark}
The above written rules remain valid if $dx$ is replaced
by $\delta(x)$ and $F$ by $1$.
\end{remark}

Working modulo $\CR$ simplifies the writing of a one-form:
\begin{lemma}
    \label{1formmodR}
(i) Every one-form $A$ can be, up to elements from $\CR$, presented as
$$
A \simeq x_a \, da + da^* \, x_{a^*} + x_b\, db + db^* \,
x_{b^*},
$$
where all $x_*$ are the elements of $\A$.

(ii) When $A$ is selfadjoint, $A$ can be
written up to $\CR$ (not in a unique way, though) as
$$
A \simeq x_a \,da - da^* \,(x_a)^* + x_b \,db - db^* \,(x_b)^*\, ,
$$
where $x_a,x_b$ are arbitrary elements of $\A$.
\end{lemma}

\begin{proof}
$(i)$ A basis for one-forms consists of the following
forms:
$a^{\alpha} b^{\beta} (b^*)^{\gamma} \,
d \big( a^{\alpha'} b^{\beta'} (b^*)^{\gamma'} \big)$,
where $\alpha, \alpha' \in \Z$ and
$\beta,\gamma,\beta',\gamma' \in \N$.

Using the Leibniz rule and the commutation rules within
the algebra (up to the $\CR$ according to Lemma
\ref{table:commrule}),
we reduce the problem to the case of the forms:
$ \big( a^{\alpha} b^{\beta} (b^*)^{\gamma} \big)\,
dx \,\big( a^{\alpha'} b^{\beta'} (b^*)^{\gamma'} \big)$,
where $x$ can be either of the generators $a,a^*,b,b^*$. If
$x=b$ or $x=b^*$, the straightforward application of the
rules of the differential calculus leads to the answer that
the one-form could be expressed as:
$ a^{\alpha} b^{\beta} (b^*)^{\gamma} \,db$ and
$db^* \,a^{\alpha} b^{\beta} (b^*)^{\gamma}.$

Similar considerations for the case $x=a,a^*$ lead to the
remaining terms.

Note that the presentation is not unique,
since there still might remain terms, which are in $\CR$,
for example: $b^* db + db^* b=d(bb^*)\in \CR$.

$(ii)$ is direct.
\end{proof}

Next we can start explicit calculation of the integrals, beginning
with the tadpole terms. 

Application of the Leibniz rule yields to a presentation of one-forms
which is different from the one of previous lemma.
Each $\delta$-one-form could be expressed, as a finite sum of the
terms
$x \delta(z) y$, where $z$ is one of the generators $a,a^*,b,b^*$
and $x,y$ are some elements of the algebra $\A\big(SU_q(2)\big)$.

\begin{prop}
For all $x,y \in \A\big(SU_q(2)\big)$ and $z\in \{a,a^*,b,b^*\}$ we
have
$$\ncint x \delta(z) y \,|D|^{-1} = \ncint yx \delta(z) \,|D|^{-1} +
\ncint x \delta(z)  \delta(y) \, |D|^{-2}- \ncint
x\delta(z)\delta^2(y) \,|\DD|^{-3}.$$
\end{prop}
\begin{proof}
This is just the application of the trace property of the
noncommutative integral, together with the identity: 
$|D|^{-1} \delta(z) |D|^{-1} = - \left[ |D|^{-1}, z \right]$.
\end{proof}
\begin{remark}
The computation of tadpole-like integrals is reduced to the integrals
$ \ncint x \delta(z)|\DD|^{-1}$  for all generators $z$ and the
integrals of higher order in $|D|^{-2}$. However, the calculations of
higher-order terms simplify a lot after we use the relations which
hold up to the ideal $\CR$: this erases parts of integral depending
on $\vert \DD \vert ^{-2}$ and 
$\vert\DD\vert^{-3}$. Thus, beside $ \ncint x \delta(z)\,|\DD|^{-1}$,
we only need to compute $\ncint x\delta(z)\delta(z') \,|\DD|^{-2}$
where $z$ and $z'$ are generators, since all the $|\DD|^{-3}$
integrals have already been explicitly computed in section 4.6 (these
integrals do not depend on $q$.)

Besides the tadpole, the only integrals that need to be computed are
$\ncint A\,|\DD|^{-2}$ and $\ncint A^2 \,|\DD|^{-2}$ where $A$ is a
$\delta$-1-form. Working modulo $\CR$ and using again Leibniz rule,
we only need to compute $\ncint x\delta(z) \, |\DD|^{-2}$ and  the
previous integrals $\ncint x\delta(z)\delta(z') \, |\DD|^{-2}$.
\end{remark}
\subsubsection{Operators $L_q$ and $M_q$}

In the notation $v^{j}_{l,m}$ of $\H$, we have already use the $j$
dependence in \eqref{J_q} with $J_q \,v^{j}_{m,l} := q^{j}
\,v^{j}_{m,l}$.

Let $L_q$ and $M_q$ be the similar diagonal operators
\begin{align*}
&L_q \,v^{j}_{m,l} := q^{2l} \,v^{j}_{m,l}\,, \\
&M_q \,v^{j}_{m,l}: = q^{2m} \,v^{j}_{m,l}\,.
\end{align*}
We immediately get
\begin{lemma}
\label{lemmaLqMq}
For $n\in \N^*$, $\ncint (L_q)^n \,|\DD^{-2}| = \ncint (M_q)^n
\,|\DD^{-2}| = \frac{2}{1-q^{2n}}\,.$
\end{lemma}

\begin{proof} We have
\begin{align*}
\Tr \big(L_q^n |\DD|^{-2-s} \big)&= \sum_{2j=0}^{\infty}
\sum_{m=0}^{2j} \sum_{l=0}^{2j+1} \langle v^j_{m,l},L_q^n
|\DD|^{-2-s}v^j_{m,l}\rangle \\
&= \sum_{2j=0}^\infty (2j+1)\tfrac{1-q^{2n(2j+2)}}{1-q^{2n}}
d_{j^+}^{-2-s}+ \sum_{2j=0}^\infty
(2j+1)\tfrac{1-q^{2n(2j+2)}}{1-q^{2n}} d_{j}^{-2-s} \\
&\sim_0 \tfrac{1}{1-q^{2n}}\big( \zeta(s+1,\tfrac{3}{2}) +
\zeta(s+1,\tfrac{1}{2} \big) \sim_e \tfrac{2}{1-q^{2n}} \zeta(s+1) \,
.
\end{align*}
where $\sim_0$ means modulo a function holomorphic at 0. This gives
the result for $L_q^n$ and a  similar computation can be done for
$M_q^n$.
\end{proof}

The interest of these operators stems in 
\begin{lemma}
   \label{uptoR}
We have $L_qM_q \in \CR$. Moreover,
\begin{align*}
& b \, \delta b^* \simeq  M_q - L_q, \quad   b^* \delta b \simeq L_q
- M_q,
\quad b b^* \simeq L_q + M_q,\\
& a \,\delta(a^*) \simeq -aa^*\simeq L_q + M_q -1,\quad a^*\, \delta
a \simeq a^*a \simeq 1 - q^2 (L_q + M_q) ,\\
& da\, d a^* \simeq L_q +M_q-1,\quad da^*\,da \simeq q^2(L_q
+M_q)-1,\\ 
& b^{n-2} (b^*)^n \, db \, db \simeq  (L_q)^n + (M_q)^n, \\
& b^{n-1} \,(b^*)^{n-1} \, db \, db^* \simeq - (L_q)^n - (M_q)^n, \\
& b^{n} (b^*)^{n-2} \, db^* \, db^* \simeq (L_q)^n + (M_q)^n.
\end{align*}
\end{lemma}

\begin{proof}
Since $L_qM_q=q^2\,a_-a_-^* \in \CR$, we compute up to the ideal $\CR$
\begin{align*}
b \, \delta b^* & = (b_+ +b_-) (b_-^*-b_+^*) \simeq -b_+b_+^*
+b_-b_-^*=M_q -L_q + L_q M_q(1-q^2)  \simeq  M_q - L_q
\end{align*}
and similarly for the other relations.
\end{proof}

\subsubsection{Automorphisms of the algebra and symmetries of
integrals}
\begin{prop}
\label{calculavecLM}
For any $n \in \N^*$,  
\begin{align*}
& \ncint (bb^*)^n \,|\DD|^{-1} =
\tfrac{-2(1+q^{2n})}{(1-q^{2n})^2}\,, \\
& \ncint (bb^*)^{n} b^* \,\delta b\, |\DD|^{-1} = 
\ncint (bb^*)^{n} b \,\delta b^*\, |\DD|^{-1} =
\tfrac{2}{1-q^{2n+2}}\,, \\
& \ncint (bb^*)^n a \, da^* \,\DD^{-1}  =
\tfrac{-2q^{4n+2}-2q^{4n}-2q^{2n+2}+6q^{2n}}{(1-q^{2n})^2(1-q^{2n+2})}\,,\\
& \ncint (bb^*)^n a^* \, da \,\DD^{-1}  =
\tfrac{6q^{2n+2}-2q^{2n}-2q^2-2}{(1-q^{2n})^2(1-q^{2n+2})}\,.
\end{align*}
\end{prop}
Note that the knowledge of these integral is enough for the
computation of any term of the form $\ncint x\delta(z) |\DD|^{-1}$,
where $z$ is a generator, since any other $\delta$-one-form will be
unbalanced.

To show this proposition, we will use few symmetries, properties of
the ideal $\CR$ and replacement of $\delta$-one-forms in terms of
$L_q,\, M_q$ as above.

Let $U$ be the following unitary operator on the Hilbert space:

\begin{equation*}
U \,v^{j\up}_{m,l} = (-1)^{m+l} \,v^{j^+\dn}_{l,m}, \;\;\;
U \,v^{j\dn}_{m,l} = (-1)^{m+l} \,v^{j^-\up}_{l,m}.
\end{equation*}
Then, by explicit computations we have
\begin{align*}
U^* a U =  a, \qquad U^* a^* U = a^*,  &\qquad U^* b U =  b^*, \qquad
U^* b^* U = b,\qquad \text{and }\quad U^*  \DD U = - \DD. 
\end{align*}

\begin{lemma}
Each noncommutative integral \eqref{I_pA^n} of an element of the
algebra or differential
forms is (up to sign) invariant under the algebra automorphism $\rho$
defined by
\begin{align}
\rho(a):=a, \;\; \rho(a^*):=a^*,\;\; \rho(b):=b^*,\;\; \rho(b^*):=b.
\label{autominv}
\end{align}
\end{lemma}

\begin{proof}
For any homogeneous polynomial $p$ and any $k\in \N$,
\begin{align*}
 \ncint p(a,a^*,b,b^*,\DD) \,\DD^{-k} &= \ncint U^* p(a,a^*,b,b^*,
\DD) \,\DD^{-n} U \\
& = (-1)^k \ncint p(U^* a U, U^* a^* U, U^* b U, U^* b^* U,U^* \DD U)
\,\DD^{-k} \\
& = (-1)^{k+d} \ncint p(\rho(a),\rho(a^*),\rho(b),\rho(b^*),\DD)
\,\DD^{-k},
\end{align*}
where $d$ is the degree of $p$ with respect to $D$.
\end{proof}

\begin{corollary}
\label{cor-01}
For any $n\in \N$,
$ \ncint(bb^*)^{n}\, b^* \,d b\, \DD^{-1}
=  \ncint(bb^*)^{n} \,b \, d b^* \,\DD^{-1}.$
\end{corollary}

\begin{lemma} 
\label{use-l1}
For any $x,y \in \Psi^0(\A)$,
\begin{align*}
& \hspace{-4cm} (i) \quad \, \ncint x y |\DD|^{-1} = \ncint yx
|\DD|^{-1} + \ncint x \delta(y) |\DD|^{-2} - \ncint x \delta^2(y)\,
|\DD|^{-2}.\\
&  \hspace{-4cm}(ii) \quad \ncint z \,x \,\DD^{-1} y\,\DD^{-1}
=\ncint z \,xy \,\DD^{-2}, \text{ where $z \in \A$ contains $b$ or
$b^*$.}
\end{align*}
\end{lemma}

\begin{proof}
$(i)$ is direct consequence of the trace property of $\ncint$ and the
fact that $OP^{-4}$ operators are trace-class.

$ii)$ We calculate:
\begin{align*}
 \ncint z \,x \,\DD^{-1} y \,\DD^{-1} &=
 \ncint z \,x\left( y \, \DD^{-1}
- \DD^{-1} [ \DD, y ] \,\DD^{-1} \right) \DD^{-1}=  \ncint z \,x y
\,\DD^{-2}
  -  \ncint z \,x \DD^{-1} [D, y] \,\DD^{-2} \\
 &= \ncint z \,x y \,\DD^{-2}.
\end{align*}
The last step is based on the observation that any integral with
$\DD^{-3}$ vanishes if the expression integrated contains $b$
or $b^*$.
\end{proof}

\begin{lemma}
For any $n \in \N$,
\begin{align*}
&\hspace{-7.5cm} (i)  \quad \, \ncint(bb^*)^{n} \, b^* \,db
\,\DD^{-1}= \tfrac{2}{1-q^{2n+2}}\,.\\
&\hspace{-7.5cm} (ii)  \quad \ncint (bb^*)^{n} d(bb^*) \,\DD^{-1}=0.\\
& \hspace{-7.5cm} (iii) \! \quad  \ncint (bb^*)^n |D|^{-1} =
\tfrac{-2(1+ q^{2n})}{(1-q^{2n})^2}\,. 
\end{align*}
\end{lemma}

\begin{proof}
$(i)$ With $n>1$, we begin with $\ncint d \left( (b\,b^*)^n \right)
\DD^{-1} = 0, $
which follows directly from the trace property of the noncommutative
integral. Expanding the expression using Leibniz rule and the
commutation
\begin{equation}
x \DD^{-1} = \DD^{-1} x + \DD^{-1} [D,x] \DD^{-1}, \label{bcom}
\end{equation}
we obtain
\begin{align*}
0  =& \sum_{k=0}^{n-1} \ncint  b^k \, db \, b^{n-k-1} (b^*)^n
\,\DD^{-1}
+ \sum_{k=0}^{n-1} \ncint b^n (b^*)^k \, db^* \, (b^*)^{n-k-1}
\,\DD^{-1} \\
=& n \left( \ncint b^{n-1} (b^*)^n \, db \,\DD^{-1} +
           \ncint b^{n} (b^*)^{n-1} \, db^* \,\DD^{-1}  \right) \\
&\qquad + \sum_{k=0}^{n-1} \ncint
\left( b^k \, db \, \DD^{-1} d( b^{n-k-1} (b^*)^n) \DD^{-1}
+ b^n (b^*)^k \, db^* \, \DD^{-1} d ((b^*)^{n-k-1}) \DD^{-1} \right) .
\end{align*}
Using Lemma \ref{use-l1}, 
\begin{align*}
0=& n \ncint(bb^*)^{n-1} ( b^* \,db + b \, db^*) \,\DD^{-1}  \\
& \quad+ \ncint \left( \oh n (n-1) b^{n-2} (b^*)^n \, db \, db
+ n^2 b^{n-1} \,(b^*)^{n-1} \, db \, db^*
+ \oh n (n-1) b^{n} (b^*)^{n-2} \, db^* \, db^* \right) \DD^{-2}.
\end{align*}
The integrals with $\DD^{-2}$ could be easily calculated when
we take restrict ourselves to calculations modulo ideal $\CR$:
\begin{align*}
 n \ncint(bb^*)^{n-1} & ( b^* \,db + b \, db^*) \,\DD^{-1} = 
 - 2 \left( n(n-1) - 2 n^2 + n(n-1) \right) \tfrac{1}{1-q^{2n}}
= 4 n \tfrac{1}{1-q^{2n}}\, .
\end{align*}
Hence $ \ncint(bb^*)^{n-1} ( b^* \,db + b \, db^*) \, \DD^{-1} =
\tfrac{4}{1-q^{2n}}$, 
which together with Corollary \ref{cor-01} proves $i)$.

$(ii)$ In a similar way, 
$\ncint (bb^*)^{n-1} d( bb^*) \, \DD^{-1}= 0 = \ncint(bb^*)^{n-1} d(
aa^*) \, \DD^{-1}$ implies:
\begin{align*}
0 = & \sum_{k=0}^{n-1} (b b^*)^{n-k-1} d(bb^*) (b b^*)^{k}\, \DD^{-1}
\\ 
  =&  n \ncint (bb^*)^{n-1} d(bb^*) \,\DD^{-1}
+ \oh n(n-1) \ncint (bb^*)^{n-2} d(bb^*) \, d(bb^*) \,\DD^{-2}\\
  = & n \ncint (bb^*)^{n-1} d(bb^*) \,\DD^{-1}, 
\end{align*}
where in the last step we used that $d(bb^*) \in \CR$. The identity
$(ii)$ now follows from the equality $aa^* = 1 - bb^*$.

$(iii)$ Using Lemma \ref{use-l1}, we get
\begin{align*}
A_n &:= \ncint (bb^*)^n |D|^{-1} = \ncint (bb^*)^n (aa^* + bb^*)
|D|^{-1} \\
& \hspace{-2.3cm} \text{and we push now $a^*$ through $|D|^{-1}$ and
from cyclicity
of the trace through $(bb^*)^n$,} \\
& \,= A_{n+1} + \ncint (bb^*)^n q^{2n} a^* a \,|D|^{-1} +
\ncint (bb^*)^n q^{2n} a \delta(a^*) \,|D|^{-2}  \\
& \hspace{-2.3cm} \text{the last term being calculated explicitly,
since up to ideal
$\CR$, $a \delta(a^*) \simeq L_q + M_q -1$,} \\
& \, = A_{n+1} (1-q^{2n+2}) + q^{2n} A_n +
4 \left( \tfrac{1}{1-q^{2n+2}} - \tfrac{1}{1-q^{2n}} \right),
\end{align*}
which leads to
\begin{equation*}
A_n (1-q^{2n}) + \tfrac{4}{1-q^{2n}} = A_{n+1} (1-q^{2n+2}) +
\tfrac{4}{1-q^{2n+2}}\,.
\end{equation*}
Assuming $A_n = \tfrac{f_n}{(1-q^{2n})^2}$ we have 
$\frac{f_n + 4}{1-q^{2n}} = \tfrac{f_{n+1}+4}{1-q^{2n+2}}\,,$ 
and taking into account that $A_0 = - 2 \tfrac{1+q^2}{(1-q^2)^2}$, 
we obtain
$A_n = - 2 \tfrac{1+ q^{2n}}{(1-q^{2n})^2}\,.$
\end{proof}

Finally, to get Proposition \ref{calculavecLM}, it remains to prove

\begin{lemma}
For $n\geq 1$,
\begin{align*}
\ncint (bb^*)^n a \, da^* \,\DD^{-1} & =
\tfrac{-2q^{4n+2}-2q^{4n}-2q^{2n+2}+6q^{2n}}{(1-q^{2n})^2(1-q^{2n+2})}\,,\\
\ncint (bb^*)^n a^* \, da \,\DD^{-1} & =
\tfrac{6q^{2n+2}-2q^{2n}-2q^2-2}{(1-q^{2n})^2(1-q^{2n+2})}\,.
\end{align*}
\end{lemma}

\begin{proof}
First, using Leibniz rule, (\ref{bcom}) and Lemma \ref{use-l1} we
have (for $n \geq 1$)
$$ \ncint (bb^*)^n a \, da^*\, \DD^{-1} =
- q^{2n} \ncint (bb^*)^n a^* \, da - \ncint (bb^*)^n da \, da^*
\,\DD^{-2}. $$

Further, we use the identity (\ref{eq:F2}):
$$ \ncint (bb^*)^n \left( a^* \, da + q^2 a \, da^*
   + q^2 b\, db^* + q^2 b^* \, db \right)\, \DD^{-1}
   = (1-q^2) \ncint (bb^*)^n \,|D|^{-1}.$$
taking into account that $F\, \DD = |\DD|$.

These equations give together a system of linear equations
\begin{align*}
\ncint (bb^*)^n a \, da^* \,\DD^{-1} + q^{2n} \ncint (bb^*)^n a^* \,
da \,\DD^{-1}
&= -4 \left( \tfrac{1}{1-q^{2n+2}} - \tfrac{1}{1-q^{2n}} \right), \\
q^2 \ncint (bb^*)^n a \, da^* \, \DD^{-1} + \ncint (bb^*)^n a^* \, da
\,\DD^{-1}
&= -2 (1-q^2) \tfrac{1+q^{2n}}{(1-q^{2n})^2} - \tfrac{4
q^2}{1-q^{2n+2}}
\end{align*}
which is solved by the expressions stated in the lemma.
\end{proof}

\subsubsection{The noncommutative integrals at $|D|^{-2}$}

We need to separate this task into two problems. First, we shall
to calculate all integrals $\ncint x\, \delta(z)|D|^{-2}$,
with $x \in \A(SU_q(2))$ and $z$ being one of the generators. The
second
problem is to calculate  $\ncint x \,\delta(y) \, \delta(z)\,
|D|^{-2}$, with
both $y$ and $z$ being the generators $\{a,a^*,b,b^*\}$.

\begin{lemma} The only a priori non-vanishing integrals of the
type $\ncint x \, \delta(z)  \,|D|^{-2}$ are for $n\in \N$:
\begin{align*}
&\ncint (bb^*)^n b^*\delta(b) \,|\DD|^{-2} =\ncint (bb^*)^n
b\delta(b^*) \,|\DD|^{-2}=0 ,\\
& \ncint (bb^*)^n a\delta(a^*) \,|\DD|^{-2} =
\tfrac{4q^{2n}(1-q^2)}{(q^{2n+2}-1)(1-q^{2n})}\, , \quad n>0\\
&\ncint (bb^*)^n a^*\delta(a) \,|\DD|^{-2} =
\tfrac{4(1-q^2)}{(1-q^{2n+2})(1-q^{2n})} \,.
\end{align*}
\end{lemma}
\begin{proof}
Since $a\delta(a^*)\simeq L_q+M_q-1$ and $(bb^*)^n\simeq
L_q^n+M_q^n$, we get 
$$
(bb^*)^n a\delta(a^*) \simeq L_q^{n+1}+M_q^{n+1}-L_q^n-M_q^n
$$
and the second result is obtained from Lemma \ref{lemmaLqMq}. The
other integrals are computed in a similar way.
\end{proof}

\begin{lemma}
The only a priori non-vanishing integrals of the
type $\ncint x \, dy \, dz \,|D|^{-2}$ are for $n\in \N$:
\begin{align*}
&\ncint (bb^*)^{n} \, (b^*)^2 db\, db \,|\DD|^{-2} =
\tfrac{4}{1-q^{2n+4}}, \\
&\ncint (bb^*)^{n} \, db\, db^*  \, |\DD|^{-2} =
\tfrac{4}{1-q^{2n+2}}, \\
&\ncint (bb^*)^{n} \, (a^* b^*) (da \,db)\,|\DD|^{-2} = 0, \\
&\ncint (bb^*)^{n} \, (a b^*) (da^* \,db)\,|\DD|^{-2} = 0, \\
&\ncint (bb^*)^{n} \, (a^* b) (da \,db^*)\,|\DD|^{-2} = 0, \\
&\ncint (bb^*)^{n} \, (a b) (da^* \,db^*)\,|\DD|^{-2} = 0,\\
&\ncint (bb^*)^{n} \, (da \,da^*)\,|\DD|^{-2}
=\tfrac{4(q^{2n+2}-q^{2n})}{(1-q^{2n+2})(1-q^{2n})}\,, \quad n>0\\
&\ncint (bb^*)^{n} \, (da^* \,da) \,|\DD|^{-2} 
=\tfrac{4(q^2-1)}{(1-q^{2n+2})(1-q^{2n})}\,.
\end{align*}
\end{lemma}

\begin{proof}
This follows from Lemma \ref{lemmaLqMq} with the equivalences up to
$\CR$ gathered in Lemma \ref{uptoR}.
\end{proof}

\section{Examples of spectral action}

It is clear from Theorem \ref{mainThmJ} that any one-form of
the form $ada$, $bdb$, $adb$, $a^*db$, etc... do not contribute to
the spectral action. Indeed, only
the balanced parts of one-forms give a possibly nonzero term
in the coefficients.
Let us now give the values of the terms $\ncint A^n\,\vert \DD
\vert^{-p}$ and the full $\zeta_{\DD_{\Abb}}(0)$ for few examples

\begin{table}[ht]
\caption{Values of noncommutative integrals}
\centering 
\begin{tabular}{c c c c c c c c} 
\hline\hline 
\\
$\Abb$ & $\ncint A\,\vert \DD \vert^{-3}$  & $\ncint A^2\,\vert \DD
\vert^{-3}$ & $\ncint A^3\,\vert \DD \vert^{-3}$ & $\ncint A\,\vert
\DD \vert^{-2}$ & $\ncint A^2\,\vert \DD \vert^{-2}$
& $\ncint A\,\vert \DD \vert^{-1}$ & $\zeta_{\DD_{\Abb}}(0)$ \\
[0.5ex] 
\hline 
\\
$a^*da$ & 2 & 2 & 2 & $\tfrac{4q^2}{q^2-1}$ &
$\tfrac{4q^2(q^2+2)}{q^4-1}$ & $\tfrac{3q^2+1}{2(q^2-1)}$ &
$\tfrac{11q^4+36q^2+13}{3(q^4-1)}$ \\
$b^*db$ & 0 & 0 & 0 & 0 & $\tfrac{-4}{q^4-1}$ & $\tfrac{-2}{q^2-1}$ &
$\tfrac{4q^2}{q^4-1}$\\
$ada^*$ & $-2$ & $2$ & $-2$ & $\tfrac{-4}{q^2-1}$ &
$\tfrac{4(2q^2+1)}{q^4-1}$ & $\tfrac{q^2+3}{2(q^2-1)}$ &
$\tfrac{13q^4+36q^2+11}{3(q^4-1)}$  \\
$bdb^*$ & 0 & 0 & 0 & 0 & $\tfrac{-4}{q^4-1}$ & $\tfrac{-2}{q^2-1}$ &
$\tfrac{4q^2}{q^4-1}$\\
\\
\hline 
\end{tabular}
\label{table:nonlin}
\end{table}

1) Clearly the spectral action depends on $q$: for instance,
\begin{align*}
&\SS(\DD_{a^*da},\Phi,\Lambda) \, = \,2\,\Phi_{3}\,\Lambda^{3}
-8\,\Phi_{2}\,\Lambda^{2}+\,\tfrac{q^2+15}{2(1-q^2)}\,\Phi_{1}\,\Lambda^{1}+\,
  \tfrac{11q^4+36q^2 +13}{3(q^4-1)}\,\Phi(0).
\end{align*}
2) Moreover, for $B:=a\,\delta a^*$ and $A:=B+B^*$, we get since $B
\simeq B^*\mod \CR$,
\begin{align}
     \label{symmetrization}
\ncint A^p |\DD|^{-k} = 2^{p} \ncint B^p |\DD|^{-k}, \quad 1\leq
p\leq k\leq 3\, .
\end{align}
Thus the spectral action of the selfadjoint one-form
$\mathbb{A}:=ada^*+(ada^*)^*$ is
\begin{align*}
\SS(\DD_{\mathbb{A}},\Phi,\Lambda) \, = \,2\,\Phi_{3}\,\Lambda^{3}+
16\,\Phi_2\,\Lambda^2+ \tfrac{q^2-33}{2(1-q^2)}\,\Phi_1\,\Lambda^1+ 
  \tfrac{122q^4+168 q^2-2}{3(q^4-1)}\,\Phi(0) .
\end{align*}
3) When $B_n:=(bb^*)^n \,b\, \delta b^*$, then by Lemma
\eqref{uptoR}, $B_n \simeq B_n^*$, so for $A_n:=B_n+B_n^*$, the
equation \eqref{symmetrization} is still valid and $\ncint B_n^p
\,|\DD|^{-k}$ are all zero but $\ncint B_n
\,|\DD|^{-1}=\tfrac{2}{1-q^{2n+2}}$ and $\ncint B_n^2
\,|\DD|^{-2}=\tfrac{4}{1-q^{4n+4}}$, so
\begin{align}
   \label{strange}
\SS(\DD_{\mathbb{A}_n},\Phi,\Lambda) \, =
\,2\,\Phi_{3}\,\Lambda^{3}-\half\, \Phi_1 \Lambda^1 +
\tfrac{8}{1+q^{2n+2}} \,\Phi(0).
\end{align}
Remark that this spectral action still exists as $q \to 1$!

Note however that the symmetrization process \eqref{symmetrization}
is not true in general, for instance if $B:=a \, \delta b$ and
$A:=B+B^*$, then $\ncint A^2
|\DD|^{-1}=\tfrac{8(q^4-q^2-1)}{(1-q^4)^2}$ while $\ncint B^2
|\DD|^{-1}=0$ or $\ncint [B,\,B^*] |\DD|^{-1}= \tfrac{4}{1-q^4}\,$.

4) The spectral action can be also independent of $q$: for instance,
if $\mathbb{A} = \tfrac{1}{1-q^2}\,\xi(\DD)$ is the $q$-dependent
selfadjoint one-form given in \eqref{eq:F3}, then,
$$
\SS(\DD_{\mathbb{A}},\Phi,\Lambda) \, = \,2\,\Phi_{3}\,\Lambda^{3}
-8\,\Phi_{2}\,\Lambda^{2}+\tfrac{15}{2}\, \Phi_1\,
\Lambda^1-\tfrac{13}{3}.
$$

\section{The commutative sphere $\mathbb{S}^3$}

Since $ SU(2) \simeq \mathbb{S}^3$, we get a concrete spinorial
representation of the algebra $\A:=C^{\infty}(\mathbb{S}^3)$ on the
same Hilbert space $\H$ and same Dirac operator $\DD$ with
\eqref{eq:reprexact} where $q=1$ which means that $q$-numbers are
trivial: $[\alpha]=\alpha$. So
\begin{align}
  \label{eq:reprexacq=1t}
\pi(a)\, \kett{j\mu n } &:= \alpha^+_{j\mu n}\,\kett{j^+ \mu^+ n^+}
+\alpha^-_{j\mu n}\,\kett{j^- \mu^+ n^+},\nonumber \\
\pi(b)\, \kett{j\mu n } &:= \beta^+_{j\mu n}\,\kett{j^+ \mu^+ n^-}
+\beta^-_{j\mu n}\,\kett{j^- \mu^+ n^-},\nonumber\\
\pi(a^*)\, \kett{j\mu n } &:= \tilde{\alpha}^+_{j\mu n}\,\kett{j^+
\mu^- n^-}+\tilde{\alpha}^-_{j\mu n}\,\kett{j^- \mu^- n^-},\nonumber
\\
\pi(b^*)\, \kett{j\mu n } &:= \tilde{\beta}^+_{j\mu n}\,\kett{j^+
\mu^- n^+}+\tilde{\beta}^-_{j\mu n}\,\kett{j^- \mu^- n^+}
\end{align}
where
\begin{align*}
\alpha^+_{j\mu n}&:=\sqrt{ j+\mu+1 }
\left( \begin{array}{cc}
\tfrac{ \sqrt{j+n+3/2}}{2j+2} &   0\\
\tfrac{ \sqrt{j-n+1/2}}{(2j+1)(2j+2} & \tfrac{\sqrt{j+n+1/2]}}{2j+1}  
\end{array} \right),\\
\alpha^-_{j\mu n}&:=\sqrt{ j-\mu }
\left( \begin{array}{cc}
\tfrac{ \sqrt{j-n+1/2}}{2j+1} &  -\tfrac{\sqrt{j+n+1/2}}{2j(2j+1)}\\
0 & \tfrac{\sqrt{j-n-1/2}}{2j}  \\
\end{array} \right),\\
\beta^+_{j\mu n}&:=\sqrt{j+\mu+1 }
\left( \begin{array}{cc}
\tfrac{ \sqrt{j-n+3/2}}{2j+2} &   0\\
-\tfrac{ \sqrt{j+n+1/2}}{(2j+1)(2j+2)} & \tfrac{\sqrt{j-n+1/2}}{2j+1}
\end{array} \right),\\
\beta^-_{j\mu n}&:=\sqrt{ j-\mu }
\left( \begin{array}{cc}
-\tfrac{\sqrt{j+n+1/2}}{2j+1} &  - \tfrac{ \sqrt{j-n+1/2}}{2j(2j+1} \\
0 & -\tfrac{\sqrt{j+n-1/2}}{2j}  
\end{array} \right)
\end{align*}
with $\tilde{\alpha}^{\pm}_{j\mu n}:=(\alpha^{\mp}_{j^{\pm}\mu^-
n^-})^*$, 
$\tilde{\beta}^{\pm}_{j\mu n}:=(\beta^{\mp}_{j^{\pm}\mu^- n^+})^*$.

Note that the representation on the vectors $v^j_{m,l}$ is not as
convenient as in \eqref{eq:rpnappr}.

One can check that the generators $\pi(a)$, $\pi(b)$ and their
adjoint commute and that $[x,[\DD',y]]=0$ for any $x,\,y \in \A$.

\subsection{Translation of Dirac operator}

In general the Dirac operator is defined in a more symmetric way than
that we did. So, although not absolutely necessary here,  we define
for the interested reader the unbounded self-adjoint translated
operator $\DD'$ on $\H$ by the constant $\lambda$ as 
$$
\DD':= \DD+\lambda.
$$
For instance, this gives for $\lambda=-\half$ in the case of
$\mathbb{S}^3$, see \cite{Homma}, 
$\DD'\, v^j_{m,l} = (2j+1) \genfrac{(}{)}{0pt}{1}{1 \quad
0}{ \,0 \,\,\, -1}  v^j_{m,l}$ so $v^j_{m,l}$ is an eigenvector of
$\vert \DD' \vert$.

As the following lemma shows, the computation of noncommutative
integrals involving $\DD$ can be reduced to the computation of
certain noncommutative integrals involving $\DD'$:

\begin{lemma}
   \label{translation}
If $\ncint' T := \Res_{s=0} \Tr \big(T |\DD'|^{-s}\big)$, then for
any 1-form $A$ on a spectral triple of dimension n,
\begin{align*}
&\ncint A \,|\DD|^{-(n-2)} =   \ncint' A\,|\DD'|^{-(n-2)} +\lambda\,
(n-2) \ncint' A \,\DD'|\DD'|^{-n} + \lambda^2\, \tfrac{(n-1)(n-2)}{2}
\ncint'A\, |\DD'|^{-n},\\
&\ncint A \,\DD^{-(n-2)} =   \ncint' A\,\DD'^{-(n-2)} +\lambda\,
(n-2) \ncint' A \,\DD'^{-(n-1)} + \lambda^2\, \tfrac{(n-1)(n-2)}{2}
\ncint'A\, \DD'^{-n}.
\end{align*}
\end{lemma}

\begin{proof}
Recall from \cite[Proposition 4.8]{MCC} that for any
pseudodifferential operator $P$,
$$
\ncint P |\DD|^{-r} = \Res_{s=0} \Tr \big(P|\DD|^{-r}|\DD'|^{-s}
\big).
$$
Moreover, by \cite[Lemma 4.3]{MCC}, for any $s\in \C$ and $N\in \N^*$
\begin{equation}
      \label{D-s}
|\DD|^{-s}= |\DD'|^{-s} + \sum_{p=1}^N K_{p,s} \,Y^p\, |\DD'|^{-s}
\mod OP^{-N-1-\Re(s)}
\end{equation}
where 
$Y= \sum_{k=1}^{N}\tfrac{(-1)^{k+1}}{k} (-2\la \DD'+\lambda^2)^k
\DD'^{-2k} \mod OP^{-N-1}$ 
and $K_{p,s}$ are complex numbers that can be explicitly computed.
Precisely, we find
$K_{p,s}= (-\tfrac{s}{2})^p \,V(p)$
where $V(p)$ is the volume of the $p$-simplex. Since the spectral
dimension is n, we work modulo $OP^{-(n+1)}$, and for $s=n-2$, we get
from (\ref{D-s}): 
$|\DD|^{-(n-2)}= |\DD'|^{-(n-2)} +\lambda (n-2)\DD'|\DD'|^{-n} +
\lambda^2 \tfrac{(n-1)(n-2)}{2}|\DD'|^{-n}  \mod OP^{-(n+1)}$.

As a consequence, we have for $P\in OP^0$ (the $OP^0$ spaces are the
same for  $\DD$ or $\DD'$),
\begin{align*}
\ncint P |\DD|^{-(n-2)} = \ncint' P|\DD'|^{-(n-2)} +\lambda (n-2)
\ncint' P \DD'|\DD'|^{-n} + \lambda^2 \tfrac{(n-1)(n-2)}{2} \ncint'P
|\DD'|^{-n}.
\end{align*}
Since $A$ and $AF$ are in $OP^{0}$, we get both formulae.
\end{proof}

\subsection{Tadpole and spectral action on $\mathbb{S}^3$} 
We consider now the commutative spectral triple
$(C^{\infty}(\mathbb{S}^3),\,\H,\, \DD)$. It is 1-summable since 
$\langle \! \langle j\mu n\,s\,|\, [F,\pi(x)] \,\kett{j\mu n\,s}=0$
when $x=a,a^*,b,b^*$ for any $j,\,\mu,\,n,\,s=\up,\dn$.

All integrals of above lemma are zero for $\mathbb{S}^3$:
\begin{prop}
  \label{notadpole}
There is no tadpole of any order on the commutative real spectral
triple $(C^{\infty}(\mathbb{S}^3),\,\H,\, \DD)$. In fact, for any
one-form or $\delta$-one-form $A$, $\ncint A \DD^{-p}=0$ for
$p\in\set{1,2,3}$.
\end{prop}

\begin{proof}
We first want to prove $\ncint A \DD^{-p}=0$ for $p\in\set{1,2,3}$ and
any one-form $A$.  Since the representation is real, that is any
matrix elements of the generators are real, so must be the trace of
$A \DD^{-p}$. Hence $\ncint A \DD^{-p}= \ncint A^* \DD^{-p}$.

The reality operator $J$ introduced in \eqref{DefJ} satisfies, when
$q=1$, the commutative relation $JxJ^{-1}=x^*$ for $x \in \A$.
Thus $ J A J^{-1} = - A^*$, so 
$ \ncint A \DD^{-p} = \ncint J \left( A^* \DD^{-p} \right) J^{-1} 
= - \ncint A^* \DD^{-p}$ and $\ncint A \DD^{-p}=0$. Similar proof for
the $\delta$-form $AF$.
\end{proof}
For any selfadjoint one-form $A$, $\DD_{A}:=\DD+\wt A=\DD$. Thus, the
spectral action for the real spectral triple
$\big(C^{\infty}(\mathbb{S}^3),\,\H,\, \DD \big)$ for $\DD_A$  is
trivialized by
\begin{align}
   \label{spectralactioncomm} 
\SS(\DD_{A},\Phi,\Lambda) \, = \,2\,\Phi_{3}\,\Lambda^{3}
-\tfrac{1}{2}\, \Phi_1\, \Lambda^1+\mathcal{O}(\Lambda^{-1}).
\end{align}
But it is more natural to compare with the spectral action of
$\DD+A$.  This is  obtained respectively from Lemma \ref{ncintD} and
general heat kernel approach \cite{newGilkey}:
\begin{align*}
\SS(\DD + A,\Phi,\Lambda) \, = \,2\,\Phi_{3}\,\Lambda^{3}   + \ncint
\vert \DD+A \vert^{-1} \,  \Phi_1\,
\Lambda^1+\mathcal{O}(\Lambda^{-1}) 
\end{align*}
since all terms of \eqref{formuleaction1} in $\Lambda^{n-k}$ are zero
for $k$ odd and $\zeta_{\DD+A}(0)=0$ when $n$ is odd: as a
verification, $\ncint \vert \DD+A \vert^{-2}$ is zero according to
\cite[Lemma 4.10]{MCC}, Lemmas \ref{ncintD} and  Proposition
\ref{notadpole}. Similarly, $\zeta_{\DD+A}(0)=0$ because in
\eqref{Zeta}, all terms with $k$ odd are zero (same proof as in
Proposition \ref{notadpole}) but for $k$ even, it is not that easy to
show that $\ncint A \DD^{-1} A \DD^{-1}= 0$.

Moreover, the curvature term does not depend on $A$:
\begin{lemma}
\label{commAF}
For any one-form $A$ on a commutative spectral triple of dimension
$n$ based on a compact Riemannian spin$^c$ manifold without boundary,
we have 
\begin{align}
  \label{curvature}
 \ncint |\DD+A|^{-(n-2)}= \ncint \vert \DD\vert^{-(n-2)} .
\end{align}
\end{lemma}

\begin{proof}
Follows from \cite[first formula page 511]{Odysseus}  with
$\rho:=A=A^*$, $N(\rho)=\rho$ (the constraint $J\rho J^{-1}=\pm \rho$
is not used.)

One can also use \cite[Proposition 1.149]{ConnesMarcolli}.
\end{proof}
From \cite[Lemma 4.10]{MCC} 
 $\ncint |\DD+A|^{-(n-2)}= \ncint \vert \DD\vert^{-(n-2)} +
\tfrac{n(n-2)}{4}\, \ncint (AF)^2|\DD|^{-3}+
\tfrac{(n-2)^2}{4}\,\ncint A^2 |\DD|^{-3}$    using $X:=A\DD+\DD
A+A^2$ and $[\vert \DD \vert, A]\in OP^0$, but  again, it is not that
easy to show that the last two terms cancelled:  for instance here,
for $B=b[\DD,b^*]$, we obtain by direct computation (using the
easiest translated Dirac operator $\DD'$)
\begin{align*}
\Tr \big(B^2 \vert \DD'\vert^{-3-s}\big)&=\Tr \big((B^*)^2 \vert
\DD'\vert^{-3-s}\big)=\tfrac12 \Tr \big(BB^* \vert
\DD'\vert^{-3-s}\big)= \tfrac12 \Tr \big(B^*B \vert
\DD'\vert^{-3-s}\big) \\
&=\tfrac 43\sum_{2j \in \N} \tfrac{j+1}{(2j+1)^{2+s}}\,,
\end{align*}
so $\ncint B^2 \vert \DD' \vert^{-3}=\tfrac23$. Similarly, one checks
that $\ncint  (BF)^2 \vert \DD \vert^{-3}= \tfrac 12 \ncint BFB^*F
\vert \DD \vert ^{-3}=-\tfrac 29$. Thus if $A:=B+B^*$, $\ncint A^2
\vert \DD\vert^{-3}=\ncint A^2 \vert \DD'\vert^{-3}=4$ and $\ncint
(AF)^2 \vert \DD \vert^{-3}=-\tfrac 43$ which yields to
\eqref{curvature}.

Thus for any one-form $A$ on the 3-sphere,
\begin{align*}
 \quad\SS (\DD +A, \Phi, \Lambda) \, = \,2\,\Phi_{3}\,\Lambda^{3}   -
\tfrac 12 \,\Phi_1\, \Lambda^1+\mathcal{O}(\Lambda^{-1},A)
\end{align*}
which as \eqref{spectralactioncomm} is not identical to
\eqref{strange}  which contains a nonzero constant term $\Lambda^0$
for $q=1$.

\section{Conclusion}
We computed in this paper the full spectral action on the $SU_q(2)$
spectral triple of \cite{DLSSV} with the reality operator $J$ (notice 
the change of definition for pseudodifferential operators.) 
The dimension spectrum being a finite set, there is only a finite number
of terms in the spectral action expansion. The tadpole hypothesis is
not satisfied on $SU_q(2)$. We saw that that the action depends on
$q$ and the limit $q\to 1$ does not exist automatically. When 
it exists, such limit does not lead to the associated action on the
commutative sphere $\mathbb{S}^3$. The sign $F$ of the Dirac operator
has special properties: first, it commutes modulo $OP^{-\infty}$ 
with elements of the algebra, and second, it can be seen as a
one-form, giving terms independent of $q$ in the spectral action.

Here, we were interested in the computation of the spectral action of
a quantum group. Naturally, it would be interesting to investigate
other related cases like the Podle\'s spheres
\cite{D'AndreaDabrowski, D'ADLV} or the Euclidean quantum spheres
\cite{Landi, Dabrowski}, especially the 4-sphere \cite{D'ADL}.

\end{document}